\documentclass[journal,draftcls,onecolumn,12pt]{IEEEtran}

\usepackage{cite}
\usepackage{graphicx}
\usepackage[cmex10]{amsmath}
\usepackage{amsfonts}
\usepackage{amssymb}

\ifCLASSOPTIONcompsoc
\usepackage[caption=false,font=normalsize,labelfon
t=sf,textfont=sf]{subfig}
\else
\usepackage[caption=false,font=footnotesize]{subfig}
\fi

\usepackage{amsthm}

\usepackage{algorithm}
\usepackage{algorithmic}

\usepackage{tikz}
\usepackage{pgflibraryshapes}
\tikzstyle{block}=[draw opacity=0.7,line width=1.4cm]

\renewcommand{\theequation}{\arabic{section}.\arabic{equation}}
\newtheorem{remark}{Remark}

\newtheorem{theorem}{Theorem}

\newtheorem{lemma}{Lemma}
\newtheorem{corollary}{Corollary}

\theoremstyle{definition}

\newcommand{\defeq}{\mbox {$  \ \stackrel{\Delta}{=} $}}

\def\argmax{\operatornamewithlimits{arg\,max}}

\theoremstyle{remark}

\begin{document}

\title{Jar Decoding: Non-Asymptotic Converse Coding Theorems, Taylor-Type Expansion, and Optimality
\thanks{This work was supported in part
by the Natural Sciences and Engineering Research Council of Canada
under Grant RGPIN203035-11, and by the Canada
Research Chairs Program.}}

\author{En-hui Yang and Jin Meng\thanks{En-hui Yang and Jin Meng are with the Dept.
of Electrical and Computer
Engineering, University of Waterloo, Waterloo, Ontario N2L 3G1,
Canada. Email: ehyang@uwaterloo.ca, j4meng@uwaterloo.ca} }

\date{April 15, 2012}
\maketitle

\newpage

\begin{abstract}
Recently, a new decoding rule called jar decoding was proposed, under which the decoder first forms a set of suitable size, called a jar, consisting of sequences from the channel input alphabet considered to be closely related to the received channel output sequence through the channel, and then takes any codeword from the jar as the estimate of the transmitted codeword; under jar decoding, a non-asymptotic achievable tradeoff between the coding rate and word error probability was also established for any discrete input memoryless channel with discrete or continuous output (DIMC).  Along the path of non-asymptotic analysis, in this paper, it is  further shown that jar decoding is actually optimal up to the second order coding performance by establishing new non-asymptotic converse coding theorems, and determining the (best) coding performance of finite block length for any block length $n$ and word error probability $\epsilon$ up to the second order. Specifically, a new converse proof technique dubbed the outer mirror image of jar is first presented and used to establish new non-asymptotic converse coding theorems for any encoding and decoding scheme. To determine the  coding performance of finite block length for any block length $n$ and error probability $\epsilon$,  a quantity $\delta_{t, n} (\epsilon)$ is then defined to measure the relative magnitude of the error probability $\epsilon$ and block length $n$ with respect to a given channel and an input distribution $t$. By combining the achievability of jar decoding and the new converses,  it is demonstrated that when $\epsilon < 1/2$, the best channel coding rate $R_n (\epsilon)$ given $n$ and $\epsilon$ has a ``Taylor-type expansion'' with respect to $\delta_{t, n} (\epsilon)$, where the first two terms of the expansion are $\max_{t} [ I(t; P) - \delta_{t, n} (\epsilon) ] $, which is equal to $ I(t^*, P) - \delta_{t^*, n} (\epsilon) $ for some optimal distribution $t^*$, and the third order term of the expansion is $O(\delta^2_{t^*, n} (\epsilon)) $ whenever $\delta_{t^*, n} (\epsilon) = \Omega(\sqrt{ \ln n /  n})$, thus implying the optimality of jar decoding up to the second order coding performance.  Finally, based on the Taylor-type expansion and the new converses,  two approximation formulas for $R_n (\epsilon)$  (dubbed ``SO'' and ``NEP'') are provided; they are further evaluated and compared against some of the best bounds known so far, as well as  the normal approximation of $R_n (\epsilon)$ revisited recently in the literature.  It turns out that while the normal approximation is all over the map, i.e. sometime below achievable bounds and sometime above converse bounds, the SO approximation is much more reliable as it is always below converses; in the meantime,  the NEP approximation is  the best among the three and always provides an accurate estimation for $R_n (\epsilon)$. An important implication arising from the Taylor-type expansion of $R_n (\epsilon) $  is that in the practical non-asymptotic regime, the optimal marginal codeword symbol distribution is not necessarily a capacity achieving distribution.
\end{abstract}


\newpage

\section{Introduction}
\label{sec:introduction}
\setcounter{equation}{0}

Recently,  a new decoding rule called jar decoding was proposed in \cite{yang-meng:jardecoding}, \cite{yang-meng:isit2012_jardecoding}, under which the decoder first forms a set of suitable size, called a jar, consisting of sequences from the channel input alphabet considered to be closely related to the received channel output sequence through the channel, and then takes any codeword from the jar as the estimate of the transmitted codeword. It was shown in \cite{yang-meng:jardecoding} and \cite{yang-meng:isit2012_jardecoding} that under jar decoding, for any binary input memoryless channel with discrete or continuous output and with uniform capacity achieving distribution (BIMC), linear codes ${\cal C}_n$ of block length $n$ with rate $R({\cal C}_n)$ and word error probability $P_e ({\cal C}_n)$ exist such that
\begin{equation}
    \label{eq1-1}
    P_e (\mathcal{C}_{n}) \leq \left( \bar{\xi}_H (X|Y,\lambda,n) + \frac{2(1-C_{BE})
      M_{\mathrm{H}} (X|Y,\lambda)}{\sqrt{n} \sigma^3_{\mathrm{H}} (X|Y,\lambda)}\right) e^{- n r_{X|Y} (\delta) }
  \end{equation}
  and
  \begin{equation}
    \label{eq1-2}
    {R}(\mathcal{C}_{n}) \geq C_{\mathrm{BIMC}} - \delta - r_{X|Y}
    (\delta) + \frac{\ln \frac{2 (1-C_{BE})
      M_{\mathrm{H}} (X|Y,\lambda)}{ \sqrt{n} \sigma^3_{\mathrm{H}} (X|Y,\lambda)} }{n}
  \end{equation}
 for any  $\delta \in (0, \Delta^* (X|Y))$,  where $C_{\mathrm{BIMC}}$ is the capacity of  the given BIMC, $\lambda = r'_{X|Y} (\delta)$, and all other quantities are defined later in Sections \ref{sec:non-asympt-conv} and \ref{sec:appr-eval}. Similar achievable results were also established in \cite{yang-meng:jardecoding} for non-linear codes for any  discrete input memoryless channel with discrete or continuous output (DIMC).

The achievability given in \eqref{eq1-1} and \eqref{eq1-2} is quite sharp. It implies \cite{yang-meng:jardecoding}, \cite{yang-meng:isit2012_jardecoding} that for any BIMC, there exist  linear codes
$\mathcal{C}_n$ of block length $n$ such that
\begin{equation}
  \label{eq1-3}
  R(\mathcal{C}_n) \geq C_{\mathrm{BIMC}} - \sigma_{\mathrm{H}} (X|Y)
    \sqrt{\frac{2 \alpha \ln n}{n}} - \left( \alpha + \frac{1}{2} \right) \frac{\ln n}{n} - O
    \left( \frac{\ln \ln n}{n} \right)
\end{equation}
while maintaining the word error probability
\begin{equation}
  \label{eq1-4}
  P_e(\mathcal{C}_n) \leq \frac{n^{-\alpha}}{2 \sqrt{\pi
          \alpha \ln n}} + O \left( n^{-\alpha}
        \frac{\ln n}{\sqrt{n}} \right) = \Theta \left(
        \frac{n^{-\alpha}}{\sqrt{\ln n}} \right)
\end{equation}
and
\begin{equation}
  \label{eq1-5}
  R(\mathcal{C}_n) \geq C_{\mathrm{BIMC}} - \frac{c}{\sqrt{n}}
    - \frac{\ln n}{2 n} + \frac{1}{n} \ln \frac{(1-C_{BE})
      M_{\mathrm{H}} (X|Y)}{\sigma^3_{\mathrm{H}} (X|Y)}
\end{equation}
while maintaining the word error probability
\begin{equation}
  \label{eq1-6}
  P_e(\mathcal{C}_n) \leq Q \left( \frac{c}{ \sigma_{\mathrm{H}} (X|Y) } \right )
    + \frac{M_{\mathrm{H}} (X|Y)}{\sigma^3_{\mathrm{H}} (X|Y)} {1
      \over \sqrt{n}},
\end{equation}
where $\sigma^2_{\mathrm{H}} (X|Y)$ and $M_{\mathrm{H}}
    (X|Y)$ are parameters related to the
    channel and specified in Section~\ref{sec:non-asympt-conv},
    \begin{equation}
      \label{eq1-7}
      Q(z) = {1\over \sqrt{2 \pi}} \int_z^{\infty} e^{-t^2/2} d t,
    \end{equation}
    and $C_{BE} < 1$  is the universal constant in the
    Berry-Esseen central limit theorem. Furthermore, when the error probability is
    maintained constant in \eqref{eq1-6},  the first two terms (i.e.,
    $C_{\mathrm{BIMC}}$ and
    $\frac{c}{\sqrt{n}}$) in \eqref{eq1-5}
 coincide with the asymptotic second order coding rate analysis in \cite{strassen-1962}, \cite{Hayashi-2009}, \cite{Yury-Poor-Verdu-2010}. Consequently, jar decoding is shown to be second order optimal asymptotically when the error probability $\epsilon$ is maintained constant with respect to block length $n$.

In the non-asymptotic regime, however, the concept of constant error probability with respect to block length $n$ is not applicable. For example, suppose that $n= 1000$ and the error probability $\epsilon$ is equal to $10^{-6}$. How would one interpret the relationship between $\epsilon$ an $n$ in this case? Does it make sense to interpret $\epsilon$ as a constant with respect $n$? Or is it better to interpret $\epsilon$ as a polynomial function of $n$, namely, $\epsilon = n^{-2}$? Since $\epsilon$ is pretty small relatively to $n$, we believe that the latter interpretation makes a lot of sense in this particular case. In general, when both the error probability $\epsilon$ and block length $n$ are finite, what really matters is their relative magnitude to each other.  Therefore, it is interesting to see if the achievability in \eqref{eq1-1} and \eqref{eq1-2} remains tight up to the second order in the non-asymptotic regime where both  the error probability $\epsilon$ and block length $n$ are finite.

In this paper, we provide an affirmative answer to the above question. Specifically, we first present a new converse proof technique dubbed the outer mirror image of jar and use the technique to establish new non-asymptotic converse coding theorems for any binary input memoryless symmetric channel with discrete or continuous output (BIMSC) and any DIMC. We then introduce a quantity $\delta_{t,n} (\epsilon)$ to measure the relative magnitude of the error probability $\epsilon$ and block length $n$ with respect to a given channel and an input distribution $t$. By combining the achievability of jar decoding (see \eqref{eq1-1} and \eqref{eq1-2} in the case of BIMSC)  with the new converses,  we further show that when $\epsilon < 1/2$, the best channel coding rate $R_n (\epsilon)$ given $n$ and $\epsilon$ has a ``Taylor-type expansion'' with respect to $\delta_{t, n} (\epsilon)$ in a neighborhood of $\delta_{t, n} (\epsilon) =0$, where the first two terms of the expansion are $\max_{t} [ I(t; P) - \delta_{t, n} (\epsilon) ] $, which is equal to $ I(t^*, P) - \delta_{t^*, n} (\epsilon) $ for some optimal distribution $t^*$, and the third order term of the expansion is $O(\delta^2_{t^*, n} (\epsilon)) $ whenever $\delta_{t^*, n} (\epsilon) = \Omega(\sqrt{ \ln n /  n})$.  Since the leading two terms in the achievability of jar decoding (see \eqref{eq1-2} in the case of BIMSC when $P_e (\mathcal{C}_{n}) = \epsilon$) coincide with the first two terms of this Taylor-type expansion of $R_n (\epsilon)$, jar decoding is indeed optimal up to the second order coding performance in the non-asymptotical regime.

Finally, based on the Taylor-type expansion of $R_n (\epsilon)$ and our new non-asymptotic converses, we also derive two approximation formulas (dubbed ``SO'' and ``NEP'') for $R_n (\epsilon)$ in the non-asymptotic regime. The SO approximation formula consists only of the first two terms in the Taylor-type expansion of $R_n (\epsilon)$. On the other hand, in addition to the first two terms in the Taylor-type expansion of $R_n (\epsilon)$, the NEP approximation formula includes some higher order terms from our non-asymptotic converses as well. (Here,  NEP stands for non-asymptotic equipartition properties established recently in \cite{yang-meng:nep}, and underlies both the achievability bounds in \eqref{eq1-1} and \eqref{eq1-2} and our non-asymptotic converses.) These formulas are further evaluated and compared against some of the best bounds known so far, as well as the normal approximation of $R_n (\epsilon)$ in \cite{Yury-Poor-Verdu-2010}. It turns out that while the normal approximation is all over the map, i.e. sometime below achievability and sometime above converse, the SO approximation is much more reliable as it is always below converses; in the meantime, the NEP approximation is  the best among the three and always provides an accurate estimation for $R_n (\epsilon)$.  An important implication arising from the Taylor-type expansion of $R_n (\epsilon) $  is that in the practical non-asymptotic regime, the optimal marginal codeword symbol distribution is not necessarily a capacity achieving distribution.

The rest of this paper is organized as follows. Non-asymptotic converses and the Taylor-type expansion of $R_n (\epsilon)$ for BIMSC and DIMC are established in Sections \ref{sec:non-asympt-conv} and \ref{sec:non-asysmpt-converse-coding-dsc},  respectively. The SO and NEP approximation  formulas are developed, numerically calculated,  and compared against the normal approximation in Section \ref{sec:appr-eval} for the binary symmetric channel (BSC),  binary erasure channel (BEC),  binary input additive Gaussian channel (BIAGC), and Z-channel. And finally conclusions are drawn in Section \ref{sec:conclusion}.

\section{Non-Asymptotic Converse and Taylor-type Expansion: BIMSC}
\label{sec:non-asympt-conv}
\setcounter{equation}{0}

Consider a BIMC $\{p(y|x): x \in \mathcal{X}, y \in \mathcal{Y}\}$, where $\mathcal{X}=\{0,1\}$ is the channel input alphabet, and $\mathcal{Y}$ is the channel output alphabet, which is arbitrary and could be discrete or continuous. Throughout this section, let $X$ denote the uniform random variable on $\mathcal{X}$ and $Y$ the corresponding channel output of the BIMC in response to $X$. Then the capacity (in nats) of the BIMC is calculated by
\begin{equation}
  \label{eq-non-bimc-1}
  C_{\mathrm{BIMC}} = \ln 2 - H(X|Y)
\end{equation}
where $H(X |Y)$ is the conditional entropy of $X$ given $Y$. Here and throughout the rest of the paper, $\ln$  stands for the logarithm with base $e$, and all information quantities are measured in nats. Further assume that the random variable $ -\ln p(0|Y) $ given $X=0$ and the random variable  $-\ln p(1|Y) $  given $X=1$ have the same distribution, where $p(0|Y)$ ($p(1|Y)$, respectively) denotes the conditional probability of $X=0$ ($X=1$, respectively) given $Y$. Such a BIMC is called a binary input memoryless symmetrical channel (BIMSC). (It can be verified  that BSC, BEC, BIAGC, and general binary input symmetric output channels all belong to the class of BIMSC.) Under this assumption, we have
\begin{equation}
  \label{eq-non-bimc-3}
  \Pr \left\{ \left. -\frac{1}{n} \ln p(X^n|Y^n) > H(X|Y) + \delta
  \right| X^n = x^n \right\} =  \Pr \left\{ -\frac{1}{n} \ln p(X^n|Y^n) > H(X|Y) + \delta
  \right\}
\end{equation}
for any $x^n \in \mathcal{X}^n$, where $Y^n$ is the output of the BIMSC in response to $X^n$, the $n$ independent copies of $X$. Throughout this paper, for any set $S$, we use $S^n$ to denote the set of all sequences of length $n$ drawn from $S$.

\subsection{Definitions}

Before stating our converse channel coding theorem for the BIMSC, let us first introduce some
definitions from \cite{yang-meng:nep}. Define
\begin{equation} \label{eq2-1}
 \lambda^* (X|Y) \defeq \sup
 \left\{\lambda \geq 0: \int p(y) \left [\sum_{x \in \mathcal{X}} p^{-\lambda +1} (x|y)  \right ] d y < \infty \right \}
\end{equation}
where $\int d y $ is understood throughout this paper to be the
summation over $\mathcal{Y}$  if $\mathcal{Y}$ is discrete.
Suppose that
\begin{equation} \label{eq2-1+}
 \lambda^* (X|Y) >0 \;.
\end{equation}
Define for any $\delta \geq 0$
\begin{equation} \label{eq2r}
 r_{X|Y} (\delta) \defeq \sup_{\lambda \geq 0} \left [ \lambda (H(X|Y) + \delta) -
 \ln \sum_{x \in \mathcal{X}} \int p(y) p^{-\lambda +1} (x|y) d y \right ].
\end{equation}
For any $\lambda \in [0, \lambda^* (X|Y))$, let $X _{\lambda}$ and $Y _{\lambda}$ be random variables
under joint distribution $p(x,y)  f_{\lambda} (x, y)$ where
\begin{equation} \label{eq2-12}
  f_{\lambda} (x, y) \defeq {p^{-\lambda} (x|y) \over \sum_{u \in \mathcal{X}} \int p(v) p^{-\lambda +1} (u|v) d v } .
\end{equation}
Further define
 \begin{equation} \label{eq2d}
 \delta(\lambda) \defeq  \mathbf{E} [-\ln p(X_{\lambda} | Y_{\lambda})]
  - H(X|Y) \;
\end{equation}
\begin{equation}
  \label{eq2e}
  \Delta^* (X|Y) \defeq \lim_{\lambda \uparrow \lambda^* (X|Y)}
\delta (\lambda)
\end{equation}
\begin{equation} \label{eq2-13}
  \sigma^2_H (X|Y, \lambda)  \defeq \mathbf{Var} [-\ln p(X_{\lambda} | Y_{\lambda})]
    = \mathbf{E} [\left|-\ln p(X_{\lambda} | Y_{\lambda}) - \mathbf{E} [-\ln p(X_{\lambda} | Y_{\lambda})] \right|^2]
  \end{equation}
\begin{equation} \label{eq2-14}
  M_H (X|Y, \lambda)  \defeq \mathbf{M_3} [-\ln p(X_{\lambda} | Y_{\lambda})]
    = \mathbf{E} [\left|-\ln p(X_{\lambda} | Y_{\lambda}) - \mathbf{E} [-\ln p(X_{\lambda} | Y_{\lambda})] \right|^3]
\end{equation}
and
\begin{equation} \label{eq2-cm-1}
  \hat{M}_H (X|Y, \lambda)  \defeq \mathbf{\hat{M}_3} [-\ln p(X_{\lambda} | Y_{\lambda})]
    = \mathbf{E}\left [-\ln p(X_{\lambda} | Y_{\lambda}) - \mathbf{E} [-\ln p(X_{\lambda} | Y_{\lambda})] \right]^3
 \end{equation}
 where $\mathbf{E} [\cdot]$, $\mathbf{Var}[\cdot]$,  $\mathbf{M_3} [\cdot]$, and $\mathbf{\hat{M}_3} [\cdot]$ are respectively expectation, variance,  third absolute
central moment, and third central moment operators on random variables, and
write $\hat{M}_H (X|Y,0) $ as $\hat{M}_H (X|Y)$, $M_H (X|Y,0) $ as $M_H (X|Y)$,  and $\sigma^2_H (X|Y, 0)$ as $\sigma^2_H (X|Y)$.  Clearly, $\sigma^2_H (X|Y)$, $M_H (X|Y)$, and $\hat{M}_H (X|Y)$ are the variance,  third absolute central moment, and third central moment of $-\ln p(X|Y)$. In particular, $\sigma^2_H (X|Y)$ is referred to as the conditional information variance of $X$ given $Y$ in \cite{yang-meng:nep}. Assume that
\begin{equation}
  \label{eq2-15}
 \sigma^2_H (X|Y) >0 \mbox{ and }  M_H (X|Y) = \mathbf{M_3} [-\ln p(X|Y)]< \infty.
\end{equation}
Then it follows from \cite{yang-meng:nep} that $r_{X|Y} (\delta)$ is strictly increasing, convex, and continuously differentiable up to at least the third order inclusive over $\delta \in [0, \Delta^* (X|Y))$, and furthermore has the following parametric expression
 \begin{equation} \label{eq2p1}
  r_{X|Y} (\delta(\lambda)) =   \lambda (H(X|Y) + \delta (\lambda)) -
 \ln \sum_{x \in \mathcal{X}} \int p(y) p^{-\lambda +1} (x|y) d y
 \end{equation}
with $\delta (\lambda)$ defined in \eqref{eq2d} and $\lambda =
r'_{X|Y} (\delta)$.
In addition, let
\begin{eqnarray}
  \lefteqn{\bar{\xi}_H (X|Y,\lambda,n) \defeq  \frac{2C_{BE} M_H(X|Y,\lambda)}{\sqrt{n} \sigma^3_H (X|Y,\lambda)} } \nonumber \\
    &&{+}\: e^{\frac{n \lambda^2 \sigma^2_H (X|Y,\lambda)}{2}}
                \left[ Q \left( \sqrt{n} \lambda \sigma_H(X|Y,\lambda) \right)
                      - Q \left( \rho^* + \sqrt{n} \lambda \sigma_H(X|Y,\lambda) \right)
                \right] \\
  \lefteqn{\underline{\xi}_H (X|Y,\lambda,n) \defeq e^{\frac{n \lambda^2 \sigma^2_H (X|Y,\lambda)}{2}}
              Q \left( \rho_* + \sqrt{n} \lambda \sigma_H(X|Y,\lambda) \right) }
\end{eqnarray}
with $Q(\rho^*) = \frac{C_{BE} M_H(X|Y,\lambda)}{\sqrt{n} \sigma^3_H (X|Y,\lambda)}$ and
$Q(\rho_*)=\frac{1}{2} - \frac{2 C_{BE} M_H(X|Y,\lambda)}{\sqrt{n} \sigma^3_H (X|Y,\lambda)}$.

The significance of the above quantities related to the channel can be
seen from Theorem 4 in \cite{yang-meng:nep}, summarized as below:
{\em
  \begin{description}
 \item[(a)] There exists a $\delta^* >0$ such that for any $\delta \in (0, \delta^*]$,
 \begin{equation} \label{eq2-4-}
 r_{X|Y} (\delta) = {1 \over 2 \sigma^2_H (X|Y) } \delta^2 +
 O(\delta^3) .
  \end{equation}

  \item[(b)] For any $\delta \in (0, \Delta^* (X|Y))$ and any positive integer $n$
\begin{eqnarray} \label{eq2-17}
  \bar{\xi}_H (X|Y,\lambda,n) e^{- n r_{X|Y} (\delta) }
  &\geq&
  { \Pr \left \{ - {1 \over n} \ln p(X^n |Y^n ) > H(X|Y) +\delta \right \}} \nonumber \\
  &\geq& \underline{\xi}_H (X|Y,\lambda,n)
  e^{- n r_{X|Y} (\delta) },
  \end{eqnarray}
  where $\lambda = r'_{X|Y} (\delta) >0$. Moreover, when $\delta = o(1)$ and $\delta = \Omega (1/\sqrt{n})$,
  \begin{eqnarray}
    \label{eq2-17-1}
    \bar{\xi}_H (X|Y,\lambda,n) &=& e^{\frac{n \lambda^2 \sigma^2_H (X|Y,\lambda)}{2}}
                                                   Q \left( \sqrt{n} \lambda \sigma_H(X|Y,\lambda) \right)\left( 1 + o(1) \right) \\
    \label{eq2-17-2}
    \underline{\xi}_H (X|Y,\lambda,n) &=& e^{\frac{n \lambda^2 \sigma^2_H (X|Y,\lambda)}{2}}
                                                   Q \left( \sqrt{n} \lambda \sigma_H(X|Y,\lambda) \right)\left( 1 - o(1) \right)
  \end{eqnarray}
  and
  \begin{equation}
    \label{eq2-17-3}
    e^{\frac{n \lambda^2 \sigma^2_H (X|Y,\lambda)}{2}}
    Q \left( \sqrt{n} \lambda \sigma_H(X|Y,\lambda) \right) = \Theta \left( \frac{1}{\sqrt{n} \lambda} \right)
  \end{equation}
  with $\lambda = r'_X (\delta) = \Theta (\delta)$.

 \item[(c)] For any  $ \delta \leq c \sqrt{\ln n \over n} $, where $c < \sigma_H (X|Y)$ is a constant,
  \begin{eqnarray} \label{eq2-17+}
    Q  \left ( {\delta \sqrt{n} \over \sigma_H (X|Y)} \right ) - {C_{BE} M_H (X|Y) \over \sqrt{n} \sigma^3_H (X|Y)}
    & \leq  &  \Pr \left \{ - {1 \over n} \ln p(X^n |Y^n ) > H(X|Y) + \delta \right \} \nonumber \\
  &  \leq &  Q  \left ( {\delta \sqrt{n} \over \sigma_H (X|Y)} \right
  ) + {C_{BE} M_H (X|Y) \over \sqrt{n} \sigma^3_H (X|Y)} .
     \end{eqnarray}
  \end{description}
}

Define for any $x^n \in \mathcal{X}^n$,
\begin{equation} \label{eq-def-bxd}
  B(x^n,\delta) \defeq \left\{ y^n: \infty > - \frac{1}{n} \ln
  {p(x^n|y^n)} > H(X|Y) + \delta \right\}
\end{equation}
and
\begin{equation} \label{eq-def-bd}
  B_{n, \delta} \defeq \cup_{x^n \in \mathcal{X}^n} B(x^n,\delta) .
\end{equation}
Since for any $y^n \in \mathcal{Y}^n$, the following set
\begin{equation} \label{eq-jar}
\left\{ x^n \in {\cal X}^n:  - \frac{1}{n} \ln
  {p(x^n|y^n)} \leq H(X|Y) + \delta \right\}
 \end{equation}
is referred to as a BIMC jar for $y^n$ in
\cite{yang-meng:jardecoding}, \cite{yang-meng:isit2012_jardecoding}, we shall call $B(x^n,\delta) $ the {\em outer mirror image of jar} corresponding to $x^n$.
Moreover, define
for any set $B \subseteq \mathcal{Y}^n$,
\begin{equation} \label{eq-def-pb}
  P(B) \defeq \Pr \left\{ Y^n \in B \right\}
\end{equation}
\begin{equation} \label{eq-def-pxb}
  P_{x^n}(B) \defeq \Pr \left\{ Y^n \in B | X^n=x^n  \right\}.
\end{equation}
It is easy to see that
\begin{eqnarray}
  \label{eq-pb}
  P_{x^n} (B(x^n,\delta)) &=&  \Pr \left\{ \left. -\frac{1}{n} \ln p(X^n|Y^n) > H(X|Y) + \delta
  \right| X^n = x^n \right\} \nonumber \\
  &=&  \Pr \left\{ -\frac{1}{n} \ln p(X^n|Y^n) > H(X|Y) + \delta
  \right\}
\end{eqnarray}
where the last equality is due to \eqref{eq-non-bimc-3}.

\subsection{Converse Coding Theorem}

We are now ready to state our non-asymptotic converse coding theorem for BIMSCs.
\begin{theorem}
  \label{thm-bimc}
  Given a BIMSC, for any  channel code $\mathcal{C}_n$ of block length $n$
  with average word error probability $P_e (\mathcal{C}_n) = \epsilon_n$,
  \begin{equation}
    \label{eq-thm-bimc-0}
    R (\mathcal{C}_n) \leq C_{\mathrm{BIMSC}} - \delta
      -
      \frac{\ln \epsilon_n - \ln P(B_{n,\delta}) + {\ln \frac{-2 \ln \epsilon_n }{\sigma^2_H(X|Y) n} }
      - \ln \left( 1 + \frac{\sqrt{\frac{-2 \ln \epsilon_n}{n}} }{\sigma_H (X|Y)} \right)}{n}
  \end{equation}
  where $\delta$ is the largest number such that
  \begin{equation}
    \label{eq-thm-bimc-0+}
    \left( 1 + \frac{2}{\sigma_H(X|Y)} \sqrt{\frac{- 2 \ln \epsilon_n }{n}} \right) \epsilon_n
      \leq \Pr \left\{ -\frac{1}{n} \ln p(X^n|Y^n) > H(X|Y) + \delta
  \right\} .
  \end{equation}
  Moreover, the following hold:
 \begin{enumerate}
  \item
  \begin{equation}
    \label{eq-thm-bimc-1}
    R (\mathcal{C}_n) \leq C_{\mathrm{BIMSC}} - \delta - \frac{\ln
      \epsilon_n - \ln P(B_{n,\delta}) + {\ln \frac{-2 \ln \epsilon_n }{\sigma^2_H(X|Y) n} }
      - \ln \left( 1 + \frac{\sqrt{\frac{-2 \ln \epsilon_n}{n}} }{\sigma_H (X|Y)} \right) }{n}
  \end{equation}
  where $\delta$ is the solution to
  \begin{equation}
    \label{eq-thm-bimc-2}
    \left( 1 + \frac{2}{\sigma_H(X|Y)} \sqrt{\frac{- 2 \ln \epsilon_n }{n}} \right) \epsilon_n
      = \underline{\xi}_H (X|Y,\lambda,n) e^{-n r_{X|Y} (\delta) }
  \end{equation}
  with $\delta(\lambda) = \delta$.
  \item When $\epsilon_n = \frac{e^{-n^{\alpha}}}{2 \sqrt{\pi n^{\alpha}}}
                                         \left( 1 - \frac{1}{2 n^{\alpha}} \right)$ for $\alpha \in (0,1)$,
      \begin{equation}
        \label{eq-thm-bimc-2+}
         R(\mathcal{C}_n) \leq C_{\mathrm{BIMSC}} - \sqrt{2} \sigma_H (X|Y) n^{-\frac{1-\alpha}{2}} + O(n^{-(1-\alpha)}) .
      \end{equation}
  \item When $\epsilon_n = \frac{n^{-\alpha}}{2 \sqrt{\pi \alpha \ln n}}
                                         \left( 1 - \frac{1}{2 \alpha \ln n} \right)$ for $\alpha > 0$,
    \begin{eqnarray}
      \label{eq-thm-bimc-3}
      R(\mathcal{C}_n) &\leq& C_{\mathrm{BIMSC}} - \sigma_{H} (X|Y) \sqrt{\frac{2 \alpha \ln n}{n}}
      + O \left( {\frac{\ln n }{n }} \right) .
    \end{eqnarray}
  \item When $\epsilon_n = \epsilon$ satisfying $\epsilon + \frac{1}{\sqrt{n}} \left( \frac{2 \sqrt{- 2 \ln \epsilon}}{\sigma_H (X|Y)} \epsilon +
            \frac{C_{BE} M_H (X|Y)}{\sigma^3_H (X|Y)} \right) < 1$,
    \begin{eqnarray}
      \label{eq-thm-bimc-6}
      R(\mathcal{C}_n) &\leq& C_{\mathrm{BIMSC}}
       - \frac{ \ln \epsilon + \ln \frac{-2 \ln \epsilon}{\sigma^2_H (X|Y) n}
       - \ln \left( 1 + \frac{\sqrt{\frac{-2 \ln \epsilon}{n}}}{\sigma_H (X|Y)} \right)}{n}
        \nonumber \\
      &&{-}\: \frac{\sigma_{H} (X|Y)}{\sqrt{n}} Q^{-1}
      \left( \epsilon + \frac{1}{\sqrt{n}} \left( \frac{2 \sqrt{- 2 \ln \epsilon}}{\sigma_H (X|Y)} \epsilon +
            \frac{C_{BE} M_H (X|Y)}{\sigma^3_H (X|Y)} \right) \right) \nonumber \\
      \\
      \label{eq-thm-bimc-7}
      &=& C_{\mathrm{BIMSC}} - \frac{\sigma_{H} (X|Y)}{\sqrt{n}} Q^{-1}
      \left( \epsilon \right) + \frac{\ln n}{n}  + O(n^{-1}) .
    \end{eqnarray}
  \end{enumerate}
\end{theorem}

\begin{IEEEproof}
Assume that the message $M$ is uniformly distributed in $\{1,2, \ldots, e^{nR(\mathcal{C}_n)}\}$,
$x^n(m)$ is the codeword corresponding to the
message $m$, and $\epsilon_{m,n}$ is the conditional error probability given message $m$. Then
\begin{equation} \label{eq-proof-bimc--2}
  \epsilon_n = \mathbf{E} [\epsilon_{M,n}].
\end{equation}
Let
\begin{equation} \label{eq-proof-bimc--1}
  \mathcal{M} \defeq \left\{ m: \epsilon_{m,n} \leq \epsilon_n
    (1+\beta_n) \right\} ,
\end{equation}
where $\beta_n > 0$ will be specified later.
By Markov inequality,
\begin{equation}
  \label{eq-proof-bimc-1}
  \Pr \{ M \in {\cal M} \} \geq \frac{\beta_n}{1+\beta_n} \mbox{ and }  |\mathcal{M}| \geq e^{nR(\mathcal{C}_n)
    + \ln \frac{\beta_n}{1+\beta_n} } .
\end{equation}
Denote the decision region for message $m \in {\cal M} $ as $D_m$. Then
\begin{eqnarray}
  \label{eq-proof-bimc-2}
  P_{x^n(m)}( B(x^n(m),\delta) \cap D_m ) &=&
  P_{x^n(m)}( B(x^n(m),\delta) ) - P( B(x^n(m),\delta) \cap D^c_m ) \nonumber \\
  &\geq& P_{x^n(m)}( B(x^n(m), \delta) ) - \epsilon_{m,n} \nonumber \\
  &\geq& P_{x^n(m)}( B(x^n(m), \delta) ) - \epsilon_n(1+\beta_n) \nonumber \\
  &=& \Pr \left\{ -\frac{1}{n} \ln p(X^n|Y^n) > H(X|Y) + \delta
  \right\}  - \epsilon_n(1+\beta_n) \nonumber \\
\end{eqnarray}
where the last equality is due to \eqref{eq-pb}.
At this point, we select $\delta$ such that
\begin{equation}
  \label{eq-proof-bimc-4}
  \Pr \left\{ -\frac{1}{n} \ln p(X^n|Y^n) > H(X|Y) + \delta
  \right\} \geq \epsilon_n (1 + 2 \beta_n).
\end{equation}
Substituting \eqref{eq-proof-bimc-4} into
\eqref{eq-proof-bimc-2}, we have
\begin{equation} \label{eq-proof-bimc-4-1}
  P_{x^n(m)}( B(x^n(m),\delta) \cap D_m ) \geq \beta_n \epsilon_n.
\end{equation}
By the fact that $D_m$ are disjoint for different $m$ and
\begin{equation} \label{eq-proof-bimc-4-2}
  \cup_{m \in \mathcal{M}} (B(x^n(m),\delta) \cap D_m) \subseteq B_{n,\delta},
\end{equation}
we have
\begin{eqnarray}
  \label{eq-proof-bimc-5+}
  P(B_{n,\delta}) &=& \int\limits_{B_{n,\delta}} p (y^n) dy^n \nonumber \\
     &\geq& \sum_{m \in \mathcal{M} } \int\limits_{B(x^n(m),\delta) \cap D_m}
     p (y^n) dy^n \nonumber \\
     &= & \sum_{m \in \mathcal{M}} \int\limits_{B(x^n(m),\delta)
       \cap D_m} \frac{p(y^n|x^n(m)) p(x^n(m))}{p(x^n(m)|y^n)} dy^n
     \nonumber \\
     &\stackrel{1)}{\geq}& \sum_{m \in \mathcal{M} } \int\limits_{B(x^n(m),\delta)
       \cap D_m} p(y^n|x^n(m)) e^{n( -C_{\mathrm{BIMSC}} + \delta)} dy^n \nonumber \\
     &=&  \sum_{m \in \mathcal{M} } e^{n( -C_{\mathrm{BIMSC}} + \delta)} \int\limits_{B(x^n(m),\delta)
       \cap D_m} p(y^n|x^n(m)) dy^n \nonumber \\
      &=&  \sum_{m \in \mathcal{M} } e^{n( -C_{\mathrm{BIMSC}} + \delta)} P_{x^n(m)}(
      B(x^n(m),\delta) \cap D_m ) \nonumber \\
      &\stackrel{2)}{\geq} & \sum_{m \in \mathcal{M} } e^{n( -C_{\mathrm{BIMSC}} + \delta)}
      \beta_n \epsilon_n = |\mathcal{M}| e^{n( -C_{\mathrm{BIMSC}} + \delta)} \beta_n \epsilon_n
\end{eqnarray}
where the inequality 1) is due to the definition of $B(x^n, \delta)$ given in \eqref{eq-def-bxd}, and the inequality 2) follows from \eqref{eq-proof-bimc-4-1}.  From \eqref{eq-proof-bimc-5+}, it follows that
\begin{equation}
  \label{eq-proof-bimc-5}
  |\mathcal{M}| \leq e^{n(C_{\mathrm{BIMSC}} - \delta) - \ln \beta_n -
    \ln \epsilon_n + \ln P(B_{n,\delta})} .
\end{equation}
Then combining
\eqref{eq-proof-bimc-1} and \eqref{eq-proof-bimc-5} yields
\begin{equation}
  \label{eq-proof-bimc-5++}
  R (\mathcal{C}_n) \leq C_{\mathrm{BIMSC}} - \delta - \frac{\ln
    \frac{\beta_n}{1+\beta_n} }{n} - \frac{\ln \beta_n}{n} - \frac{\ln
  \epsilon_n - \ln P(B_{n,\delta})}{n}
\end{equation}
By letting $\beta_n=\frac{1}{\sigma_H (X|Y)} \sqrt{\frac{-2 \ln \epsilon_n}{n}} $, \eqref{eq-thm-bimc-0} and
\eqref{eq-thm-bimc-0+} directly come from \eqref{eq-proof-bimc-4} and
\eqref{eq-proof-bimc-5++}.
\begin{enumerate}
\item
By \eqref{eq2-17} shown in \cite{yang-meng:nep},
selecting $\delta$ to be the solution to
\eqref{eq-thm-bimc-2} will make \eqref{eq-proof-bimc-4} satisfied, and
therefore \eqref{eq-thm-bimc-1} is proved.
\item Towards proving \eqref{eq-thm-bimc-2+}, we want to show that
  by making $\delta = \sqrt{2} \sigma_H (X|Y) n^{-\frac{1-\alpha}{2}} - \eta n^{-(1-\alpha)}$ for some constant $\eta$,
  \begin{equation}
    \label{eq-proof-bimc-10+}
    \Pr \left\{ -\frac{1}{n} p(X^n|Y^n) > H(X|Y) + \delta \right\} \geq
    \left( 1 + \frac{2}{\sigma_H(X|Y)} \sqrt{\frac{-2 \ln \epsilon_n}{n}} \right) \epsilon_n
  \end{equation}
  with $\epsilon_n = \frac{e^{-n^{\alpha}}}{2 \sqrt{\pi n^{\alpha}}} \left( 1 - \frac{1}{2 n^{\alpha}} \right)$.
   Then the proof follows essentially the same approach as that of \eqref{eq-thm-bimc-3}, shown below in details.
\item Apply the trivial bound $P(B_{n,\delta}) \leq 1$.
  Then to show \eqref{eq-thm-bimc-3}, we only have to show that
  $\delta = \sigma_H(X|Y) \sqrt{\frac{2 \alpha \ln n}{n}} - \frac{\eta \ln n}{n} $ for some constant $\eta$
  can make
  \begin{eqnarray}
    \label{eq-proof-bimc-11++}
    \lefteqn{\Pr \left\{ -\frac{1}{n} p(X^n|Y^n) > H(X|Y) + \delta \right\}} \nonumber \\
      &\geq& \underline{\xi}_H (X|Y,\lambda,n)  e^{- n r_{X|Y} (\delta)} \nonumber \\
      &\geq& \left(1 + \eta_0 \sqrt{\frac{\ln n}{n}} \right)
      \frac{n^{-\alpha}}{2 \sqrt{\pi \alpha \ln n}} \left( 1 - \frac{1}{2 \alpha \ln n} \right)
      \nonumber \\
      &\geq& \left( 1 + \frac{2}{\sigma_H(X|Y)} \sqrt{\frac{-2 \ln \epsilon_n}{n}} \right) \epsilon_n
  \end{eqnarray}
  satisfied, where $\lambda = r'_{X|Y} (\delta)$ and
  \begin{equation}
     \frac{2}{\sigma_H(X|Y,\lambda)} \sqrt{\frac{-2 \ln \epsilon_n}{n}} = \Theta \left( \sqrt{\frac{\ln n}{n}} \right) \leq \eta_0 \sqrt{\frac{\ln n}{n}}
  \end{equation}
  for some constant $\eta_0$.
  Towards this, recall \eqref{eq2-4-} \eqref{eq2-17-2} and \eqref{eq2-17-3},
  \begin{eqnarray}
    \label{eq-proof-bimc-11}
    e^{- n r_{X|Y} (\delta)}
    &=& e^{- n r_{X|Y} \left( \sigma_H(X|Y) \sqrt{\frac{2 \alpha \ln n}{n}}  - \frac{\eta \ln n}{n} \right)} \nonumber \\
    &=& e^{- n \left[ \frac{1}{2 \sigma^2_H (X|Y)} \left( \sigma_H(X|Y) \sqrt{\frac{2 \alpha \ln n}{n}}  - \frac{\eta \ln n}{n} \right)^2 +  O \left( \sqrt{\frac{\ln^3 n}{n^3}} \right) \right]}
            \nonumber \\
    &=& e^{ - \alpha \ln n + \frac{\eta}{\sigma_H (X|Y)}\sqrt{\frac{2 \alpha \ln^3 n}{n}} - O \left( \sqrt{\frac{\ln^3 n}{n}} \right) } \nonumber \\
    &\geq& e^{ - \alpha \ln n + \left( \frac{\sqrt{2 \alpha} \eta}{\sigma_H (X|Y)} - \eta_1 \right)  \sqrt{\frac{\ln^3 n}{n}} }
  \end{eqnarray}
  for some constant $\eta_1$, and
  \begin{eqnarray}
    \label{eq-proof-bimc-11+}
    \lefteqn{\underline{\xi}_H (X|Y,\lambda,n)} \nonumber \\
    &=& e^{\frac{n \lambda^2 \sigma^2_{H} (X|Y,\lambda)}{2}} Q \left( \rho_* + \sqrt{n} \lambda \sigma_H (X|Y,\lambda) \right) \nonumber \\
    &\geq& e^{\frac{n \lambda^2 \sigma^2_{H} (X|Y,\lambda)}{2}}
                \frac{e^{-\frac{\left( \rho_* + \sqrt{n} \lambda \sigma_H (X|Y,\lambda) \right)^2}{2}}}{\sqrt{2 \pi} (\rho_* + \sqrt{n} \lambda \sigma_H (X|Y,\lambda) )}
                \left[ 1 - \frac{1}{(\rho_* + \sqrt{n} \lambda \sigma_H (X|Y,\lambda) )^2} \right] \nonumber \\
    &=& \frac{e^{-\frac{ \rho^2_* + 2 \rho_* \sqrt{n} \lambda \sigma_H (X|Y,\lambda) }{2}}}{\sqrt{2 \pi} (\rho_* + \sqrt{n} \lambda \sigma_H (X|Y,\lambda) )}
            \left[ 1 - \frac{1}{(\rho_* + \sqrt{n} \lambda \sigma_H (X|Y,\lambda) )^2} \right]
    \nonumber \\
    &\geq& \frac{1}{2 \sqrt{\pi \alpha \ln n}} \left( 1 - \frac{1}{2 \alpha \ln n} \right) \left( 1 - \Theta \left( \sqrt{\frac{\ln n}{n}} \right) \right)
    \nonumber \\
    &\geq& \frac{1}{2 \sqrt{\pi \alpha \ln n}} \left( 1 - \frac{1}{2 \alpha \ln n} \right) \left( 1 - \eta_2 \sqrt{\frac{ \ln n}{n}} \right)
  \end{eqnarray}
  for another constant $\eta_2$,
  where $\rho_* = Q^{-1} \left( \frac{1}{2} - \frac{2 C_{BE} M_H(X|Y,\lambda)}{\sqrt{n} \sigma^3_H (X|Y,\lambda)} \right)
                        = \Theta \left( \frac{1}{\sqrt{n}}\right)$, and we utilize the fact that
                        \begin{eqnarray}
                          \lambda &=& r'_{X|Y} (\delta) \nonumber \\
                                       &=& \frac{\delta}{\sigma^2_H (X|Y)} + O (\delta^2) \\
                          \sigma_H (X|Y,\lambda) &=& \sigma_H (X|Y) \pm O(\lambda) .
                        \end{eqnarray}
  Then \eqref{eq-proof-bimc-11++} is satisfied by choosing a constant $\eta$ such that
   \begin{eqnarray} \label{eq-proof-bimc-11-1}
   \lefteqn{  e^{ \left( \frac{\sqrt{2 \alpha} \eta}{\sigma_H (X|Y)} - \eta_1 \right)  \sqrt{\frac{\ln^3 n}{n}}}
    \left( 1 - \eta_2 \sqrt{\frac{ \ln n}{n}} \right)} \nonumber \\
    &\geq&   \left [ 1 +  \left( \frac{\sqrt{2 \alpha} \eta}{\sigma_H (X|Y)} - \eta_1 \right)  \sqrt{\frac{\ln^3 n}{n}} \right ]  \left( 1 - \eta_2 \sqrt{\frac{ \ln n}{n}} \right)
    \nonumber \\
    &\geq& 1 + \eta_0 \sqrt{\frac{ \ln n}{n}}
   \end{eqnarray}
   for some constants $\eta_0$, $\eta_1$ and $\eta_2$.
\item According to
  \eqref{eq-proof-bimc-4}, we should select $\delta$ such that
  \begin{equation}
    \label{eq-proof-bimc-8}
    \Pr \left\{ -\frac{1}{n} \ln p(X^n|Y^n) > H(X|Y) + \delta \right\}  \geq
      \left(1+\frac{2}{\sigma_H(X|Y)} \sqrt{\frac{-2 \ln \epsilon}{n}} \right) \epsilon.
  \end{equation}
  Then by \eqref{eq2-17+},
  \begin{equation}
    \label{eq-proof-bimc-9}
    \delta = \frac{\sigma_{H} (X|Y)}{\sqrt{n}} Q^{-1}
      \left( \epsilon + \frac{1}{\sqrt{n}} \left( \frac{2 \sqrt{- 2 \ln \epsilon}}{\sigma_H (X|Y)} \epsilon +
            \frac{M_H (X|Y)}{\sigma^3_H (X|Y)} \right) \right)
  \end{equation}
  will guarantee \eqref{eq-proof-bimc-8}.
  Consequently,
  \eqref{eq-thm-bimc-6} is proved by substituting \eqref{eq-proof-bimc-9} and $\epsilon_n = \epsilon$ into \eqref{eq-proof-bimc-5++}
  and applying the trivial bound $P(B_{n,\delta}) \leq 1$, and
  \eqref{eq-thm-bimc-7} follows the fact that
  \begin{equation} \label{eq-proof-bimc-12}
    Q^{-1}
      \left( \epsilon + \frac{1}{\sqrt{n}} \left( \frac{2 \sqrt{- 2 \ln \epsilon}}{\sigma_H (X|Y)}  \epsilon +
            \frac{C_{BE} M_H (X|Y)}{\sigma^3_H (X|Y)} \right) \right)
    = Q^{-1} (\epsilon) - O \left( \frac{1}{\sqrt{n}} \right).
  \end{equation}
\end{enumerate}
\end{IEEEproof}

\begin{remark} \label{re1}
It is clear that the above converse proof technique depends heavily on the concept of the outer mirror image of jar corresponding to codewords. To facilitate its future reference, it is beneficial to loosely call such a converse proof technique the outer mirror image of jar.
\end{remark}

\begin{remark} \label{re2}
  In general, the evaluation of $P(B_{n,\delta})$ may not be feasible,
  in which case the trivial bound $P(B_{n,\delta}) \leq 1$ can be
  applied without affecting the second order performance in the non-exponential
error probability regime, as shown above.  However, there are cases where $P(B_{n,\delta})$
  can be tightly bounded (e.g. BEC, shown in section
  \ref{sec:appr-eval}).
\end{remark}

\begin{remark} \label{re3}
  For the bound \eqref{eq-thm-bimc-6}, when $\epsilon$ is small with respect to
  $\frac{1}{\sqrt{n}}$,  $\frac{C_{BE} M_H(X|Y)}{\sqrt{n} \sigma^3_H(X|Y)}$ (the
  estimation error that comes from Berry-Esseen central limit theorem)
  will be dominant; in this case,  \eqref{eq-thm-bimc-6} is loose.
\end{remark}

\begin{remark} \label{re3+}
  The choice $\beta_n = \frac{1}{\sigma_H (X|Y)} \sqrt{\frac{-2 \ln \epsilon_n}{n}}$
  in the proof of Theorem \ref{thm-bimc} is not arbitrary.
  Actually, it is  optimal when $\delta$ is small in the sense of minimizing the  upper bound \eqref{eq-proof-bimc-5++}
  in which $\delta$ depends on $\beta_n$ through \eqref{eq-proof-bimc-4}. To derive the expression for $\beta_n$, the following
  approximations can be adopted when $\delta$ is small:
  \begin{eqnarray}
    \label{eq-re3+-1}
    \frac{d \delta}{d \beta_n} &\approx& - \frac{2 \beta_n \sigma^2_H (X|Y)}{n \delta} \\
    \label{eq-re3+-2}
    \delta^2 &\approx& \frac{-2\sigma^2_H (X|Y) \ln \epsilon_n}{n} \\
    \ln \frac{\beta_n}{1 + \beta_n} &\approx& \ln \beta_n
  \end{eqnarray}
  where \eqref{eq-re3+-1} and \eqref{eq-re3+-2} can be developed from
  \eqref{eq2-4-} and \eqref{eq2-17}.
\end{remark}

By reviewing the proof of Theorem \ref{thm-bimc}, it is not hard to
reach the following corollary.
\begin{corollary}
  \label{col-bimc}
  Given a BIMSC,   for any channel code  $\mathcal{C}_n$ of block length $n$ with maximum error probability
  $P_m (\mathcal{C}_n) = \epsilon_n$,
  \begin{equation}
    \label{eq-thm-maxbimc-3}
    R (\mathcal{C}_n) \leq C_{\mathrm{BIMSC}} - \delta - \frac{\ln
      \epsilon_n + \ln  \frac{1}{\sigma_H(X|Y)} \sqrt{\frac{-2 \ln \epsilon_n}{n}}  - \ln P(B_{n,\delta})}{n}
  \end{equation}
  where $\delta$ is the largest number such that
  \begin{equation}
    \label{eq-thm-maxbimc-4}
    \left( 1 + \frac{1}{\sigma_H(X|Y)} \sqrt{\frac{-2 \ln \epsilon_n}{n}} \right) \epsilon_n
    \leq \Pr \left\{ -\frac{1}{n} \ln p(X^n|Y^n) > H(X|Y) + \delta \right\}.
  \end{equation}
  Moreover, the following hold:
  \begin{enumerate}
  \item
  \begin{equation}
    \label{eq-thm-maxbimc-1}
    R (\mathcal{C}_n) \leq C_{\mathrm{BIMSC}} - \delta - \frac{\ln
      \epsilon_n + \ln \frac{1}{\sigma_H(X|Y)} \sqrt{\frac{-2 \ln \epsilon_n}{n}}  - \ln P(B_{n,\delta})}{n}
  \end{equation}
  where $\delta$ is the solution to
  \begin{equation}
    \label{eq-thm-maxbimc-2}
    \left( 1+ \frac{1}{\sigma_H(X|Y)} \sqrt{\frac{- 2 \ln \epsilon_n}{n}} \right) \epsilon_n = \underline{\xi}_H (X|Y,\lambda,n) e^{-n
      r_{X|Y} (\delta) }
  \end{equation}
  with $\delta (\lambda) = \delta$.
  \item
    When $\epsilon_n=\epsilon$ satisfying $\epsilon + \frac{1}{\sqrt{n}} \left( \frac{\sqrt{-2 \ln \epsilon}}{\sigma_H(X|Y)} \epsilon +
            \frac{C_{BE} M_H (X|Y)}{\sigma^3_H (X|Y)} \right) <1$,
    \begin{eqnarray}
      \label{eq-thm-maxbimc-6}
      R(\mathcal{C}_n) &\leq& C_{\mathrm{BIMSC}} - \frac{\ln \epsilon + \ln \frac{1}{\sigma_H(X|Y)} \sqrt{\frac{-2 \ln \epsilon}{n}} }{n}
        \nonumber \\
      &&{-}\: \frac{\sigma_{H} (X|Y)}{\sqrt{n}} Q^{-1}
      \left( \epsilon + \frac{1}{\sqrt{n}} \left( \frac{\sqrt{-2 \ln \epsilon}}{\sigma_H(X|Y)} \epsilon +
            \frac{C_{BE} M_H (X|Y)}{\sigma^3_H (X|Y)} \right) \right) \\
      &=& C_{\mathrm{BIMSC}} - \frac{\sigma_{H} (X|Y)}{\sqrt{n}} Q^{-1}
      \left( \epsilon \right) + \frac{\ln n}{2 n} + O(n^{-1})
    \end{eqnarray}
  \end{enumerate}
\end{corollary}

Remarks \ref{re2}, \ref{re3} and \ref{re3+} also apply to Corollary \ref{col-bimc}.

\subsection{Taylor-type Expansion}

Fix a BIMSC. For any block length $n$ and average error probability $\epsilon$, let $R_n (\epsilon)$ be the best coding rate achievable with block length $n$ and average error probability $\leq \epsilon$, i.e.,
\begin{equation} \label{eq-bt-1}
 R_n (\epsilon) \defeq \max \{ R({\cal C}_n): {\cal C}_n \mbox{ is a channel code of block length $n$ with } P_e ({\cal C}_n) \leq \epsilon \} .
\end{equation}
In this subsection, we combine the non-asymptotic achievability given in \eqref{eq1-1} \eqref{eq1-2} with the non-asymptotic converses given in \eqref{eq-thm-bimc-0} to
\eqref{eq-thm-bimc-2}  to derive a Taylor-type expansion of $R_n (\epsilon)$ in the non-asymptotic regime where both $n$ and $\epsilon$ are finite. As mentioned early, when  both $n$ and $\epsilon$ are finite, what really matters is the relative magnitude of $\epsilon$ and $n$. As such, we begin with introducing a quantity $\delta_n (\epsilon)$ to measure the relative magnitude of $\epsilon$ and $n$ with respect to the given BIMSC.

A close look at the non-asymptotic achievability given in \eqref{eq1-1} \eqref{eq1-2} and the non-asymptotic converses given in \eqref{eq-thm-bimc-0} to
\eqref{eq-thm-bimc-2} reveals that
\begin{displaymath}
  \Pr \left\{ -\frac{1}{n} \ln p(X^n|Y^n) > H(X|Y) + \delta \right\}
\end{displaymath}
is crucial in both cases. According to \eqref{eq2-17-1} and \eqref{eq2-17-2},
\begin{eqnarray}
  \label{eq-so-1}
  \Pr \left\{ -\frac{1}{n} \ln p(X^n|Y^n) > H(X|Y) + \delta \right\} &\approx&
  e^{\frac{n \lambda^2 \sigma^2_H(X|Y,\lambda)}{2}} Q \left( \sqrt{n} \lambda \sigma_H(X|Y,\lambda) \right)
  e^{- n r_{X|Y} (\delta)} \nonumber \\
  &\defeq& g_{X|Y,n} (\delta)
\end{eqnarray}
where $\lambda = r'_{X|Y} (\delta)$. Consequently, we would like to define $\delta_n(\epsilon)$ as
the solution to
\begin{equation}
  \label{eq-so-2}
  g_{X|Y,n} (\delta)= \epsilon
\end{equation}
given $n$ and $\epsilon \leq 1/2$, where the uniqueness of the solution in certain range is shown in Lemma~\ref{le1}.
\begin{lemma} \label{le1}
  There exists $\delta^+ > 0$ such that for any $n >0$,
  $g_{X|Y,n} (\delta)$ is a strictly decreasing function of $\delta$ over $\delta \in [0,\delta^+]$.
\end{lemma}

\begin{IEEEproof}
  Since $\lambda = r'_{X|Y}(\delta)$, it follows from \eqref{eq2d} and  \eqref{eq2p1} that   $g_{X|Y,n} (\delta) = g_{X|Y,n} (\delta (\lambda))$
  is a function of $\lambda$ through $\delta = \delta ( \lambda )$. (For details about the properties of $\delta ( \lambda )$ and  $r_{X|Y}(\delta)$, please see \cite{yang-meng:nep}.) Moreover, by the fact that $\delta (0) = 0$ and
  $\delta (\lambda)$ is a strictly increasing function of $\lambda$,
  the proof of this lemma is yielded by analyzing the derivative of $g_{X|Y,n} (\delta (\lambda))$ with respect to $\lambda$
  around  $\lambda = 0$. Towards this,
  \begin{eqnarray}
    \label{eq-so-4}
    \lefteqn{\frac{d g_{X|Y,n} (\delta (\lambda))}{d \lambda}} \nonumber \\
    &=& \frac{d}{d \lambda}
            \left( e^{\frac{n \lambda^2 \sigma^2_H(X|Y,\lambda)}{2}} Q \left( \sqrt{n} \lambda \sigma_H(X|Y,\lambda) \right) \right)
            e^{- n r_{X|Y} (\delta(\lambda))} \nonumber \\
     &&{-}\: e^{\frac{n \lambda^2 \sigma^2_H(X|Y,\lambda)}{2}} Q \left( \sqrt{n} \lambda \sigma_H(X|Y,\lambda) \right)
            e^{- n r_{X|Y} (\delta(\lambda))} \frac{d}{d \lambda} \left( n r_{X|Y} (\delta (\lambda)) \right) \nonumber \\
    &=& e^{- n r_{X|Y} (\delta(\lambda))}
            \left\{
               \left[ x e^{\frac{x^2}{2}}
                       Q(x) - \frac{1}{\sqrt{2 \pi}} \right] \frac{d x}{d \lambda}
                       -e^{\frac{x^2}{2}} Q(x) n \left. \frac{d r_{X|Y} (\delta)}{d \delta} \right|_{\delta = \delta (\lambda)}
                      \frac{d \delta (\lambda)}{d \lambda}
            \right\}
  \end{eqnarray}
  where $x = \sqrt{n} \lambda \sigma_H(X|Y,\lambda)$.
  On one hand,
  \begin{eqnarray}
    \label{eq-so-5}
    \frac{d x}{d \lambda} &=&
    \sqrt{n} \left(\sigma_H(X|Y,\lambda) + \lambda \frac{d \sigma_H(X|Y,\lambda)}{d \lambda} \right) \nonumber \\
    &=&  \sqrt{n} \left(\sigma_H(X|Y,\lambda) + \frac{\lambda}{2 \sigma_H(X|Y,\lambda)} \frac{d \sigma^2_H(X|Y,\lambda)}{d \lambda} \right) .
  \end{eqnarray}
  On the other hand,
  \begin{eqnarray}
    \label{eq-so-6}
    \left. \frac{d r_{X|Y} (\delta)}{d \delta} \right|_{\delta = \delta (\lambda)}  &=& \lambda \\
    \label{eq-so-7}
    \frac{d \delta (\lambda)}{d \lambda} &=& \sigma^2_H (X|Y,\lambda)
  \end{eqnarray}
  which further implies
  \begin{eqnarray}
    \label{eq-so-8}
        e^{\frac{x^2}{2}} Q(x) n \left. \frac{d r_{X|Y} (\delta)}{d \delta} \right|_{\delta = \delta (\lambda)}
                      \frac{d \delta (\lambda)}{d \lambda} &=& e^{\frac{x^2}{2}} Q(x) n \lambda \sigma^2_H (X|Y,\lambda)
        \nonumber \\
        &=& \sqrt{n} \sigma_H(X|Y,\lambda) x e^{\frac{x^2}{2}} Q(x).
  \end{eqnarray}
  Substituting \eqref{eq-so-5} and \eqref{eq-so-8} into \eqref{eq-so-4}, we have
  \begin{eqnarray}
    \label{eq-so-9}
    \lefteqn{\frac{d g_{X|Y,n} (\delta (\lambda))}{d \lambda}} \nonumber \\
    &=& e^{- n r_{X|Y} (\delta(\lambda))}
            \left\{
               \left[ x e^{\frac{x^2}{2}} Q(x)
                       - \frac{1}{\sqrt{2 \pi}} \right]
                      \left( \frac{ \sqrt{n} \lambda \frac{d \sigma^2_H(X|Y,\lambda)}{d \lambda} }{2 \sigma_H(X|Y,\lambda)} \right)
                       - \frac{\sqrt{n} \sigma_H(X|Y,\lambda)}{\sqrt{2 \pi}}
            \right\} \nonumber \\
    &=& e^{- n r_{X|Y} (\delta(\lambda))}
           \frac{\sqrt{n} \sigma_H(X|Y,\lambda)}{\sqrt{2 \pi}}
           \left\{
               \left[ \sqrt{2 \pi} x e^{\frac{x^2}{2}} Q(x)
                       - 1 \right]
                      \left( \frac{ \lambda \frac{d \sigma^2_H(X|Y,\lambda)}{d \lambda} }{2 \sigma^2_H(X|Y,\lambda)} \right)
                       - 1
            \right\}.
  \end{eqnarray}
  Note that
  \begin{eqnarray}
    \label{eq-so-10}
    \sqrt{2 \pi} x e^{\frac{x^2}{2}} Q(x)  &<& \sqrt{2 \pi} x e^{\frac{x^2}{2}} \frac{1}{\sqrt{2 \pi} x} e^{-\frac{x^2}{2}} \nonumber \\
    &=& 1.
  \end{eqnarray}
  If $\frac{d \sigma^2_H(X|Y,\lambda)}{d \lambda} \geq 0$, then
  \begin{equation}
    \label{eq-so-11}
    \left[ \sqrt{2 \pi} x e^{\frac{x^2}{2}} Q(x)
                       - 1 \right]
                      \left( \frac{ \lambda \frac{d \sigma^2_H(X|Y,\lambda)}{d \lambda} }{2 \sigma^2_H(X|Y,\lambda)} \right)
    \leq 0,
  \end{equation}
  which further implies that $\frac{d g_{X|Y,n} (\delta (\lambda))}{d \lambda} < 0$. In the meantime, if
  $\frac{d \sigma^2_H(X|Y,\lambda)}{d \lambda} < 0$,
  \begin{eqnarray}
    \label{eq-so-12}
    \lefteqn{\left[ \sqrt{2 \pi} x e^{\frac{x^2}{2}} Q(x) - 1 \right]
                      \left( \frac{ \lambda \frac{d \sigma^2_H(X|Y,\lambda)}{d \lambda} }{2 \sigma^2_H(X|Y,\lambda)} \right)
                       - 1} \nonumber \\
    &<&  \left[ \sqrt{2 \pi} x e^{\frac{x^2}{2}} \frac{x}{\sqrt{2 \pi} (1+x^2) } e^{-\frac{x^2}{2}} - 1 \right]
             \left( \frac{ \lambda \frac{d \sigma^2_H(X|Y,\lambda)}{d \lambda} }{2 \sigma^2_H(X|Y,\lambda)} \right)
                       - 1 \nonumber \\
    &=& - \frac{1}{1+x^2}
           \left( \frac{ \lambda \frac{d \sigma^2_H(X|Y,\lambda)}{d \lambda} }{2 \sigma^2_H(X|Y,\lambda)} \right)
                       - 1 \nonumber \\
    &=& - \frac{ \lambda \frac{d \sigma^2_H(X|Y,\lambda)}{d \lambda} }
                  {2 \sigma^2_H(X|Y,\lambda) \left( 1+ n \lambda^2 \sigma^2_H(X|Y,\lambda) \right)}  - 1.
  \end{eqnarray}
  To continue, let us evaluate $\frac{d \sigma^2_H(X|Y,\lambda)}{d \lambda}$. From \eqref{eq2-12}, \eqref{eq2d},  and \eqref{eq2-13}, it is not hard to verify that   \begin{eqnarray}
    \label{eq-so-13}
    \frac{d \sigma^2_H(X|Y,\lambda)}{d \lambda}
    =\sum_{x \in \mathcal{X}} \int p(x,y) \frac{\partial f_{\lambda} (x,y)}{\partial \lambda} \ln^2 p(x|y) dy
    - 2 \sigma^2_H (X|Y,\lambda) \left( H(X)+\delta(\lambda) \right)
  \end{eqnarray}
  where
  \begin{equation}
    \label{eq-so-14}
     \frac{\partial f_{\lambda} (x,y)}{\partial \lambda} = [- \ln p(x|y) - (H(X|Y) + \delta(\lambda)) ] f_{\lambda} (x,y) .
  \end{equation}
  Plugging \eqref{eq-so-14} into \eqref{eq-so-13} yields
  \begin{eqnarray}
    \label{eq-so-3-}
   \lefteqn{ \frac{d \sigma^2_H(X|Y,\lambda)}{d \lambda}} \nonumber \\
   & = & \mathbf{E}
    \left( - \ln^3 p(X_{\lambda} | Y_{\lambda}) \right) - 3 \sigma^2_H(X|Y,\lambda)  (H(X|Y) + \delta)
            -(H(X|Y) + \delta)^3  \nonumber \\
    & = & \hat{M}_H (X|Y, \lambda)  .
  \end{eqnarray}
  Combining \eqref{eq-so-9}, \eqref{eq-so-11}, \eqref{eq-so-12}, and \eqref{eq-so-3-} together, we have
  \begin{eqnarray}
   \lefteqn{\frac{d g_{X|Y,n} (\delta (\lambda))}{d \lambda}} \nonumber \\
     &\leq & e^{- n r_{X|Y} (\delta(\lambda))}
           \frac{\sqrt{n} \sigma_H(X|Y,\lambda)}{\sqrt{2 \pi}}
           \left (
    \left| - \frac{  \lambda \hat{M}_H (X|Y, \lambda)   }
                  {2 \sigma^2_H(X|Y,\lambda) \left( 1+ n \lambda^2 \sigma^2_H(X|Y,\lambda) \right)} \right| -1 \right ) \label{eq-so-14+} \\
  &  \leq &  e^{- n r_{X|Y} (\delta(\lambda))}
           \frac{\sqrt{n} \sigma_H(X|Y,\lambda)}{\sqrt{2 \pi}}
           \left (
 \left| - \frac{  \lambda \hat{M}_H (X|Y, \lambda)  }
                  {2 \sigma^2_H(X|Y,\lambda) } \right| -1 \right) \label{eq-so-15+}
       \end{eqnarray}
  In view of the continuity of $ \sigma^2_H(X|Y,\lambda)$ and  $\hat{M}_H (X|Y, \lambda) $ as functions of $\lambda$, it is easy to see that there is a $\lambda^+ >0$ such that for any $\lambda \in [0, \lambda^+]$,
  \[ \left| - \frac{  \lambda \hat{M}_H (X|Y, \lambda)   }
                  {2 \sigma^2_H(X|Y,\lambda) } \right| -1  <0 \]
 and hence
  \[   \frac{d g_{X|Y,n} (\delta (\lambda))}{d \lambda} < 0 \]
  for any $n \geq 0$.   This completes the proof of Lemma~\ref{le1} with $\delta^+ = \delta (\lambda^+) $.
\end{IEEEproof}

\begin{remark}
From \eqref{eq-so-14+}, it is clear that when $n$ is large,
\[ \left| - \frac{  \lambda \hat{M}_H (X|Y, \lambda)  }
                  {2 \sigma^2_H(X|Y,\lambda) \left( 1+ n \lambda^2 \sigma^2_H(X|Y,\lambda) \right)} \right| -1 <0 \]
 and hence
  \[   \frac{d g_{X|Y,n} (\delta (\lambda))}{d \lambda} < 0 \]
even for $\lambda \geq \lambda^+$. Nonetheless,
  as can be seen later, we are concerned only with  the case where
  $\delta_n (\epsilon)$ is around $0$.  Consequently,
  the exact value of $\delta^+$ is not important to us.
\end{remark}

\begin{remark}

In view of Lemma~\ref{le1} and the definition of $\delta_n (\epsilon)$  in \eqref{eq-so-1} and
\eqref{eq-so-2}, it follows that $\delta_n ({1\over 2}) =0$ for any $n$ and any BIMSC. However, when $\epsilon < 1/2$, $\delta_n (\epsilon)$ depends not only on $n$ and $\epsilon$, but also on the BIMSC itself through the function $r_{X|Y} (\delta)$.  Given $n$ and $\epsilon < 1/2$, the value of $\delta_n (\epsilon)$ fluctuates a lot from one BIMSC to another through the behavior of $r_{X|Y} (\delta)$ around $\delta =0$, which depends on both the second and third order derivatives of $r_{X|Y} (\delta)$. Given a BIMSC, if $r_{X|Y} (\delta)$ is approximated as in \eqref{eq2-4-}, then $\delta_n (\epsilon) $ is in the order of $\sqrt{-\ln \epsilon \over n}$. Of course, such an approximation is accurate only when $\delta$ or $\sqrt{-\ln \epsilon \over n}$ is sufficiently small.
\end{remark}

With respect to  $\delta_n (\epsilon)$,  $R_n (\epsilon)$ has a nice Taylor-type expansion, as shown in Theorem \ref{thm-BIMSC-second-order}.

\begin{theorem}
  \label{thm-BIMSC-second-order}
  Given a BIMSC, for  any $n$ and $\epsilon$ satisfying $g_{X|Y,n} (\delta^+ /2) \leq \epsilon <1$,
  \begin{equation}
    \label{eq-so-17}
    \left| R_n (\epsilon) - \left( C_{\mathrm{BIMSC}} - \delta_n (\epsilon) \right) \right|
    \leq o \left( \delta_n (\epsilon) \right)
  \end{equation}
  where
  \begin{eqnarray}
    \label{eq-so-18}
    o \left( \delta_n (\epsilon) \right)
      &=& r_{X|Y} (\delta_n (\epsilon)) + \frac{ \ln n + d_1}{n}
  \end{eqnarray}
  if $\epsilon \leq \frac{1}{3} $, and
  \begin{equation}
    \label{eq-so-16}
    \left| R_n (\epsilon) - \left( C_{\mathrm{BIMSC}} - \frac{\sigma_H (X|Y)}{\sqrt{n}} Q^{-1} (\epsilon) \right) \right|
    \leq \frac{\ln n + d_2}{n}
  \end{equation}
   otherwise, where $d_1$ and $d_2$ are channel parameters independent of both $n$ and $\epsilon$.
\end{theorem}

\begin{IEEEproof}
  When $\epsilon > \frac{1}{3}$, \eqref{eq-so-16} can be easily proved by combining \eqref{eq1-5}, \eqref{eq1-6} and
  \eqref{eq-thm-bimc-6}. Therefore, it suffices for us to show \eqref{eq-so-17} and \eqref{eq-so-18} for $\epsilon \leq \frac{1}{3}$.
  By \eqref{eq1-1} and definition of $\bar{\xi}_H(X|Y,\lambda,n)$, for any BIMSC
  there exists a channel code $\mathcal{C}_n$ such that
  \begin{eqnarray}
    \label{eq-so-19}
    P_e (\mathcal{C}_n) &\leq&
      \left( \bar{\xi}_H (X|Y,\lambda,n) + \frac{2 (1-C_{BE}) M_H(X|Y,\lambda)}{\sqrt{n} \sigma^3_H (X|Y,\lambda)} \right)
      e^{- n r_{X|Y} (\delta)} \nonumber \\
      &\leq& g_{X|Y,n} (\delta) + \frac{2M_H(X|Y,\lambda)}{\sqrt{n} \sigma^3_H (X|Y,\lambda)}
      e^{- n r_{X|Y} (\delta)}
  \end{eqnarray}
   and
   \begin{equation}
     \label{eq-so-20}
     R(\mathcal{C}_n) \geq C_{\mathrm{BIMSC}} - \delta +
       \frac{\ln \left[ \frac{2 (1-C_{BE}) M_H(X|Y,\lambda)}{\sqrt{n} \sigma^3_H (X|Y,\lambda)} e^{- n r_{X|Y} (\delta)} \right] }{n}
   \end{equation}
   which implies that for any $\delta$ such that
   \begin{equation}
     \label{eq-so-21}
      g_{X|Y,n} (\delta) + \frac{2M_H(X|Y,\lambda)}{\sqrt{n} \sigma^3_H (X|Y,\lambda)}
      e^{- n r_{X|Y} (\delta)} \leq \epsilon
   \end{equation}
   the following inequality holds
   \begin{equation}
     \label{eq-so-21+}
     R_n(\epsilon) \geq C_{\mathrm{BIMSC}} - \delta +
       \frac{\ln \left[ \frac{2 (1-C_{BE}) M_H(X|Y,\lambda)}{\sqrt{n} \sigma^3_H (X|Y,\lambda)} e^{- n r_{X|Y} (\delta)} \right] }{n}
   \end{equation}
    where $\lambda = r'_{X|Y} (\delta)$. Now let $\bar{\delta} = \delta_n (\epsilon) + \frac{\eta}{n}$ for some constant
    $\eta >0$,  which will be specified later, and $\bar{\lambda} = r'_{X|Y} (\bar{\delta})$. By convexity of $r_{X|Y} (\delta)$,
    \begin{equation}
      \label{eq-so-22}
      r_{X|Y} (\bar{\delta}) \geq r_{X|Y} (\delta_n (\epsilon)) + \lambda_n (\epsilon) \frac{\eta}{n}
    \end{equation}
    where $\lambda_n (\epsilon) = r'_{X|Y} (\delta_n (\epsilon))$. Then
    \begin{eqnarray}
      \label{eq-so-23}
      \lefteqn{g_{X|Y,n} (\bar{\delta}) + \frac{2M_H(X|Y,\bar{\lambda})}{\sqrt{n} \sigma^3_H (X|Y,\bar{\lambda})}
      e^{- n r_{X|Y} (\bar{\delta})}} \nonumber \\
      &\stackrel{1)}{\leq} &
      \left( e^{\frac{n \bar{\lambda}^2 \sigma^2_H(X|Y,\bar{\lambda})}{2}}
              Q \left( \sqrt{n} \bar{\lambda} \sigma_H (X|Y,\bar{\lambda}) \right)
              + \frac{2M_H(X|Y,\bar{\lambda})}{\sqrt{n} \sigma^3_H (X|Y,\bar{\lambda})}
      \right) e^{- n \left( r_{X|Y} (\delta_n (\epsilon)) + \lambda_n (\epsilon) \frac{\eta}{n} \right)} \nonumber \\
      &=&
              \left( 1
              + \frac{\frac{2M_H(X|Y,\bar{\lambda})}{\sqrt{n} \sigma^3_H (X|Y,\bar{\lambda}) }}
                        {e^{\frac{n \bar{\lambda}^2 \sigma^2_H(X|Y,\bar{\lambda})}{2}}
              Q \left( \sqrt{n} \bar{\lambda} \sigma_H (X|Y,\bar{\lambda}) \right) } \right)
             e^{\frac{n \bar{\lambda}^2 \sigma^2_H(X|Y,\bar{\lambda})}{2}}
              Q \left( \sqrt{n} \bar{\lambda} \sigma_H (X|Y,\bar{\lambda})) \right)
              \nonumber \\
      &&{\times}\: e^{- n r_{X|Y} (\delta_n (\epsilon)) - \eta \lambda_n (\epsilon) }
              \nonumber \\
      &\stackrel{2)}{\leq}& \left( 1
              + \frac{2M_H(X|Y,\bar{\lambda}) \sqrt{2 \pi} \bar{\lambda}
                        \left(1 + \frac{1}{n \bar{\lambda}^2 \sigma^2_H (X|Y,\bar{\lambda})} \right) }
                        {\sigma^2_H (X|Y,\bar{\lambda})} \right)
                  \nonumber \\
      &&{\times}\: e^{\frac{n \lambda^2_n (\epsilon) \sigma^2_H(X|Y,\lambda_n (\epsilon))}{2}}
              Q \left( \sqrt{n} \lambda_n (\epsilon) \sigma_H (X|Y,\lambda_n (\epsilon)) \right)
              e^{- n r_{X|Y} (\delta_n (\epsilon)) - \eta \lambda_n (\epsilon) } \nonumber \\
      &=& g_{X|Y,n} \left( \delta_n (\epsilon) \right)
             e^{- \eta \lambda_n (\epsilon) } \left( 1 +
             \frac{2 \sqrt{2 \pi} M_H(X|Y,\bar{\lambda}) \left(1 + \frac{1}{n \bar{\lambda}^2 \sigma^2_H (X|Y,\bar{\lambda})} \right)}{ \sigma^2_H (X|Y,\bar{\lambda})} \bar{\lambda}
             \right) \nonumber \\
      &\stackrel{3)}{=}& \epsilon e^{- \eta \lambda_n (\epsilon) } \left( 1 +
             \frac{2 \sqrt{2 \pi} M_H(X|Y,\bar{\lambda}) \left(1 + \frac{1}{n \bar{\lambda}^2 \sigma^2_H (X|Y,\bar{\lambda})} \right)}{ \sigma^2_H (X|Y,\bar{\lambda})} \left( \lambda_n (\epsilon) + \frac{1}{\sigma^2_H (X|Y,\tilde{\lambda})} \frac{\eta}{n} \right)
             \right) \nonumber \\
      &\stackrel{4)}{\leq}& \epsilon
                 \frac{1 +
             \frac{2 \sqrt{2 \pi} M_H(X|Y,\bar{\lambda}) \left(1 + \frac{1}{n \bar{\lambda}^2 \sigma^2_H (X|Y,\bar{\lambda})} \right)}{ \sigma^2_H (X|Y,\bar{\lambda})} \left( \lambda_n (\epsilon) + \frac{1}{\sigma^2_H (X|Y,\tilde{\lambda})} \frac{\eta}{n} \right)}
                    {1 + \eta \lambda_n (\epsilon) + \frac{1}{2} \eta^2 \lambda^2_n (\epsilon)}\;.
    \end{eqnarray}
  In the derivation of \eqref{eq-so-23}, the inequality 1) is due to \eqref{eq-so-22}; the inequality 2) follows from   the fact that $e^{\frac{x^2}{2}} Q(x)$ is a strictly decreasing function of $x$,
    $\lambda \sigma_H (X|Y,\lambda)$ is strictly increasing with respect to $\lambda$ as shown below
    \begin{eqnarray}
      \label{eq-so-22+}
      \frac{d \lambda \sigma_H (X|Y,\lambda)}{ d \lambda}
      &=& \sigma_H (X|Y,\lambda) + \lambda  \frac{d \sigma_H (X|Y,\lambda)}{d \lambda}
      \nonumber \\
      &=&  \sigma_H (X|Y,\lambda) \left( 1 +
              \lambda \frac{ \frac{d \sigma^2_H (X|Y,\lambda)}{d \lambda} }{2 \sigma^2_H (X|Y,\lambda)}
              \right)
      \nonumber \\
      & = &
       \sigma_H (X|Y,\lambda) \left( 1 +
              \lambda \frac{ \hat{M}_H (X|Y, \lambda) }{2 \sigma^2_H (X|Y,\lambda)}
              \right)
      \nonumber \\
                &>& 0
    \end{eqnarray}
    for $\lambda \in [0, \lambda^+]$,  and
    \begin{equation}
      \label{eq-so-23-}
      e^{\frac{x^2}{2}} Q(x) \geq \frac{x}{\sqrt{2 \pi}(1+x^2)};
    \end{equation}
   the equality 3) is attributable to
    \begin{equation}
      \label{eq-so-23--}
      \bar{\lambda} = \lambda_n (\epsilon) + \left. \frac{d \lambda}{d \delta} \right|_{\lambda = \tilde{\lambda}} \frac{\eta}{n}
                           = \lambda_n (\epsilon) + \frac{1}{\sigma^2_H (X|Y,\tilde{\lambda})} \frac{\eta}{n}
    \end{equation}
    for some $\tilde{\lambda} \in [\lambda_n (\epsilon), \bar{\lambda}]$; and finally, the inequality 4) follows from the inequality
    \[ e^x > 1 + x + {x^2 \over 2} \]
  for any $x >0$.  In order to satisfy \eqref{eq-so-21},  let us now choose $\eta$ such that
    \begin{equation}
      \label{eq-so-23+}
      \eta  \lambda_n (\epsilon) \geq \frac{2 \sqrt{2 \pi} M_H(X|Y,\bar{\lambda}) \left(1 + \frac{1}{n \bar{\lambda}^2 \sigma^2_H (X|Y,\bar{\lambda})} \right)}{ \sigma^2_H (X|Y,\bar{\lambda})}   \lambda_n (\epsilon)
    \end{equation}
    and
    \begin{equation}
      \label{eq-so-24}
      \frac{1}{2} \eta^2 \lambda^2_n (\epsilon) \geq \frac{2 \sqrt{2 \pi} M_H(X|Y,\bar{\lambda}) \left(1 + \frac{1}{n \bar{\lambda}^2 \sigma^2_H (X|Y,\bar{\lambda})} \right)}{ \sigma^2_H (X|Y,\bar{\lambda})}
       \frac{1}{\sigma^2_H (X|Y,\tilde{\lambda})} \frac{\eta}{n} ,
    \end{equation}
    i.e.
    \begin{equation}
      \label{eq-so-24+}
      \eta = \frac{2 \sqrt{2 \pi} M_H(X|Y,\bar{\lambda}) \left(1 + \frac{1}{n \bar{\lambda}^2 \sigma^2_H (X|Y,\bar{\lambda})} \right)}{ \sigma^2_H (X|Y,\bar{\lambda})}
      \max \left\{ 1, \frac{2}{n \lambda^2_n (\epsilon)\sigma^2_H (X|Y,\tilde{\lambda})} \right\} .
    \end{equation}
    To see $\eta$ is bounded, note that
    $\frac{M_H (X|Y,\lambda)}{\sigma^2_H (X|Y,\lambda)}$ is always bounded for
    $\lambda \in [0,\lambda^+]$.
    On the other hand, for $\epsilon \leq \frac{1}{3}$,
    $\sqrt{n} \lambda_n (\epsilon) \sigma_H(X|Y,\lambda_n (\epsilon)) > c$ for some constant $c$,
    as $\sqrt{n} \lambda_n (\epsilon) \sigma_H(X|Y,\lambda_n (\epsilon)) \rightarrow 0$
    implies that $\epsilon = g_{X|Y,n} (\delta_n (\epsilon)) \rightarrow \frac{1}{2}$,
    and the same argument can be applied
    to $\sqrt{n} \lambda_n (\epsilon) \sigma^2_H (X|Y,\tilde{\lambda})$. Therefore,
    \begin{eqnarray}
      \label{eq-so-25}
      \eta &\leq& 2 \sqrt{2 \pi} \max_{\lambda \in [0,\lambda^+]} \left[ \frac{M_H (X|Y,\lambda)}{\sigma^2_H (X|Y,\lambda)}\right] \left( 1 + c^{-2} \right) \max \left\{ 1, 2 c^{-2} \right\} .
    \end{eqnarray}
    Then combining \eqref{eq-so-21},  \eqref{eq-so-21+}, \eqref{eq-so-22}, \eqref{eq-so-23}, \eqref{eq-so-23+}
    and \eqref{eq-so-24}  yields
    \begin{eqnarray}
      \label{eq-so-25+}
      R_n (\epsilon) &\geq& C_{\mathrm{BIMSC}} - \bar{\delta} +
      \frac{\ln \left[ \frac{2 (1-C_{BE}) M_H(X|Y,\bar{\lambda})}{\sqrt{n} \sigma^3_H (X|Y,\bar{\lambda})}
      e^{- n r_{X|Y} (\bar{\delta})} \right] }{n} \nonumber \\
      &=& C_{\mathrm{BIMSC}} - \bar{\delta} - r_{X|Y} (\bar{\delta}) +
      \frac{\ln \left[ \frac{2 (1-C_{BE}) M_H(X|Y,\bar{\lambda})}{\sigma^3_H (X|Y,\bar{\lambda})}
       \right] - \frac{1}{2} \ln n }{n} \nonumber \\
      &\stackrel{1)}{\geq}&  C_{\mathrm{BIMSC}} - \delta_n (\epsilon) - r_{X|Y} (\delta_n (\epsilon)) - \bar{\lambda} \frac{\eta}{n}
              + \frac{\ln \left[ \frac{2 (1-C_{BE}) M_H(X|Y,\bar{\lambda})}{\sigma^3_H (X|Y,\bar{\lambda})}
       \right] - \eta - \frac{1}{2} \ln n }{n}
      \nonumber \\
      &\geq&  C_{\mathrm{BIMSC}} - \delta_n (\epsilon) - r_{X|Y} (\delta_n (\epsilon)) \nonumber \\
      &&{+}\:
      \frac{ -\lambda^+ \eta +
               \ln \left[ 2 (1-C_{BE}) \min_{\lambda} \left( \frac{2M_H(X|Y,{\lambda})}{ \sigma^3_H (X|Y,{\lambda})} \right)
                   \right] - \eta - \frac{1}{2} \ln n }
            {n} \nonumber \\
      &=& C_{\mathrm{BIMSC}} - \delta_n (\epsilon) - r_{X|Y} (\delta_n (\epsilon)) - \frac{\frac{1}{2} \ln {n} + \bar{d}_1}{n},
    \end{eqnarray}
    where $\bar{d}_1$ is independent of both $n$ and  $\epsilon$. In the derivation of \eqref{eq-so-25+}, the inequality 1) follows from the convexity of $r_{X|Y} (\delta) $ and the fact that
    \[ r_{X|Y} ( \bar{\delta}) \leq r_{X|Y} (\delta_n (\epsilon))  + \bar{\lambda} {\eta \over n} .\]

 We now proceed to establish an upper bound on $R_n (\epsilon)$.
    Towards this end, recall \eqref{eq-thm-bimc-1} and \eqref{eq-thm-bimc-2} where
    we make a small modification by choosing $\beta_n = \lambda = r'_{X|Y} (\delta)$
    in the proof of Theorem \ref{thm-bimc}. Then
    for any $\delta$ such that
    \begin{equation}
      \label{eq-so-29}
      \left( 1 + 2 \lambda \right) \epsilon
      \leq \underline{\xi}_H (X|Y,\lambda,n) e^{- n r_{X|Y} (\delta)}
    \end{equation}
     we have
     \begin{eqnarray}
       \label{eq-so-30}
       R_n(\epsilon) &\leq& C_{\mathrm{BIMSC}} - \delta - \frac{\ln
      \epsilon - \ln P(B_{n,\delta}) + {2 \ln \lambda }
      - \ln \left( 1 + \lambda \right) }{n} \nonumber \\
      &\leq& C_{\mathrm{BIMSC}} - \delta + \frac{ - \ln \epsilon - 2 \ln \lambda + \lambda }{n}
     \end{eqnarray}
      where the trivial bound $P(B_{n,\delta}) \leq 1$ is applied. Now let
     $\underline{\delta} = \delta_n (\epsilon) - \frac{\eta'}{n}$ for some constant $\eta' >0$, which will be specified later,  and
     $\underline{\lambda} = r'_{X|Y} (\underline{\delta})$. Then
     \begin{eqnarray}
       \label{eq-so-31}
       \lefteqn{\underline{\xi}_H (X|Y,\underline{\lambda},n) e^{- n r_{X|Y} (\underline{\delta})}} \nonumber \\
        &\stackrel{1)}{\geq} & e^{\frac{n \underline{\lambda}^2 \sigma^2_H (X|Y,\underline{\lambda})}{2}}
               Q \left( \rho_* + \sqrt{n} \underline{\lambda} \sigma_H (X|Y, \underline{\lambda}) \right)
               e^{- n r_{X|Y} (\delta_n (\epsilon))+ \underline{\lambda} \eta' } \nonumber \\
        &=& e^{\frac{n \underline{\lambda}^2 \sigma^2_H (X|Y,\underline{\lambda})}{2}}
                    Q \left( \sqrt{n} \underline{\lambda} \sigma_H (X|Y,\underline{\lambda}) \right)
              \frac{Q \left( \rho_* + \sqrt{n} \underline{\lambda} \sigma_H (X|Y, \underline{\lambda}) \right)}
                     {Q \left( \sqrt{n} \underline{\lambda} \sigma_H (X|Y,\underline{\lambda}) \right)}
                     e^{- n r_{X|Y} (\delta_n (\epsilon))+ \underline{\lambda} \eta'} \nonumber \\
        &\stackrel{2)}{\geq}&  g_{X|Y,n} (\delta_n (\epsilon))
               \frac{Q \left( \rho_* + \sqrt{n} \underline{\lambda} \sigma_H (X|Y, \underline{\lambda}) \right)}
                     {Q \left( \sqrt{n} \underline{\lambda} \sigma_H (X|Y,\underline{\lambda}) \right)}
                     e^{\underline{\lambda} \eta' } \nonumber \\
        &\stackrel{3)}{\geq}&  (1 +2 \underline{\lambda}) \epsilon
               \frac{Q \left( \rho_* + \sqrt{n} \underline{\lambda} \sigma_H (X|Y, \underline{\lambda}) \right)}
                     {Q \left( \sqrt{n} \underline{\lambda} \sigma_H (X|Y,\underline{\lambda}) \right)}
                     e^{\underline{\lambda} (\eta'-2) }.
     \end{eqnarray}
 In the derivation of \eqref{eq-so-31}, the inequality 1) is due to the convexity of $r_{X|Y} (\delta)$ and the fact that
     \[ r_{X|Y} ( \underline{\delta}) \leq r_{X|Y} (\delta_n (\epsilon))  - \underline{\lambda} {\eta' \over n} ;\]
 the inequality 2) follows again from the fact that $e^{\frac{x^2}{2}} Q(x)$ is a strictly decreasing function of $x$  and
     $\lambda \sigma_{H} (X|Y,\lambda)$ is increasing with respect to $\lambda$; and finally the inequality 3) is attributable to the inequality $e^x \geq 1 + x $ for any $ x \geq 0$.

 In order for \eqref{eq-so-29} to be satisfied, we now   choose $\eta'$ such that
     \begin{eqnarray}
       \label{eq-so-33+}
       \eta' &=& 2 + \frac{1}{\underline{\lambda}} \ln
              \frac{Q \left( \sqrt{n} \underline{\lambda} \sigma_H (X|Y,\underline{\lambda}) \right) }
                     {Q \left( \rho_* + \sqrt{n} \underline{\lambda} \sigma_H (X|Y, \underline{\lambda}) \right)}
              \nonumber \\
              &=& 2 + \frac{1}{\underline{\lambda}} \ln \left[ 1 +
              \rho_* \frac{ \frac{1}{\sqrt{2 \pi}} e^{- \frac{(\tilde{\rho} + \sqrt{n} \underline{\lambda} \sigma_H (X|Y, \underline{\lambda}) )^2}{2}} }
                     {Q \left( \rho_* + \sqrt{n} \underline{\lambda} \sigma_H (X|Y, \underline{\lambda}) \right)}
              \right]
     \end{eqnarray}
     where $0 \leq \tilde{\rho} \leq \rho_*$. One can verify that
     \begin{eqnarray}
       \label{eq-so-34-}
       \eta' &\leq& 2 + \frac{\rho_*}{\underline{\lambda}}
               \frac{ \frac{1}{\sqrt{2 \pi}} e^{- \frac{(\tilde{\rho} + \sqrt{n} \underline{\lambda} \sigma_H (X|Y, \underline{\lambda}) )^2}{2}} }
                     {Q \left( \rho_* + \sqrt{n} \underline{\lambda} \sigma_H (X|Y, \underline{\lambda}) \right)}
       \nonumber \\
       &\leq&  2 +\frac{\rho_*}{\underline{\lambda}}
       \frac{ 1 + ( \rho_* + \sqrt{n} \underline{\lambda} \sigma_H (X|Y, \underline{\lambda}))^2}
              {\rho_* + \sqrt{n} \underline{\lambda} \sigma_H (X|Y, \underline{\lambda})}
       e^{\sqrt{n} \underline{\lambda} \sigma_H (X|Y,\underline{\lambda}) (\rho_* - \tilde{\rho}) + \frac{\rho^2_* - \tilde{\rho}^2}{2} }
    \end{eqnarray}
 where the last inequality is due to \eqref{eq-so-23-}.    From the definition of $\rho_*$, it is not hard to see that $\rho_* = \frac{\eta''}{\sqrt{n}}$ for some constant $\eta''$ depending only on channel parameters. Meanwhile, we have $\sqrt{n} \underline{\lambda} \sigma_H (X|Y,\underline{\lambda}) > c$ as discussed above.  Then
    \begin{eqnarray}
      \label{eq-so-34--}
      \eta' &\leq&  2+ \frac{\eta''}{\sqrt{n} \underline{\lambda}}
       \left( c^{-1} + \frac{\eta''}{\sqrt{n}} + \sqrt{n} \underline{\lambda} \sigma_H (X|Y, \underline{\lambda}) \right)
       e^{\eta'' \lambda^+ \max_{\lambda \in [0, \lambda^+ ]} \sigma_H (X|Y,\lambda)+ \frac{(\eta'')^2}{2 n}}
     \nonumber \\
     &\leq& 2+  \left( c^{-2}
      + c^{-1} \eta'' + 1  \right) \eta''
      \left[ \max_{\lambda \in [0, \lambda^+ ]} \sigma_H (X|Y,\lambda) \right]
      e^{\eta'' \lambda^+ \max_{\lambda \in [0, \lambda^+ ]} \sigma_H (X|Y,\lambda)+ (\eta'')^2}
     \nonumber \\
    \end{eqnarray}
  which is independent of both  $n$ and $\epsilon$.  Now
    combining \eqref{eq-so-31} and \eqref{eq-so-33+}, we have
     \begin{equation}
       \label{eq-so-34}
       \underline{\xi}_H (X|Y,\underline{\lambda},n) e^{- n r_{X|Y} (\underline{\delta})} \geq  (1 +2  \underline{\lambda} )\epsilon
     \end{equation}
     and consequently,
     \begin{eqnarray}
       \label{eq-so-37}
       R_n(\epsilon)
      &\leq&
      C_{\mathrm{BIMSC}} - \underline{\delta} + \frac{ - \ln \epsilon - 2 \ln  \underline{\lambda} + \underline{\lambda} }{n}
      \nonumber \\
      &\stackrel{1)}{\leq} & C_{\mathrm{BIMSC}} - \delta_n (\epsilon) + r_{X|Y} (\delta_n (\epsilon)) \nonumber \\
      &&{+}\: \frac{ \ln \left [ \sqrt{2 \pi} \sqrt{n} \lambda_n (\epsilon) \sigma_H (X|Y,\lambda_n (\epsilon)) \left ( 1 + {1 \over n \lambda^2_n (\epsilon) \sigma^2_H (X|Y,\lambda_n (\epsilon)) } \right ) \right ] }{n} \nonumber \\
      & & \mbox{ } + {
                   - 2\ln \underline{\lambda} + \lambda^+ + \eta'  \over n} \nonumber \\
      &=& C_{\mathrm{BIMSC}} - \delta_n (\epsilon) + r_{X|Y} (\delta_n (\epsilon))  + { \ln \left ( 1 + {1 \over n \lambda^2_n (\epsilon) \sigma^2_H (X|Y,\lambda_n (\epsilon)) } \right ) \over n} \nonumber \\
      &&{+}\: \frac{ \ln n + \ln \sqrt{2 \pi} \sigma_H (X|Y,\lambda_n (\epsilon))
                   + \ln \frac{\lambda_n (\epsilon)}{ \underline{\lambda}}
                   - \ln \sqrt{n} \underline{\lambda} + \lambda^+ + \eta' }{n} \nonumber \\
      &\stackrel{2)}{\leq} & C_{\mathrm{BIMSC}} - \delta_n (\epsilon) + r_{X|Y} (\delta_n (\epsilon)) + \frac{\ln n + \underline{d}_1 }{n}
     \end{eqnarray}
     where $\underline{d}_1$ is another constant depending only on the channel. In the derivation of \eqref{eq-so-37}, the inequality 1) is due to \eqref{eq-so-23-}
  and the definition of $\delta_n (\epsilon)$ in  \eqref{eq-so-2}; and the inequality 2) follows from the fact that
   \[ {\lambda_n (\epsilon) \over \underline{\lambda} } = 1 + {1 \over   \sigma^2_H (X|Y, \hat{\lambda})    } {\eta' \over n \underline{\lambda}} \]
  for some $\hat{\lambda} \in [ \underline{\lambda}, \lambda_n (\epsilon) ] $ and
  \[ \sqrt{n} \underline{\lambda} \sigma_H (X|Y,\underline{\lambda}) > c.\]
 Then the theorem is proved by combining \eqref{eq-so-25+} and
     \eqref{eq-so-37} and making $d_1 = \max \{ \bar{d}_1, \underline{d}_1 \}$.
\end{IEEEproof}

\begin{remark}
  The condition $\epsilon \leq \frac{1}{3}$ for \eqref{eq-so-17} and \eqref{eq-so-18}
  can be relaxed
  as we only require that $\sqrt{n} \delta_n (\epsilon)$ or equivalently $\sqrt{n} \lambda$  be lower bounded by a constant,
  which is true when $\epsilon \leq d$ for any constant $d < \frac{1}{2}$.  In addition, when $\epsilon \leq g_{X|Y, n} (\delta^+ /2)$, $\epsilon$ is an exponential function of $n$, in which case the maximum achievable rate is below the channel capacity by a positive constant even when $n$ goes to $\infty$. As such, from a practical point of view, the case $\epsilon \leq g_{X|Y, n} (\delta^+ /2)$ is not interesting, especially when one can approach the channel capacity very closely as shown in the achievability given in \eqref{eq1-1} and \eqref{eq1-2}.
  \end{remark}

\begin{remark}
In the definition of $R_n (\epsilon )$, the average error probability is used. If the maximal error probability is used instead, Theorem \ref{thm-BIMSC-second-order}
remains valid. This can be proved similarly by first using the standard technique of removing bad codewords from the code in the
  achievability given in \eqref{eq1-1} and \eqref{eq1-2} to establish similar achievability with maximal error probability and then combining it with Corollary \ref{col-bimc}.

\end{remark}

\begin{remark}
In view of Theorem~\ref{thm-BIMSC-second-order}, it is now clear that jar decoding is indeed optimal up to the second order coding performance in the non-asymptotical regime. Since the achievability given in \eqref{eq1-1} and \eqref{eq1-2} was established for linear block codes, it follows from Theorem~\ref{thm-BIMSC-second-order} that linear block coding is also  optimal up to the second order coding performance in the non-asymptotical regime for any BIMSC.
In addition, in the Taylor-type expansion of $R_n (\epsilon)$, the third order term  is $O(\delta^2_n (\epsilon)) $ whenever $\delta_n (\epsilon) = \Omega(\sqrt{ \ln n /  n})$ since it follows from   \eqref{eq2-4-}    that $r_{X|Y} (\delta_n (\epsilon)) = O(\delta^2_n (\epsilon)) $.
\end{remark}

\subsection{Comparison with Asymptotic Analysis} \label{sec2-d}

It is instructive to compare Theorem \ref{thm-BIMSC-second-order} with the second order asymptotic performance analysis as $n$ goes to $\infty$.

{\em Asymptotic analysis with constant $0< \epsilon <1$ and $n \to \infty$}: Fix $0 < \epsilon < 1$. It was shown in \cite{strassen-1962}, \cite{Hayashi-2009},  \cite{Yury-Poor-Verdu-2010} that for a BIMSC with a discrete output alphabet
   \begin{equation} \label{eq-comp-1}
   R_n (\epsilon) =  C_{\mathrm{BIMSC}} - \frac{\sigma_H (X|Y)}{\sqrt{n}} Q^{-1} (\epsilon) + O \left ({\ln n \over n} \right )
   \end{equation}
for sufficiently large $n$. The expression $C_{\mathrm{BIMSC}} - \frac{\sigma_H (X|Y)}{\sqrt{n}} Q^{-1} (\epsilon) $ was referred to as the normal approximation for $R_n (\epsilon)$. Clearly, when $\epsilon > 1/3$, \eqref{eq-comp-1} is essentially the same as
\eqref{eq-so-16}. Let us now look at the case $\epsilon \leq 1/3$.
In this case, by using the Taylor expansion of $r_{X|Y} (\delta)$ around $\delta =0$
\begin{eqnarray} \label{eq-comp-2}
r_{X|Y} (\delta) & = & {1 \over 2 \sigma^2_H (X|Y) } \delta^2 +  \frac{ \left. - \frac{d \sigma^2_H (X|Y,\lambda)}{d \lambda} \right|_{\lambda=0}}{6 \sigma^6_H (X|Y)}
 \delta^3 + O(\delta^4)  \nonumber \\
 & = & {1 \over 2 \sigma^2_H (X|Y) } \delta^2 +  \frac{ -  \hat{M}_H (X|Y) }{6 \sigma^6_H (X|Y)}
 \delta^3 + O(\delta^4)
 \end{eqnarray}
  it can be verified that
 \begin{equation} \label{eq-comp-3}
  \delta_n (\epsilon) = \frac{\sigma_H (X|Y)}{\sqrt{n}} Q^{-1} (\epsilon)  + O \left ( {1 \over n} \right ) .
  \end{equation}
Thus the Taylor-type expansion of $R_n (\epsilon )$ in Theorem~\ref{thm-BIMSC-second-order} implies the second order asymptotic analysis with constant $0< \epsilon <1$ and $n \to \infty$ shown in \eqref{eq-comp-1}.

{\em Asymptotic analysis with $n \to \infty$ and non-exponentially decaying $\epsilon$}: Suppose now $\epsilon$ is a function of $n$ and goes to $0$ as $n \to \infty$, but at a non-exponential speed. In this case, as $n \to \infty$, $\delta_n (\epsilon)$ goes to $0$ at the speed of $\Theta \left( \sqrt{-\ln \epsilon \over n} \right)$, and $\sqrt{n} \lambda_n (\epsilon) $ goes to $\infty$. By ignoring the third and higher order terms in the Taylor expansion of $r_{X|Y} (\delta)$, one has the following approximations:
\begin{equation} \label{eq-comp-4}
    g_{X|Y, n } (\delta_n (\epsilon)) \approx {1 \over \sqrt{2 \pi}   \sqrt{n} \lambda_n (\epsilon) \sigma_H (X|Y,\lambda_n (\epsilon)) }     e^{-n {\delta_n^2 (\epsilon) \over 2 \sigma^2_H (X|Y)}}
    \end{equation}
 and
 \[ Q(x) \approx {1 \over \sqrt{2 \pi} x} e^{- {x^2 \over 2}} \mbox{ for large } x. \]
 By these approximations, it is not hard to verify that in this case
  \[  \lim_{n \to \infty} {\delta_n (\epsilon) \over \frac{\sigma_H (X|Y)}{\sqrt{n}} Q^{-1} (\epsilon)  } =1. \]
 Therefore, from Theorem~\ref{thm-BIMSC-second-order}, it follows that when $\epsilon $ goes to $0$  at a non-exponential speed as $n \to \infty$, $ \frac{\sigma_H (X|Y)}{\sqrt{n}} Q^{-1} (\epsilon)  $ is still the second order term of $R_n (\epsilon)$ in the asymptotic analysis with $n \to \infty$. Indeed, this can also be verified by looking at the specific case given by \eqref{eq1-3}, \eqref {eq1-4},  and \eqref{eq-thm-bimc-3} when $\epsilon$ goes to $0$ at a polynomial speed as $n \to \infty$. To the best of our knowledge, the second order asymptotic analysis with $n \to \infty$ and non-exponentially decaying $\epsilon$ has not been addressed before in the literature.

 {\em Divergence of $\delta_n (\epsilon)$ from $ \frac{\sigma_H (X|Y)}{\sqrt{n}} Q^{-1} (\epsilon)  $}:  The agreement between  $\delta_n (\epsilon)$ and   $ \frac{\sigma_H (X|Y)}{\sqrt{n}} Q^{-1} (\epsilon)  $ terminates when the third order term
 \[ \frac{  -  \hat{M}_H (X|Y) }{6 \sigma^6_H (X|Y)}
 \delta^3 \]
 in the Taylor expansion of $r_{X|Y} (\delta)$ shown in \eqref{eq-comp-2} can not be ignored. This happens when $\delta$ is not small, which is typical in practice for finite block length $n$,  or
 \begin{equation} \label{eq-comp-6}
 \zeta_{X|Y} \defeq  \frac{ - \hat{M}_H (X|Y) }{6 \sigma^6_H (X|Y)}
 \end{equation}
 is large.  In this case, $ \frac{\sigma_H (X|Y)}{\sqrt{n}} Q^{-1} (\epsilon)  $ will be smaller than $\delta_n (\epsilon)$ by a relatively large margin if   $ \zeta_{X|Y} <0$, and larger than
  $\delta_n (\epsilon)$ by a relatively large margin if   $ \zeta_{X|Y} >0$. As such, the normal approximation would fail to provide a reasonable estimate for $R_n (\epsilon)$.  This will be further confirmed by numerical results shown in Section \ref{sec:appr-eval} for well known channels such as the BEC, BSC, and BIAGN for finite $n$.

\section{Non-Asymptotic Converse and Taylor-Type Expansion: DIMC}
\label{sec:non-asysmpt-converse-coding-dsc}
\setcounter{equation}{0}

We now extend Theorems~\ref{thm-bimc} and \ref{thm-BIMSC-second-order}
to the case of DIMC $ P = \{p(y|x), x \in \mathcal{X}, y \in
\mathcal{Y}\}$, where  $\mathcal{X}$  is discrete,  but $\mathcal{Y}$ is
arbitrary (discrete or continuous).

\subsection{Definitions}

Let $\mathcal{P}$ denote the set of all distributions over $\mathcal{X}$. Let
$\mathcal{P}_n$ denote the set of types on $\mathcal{X}^n$ with
denominator $n$ \cite{csiszar:type}, and $t(x^n)$ be the type of $x^n$.
Moreover, for $t \in \mathcal{P}_n$, let
\begin{equation} \label{eq-def-xt}
 \mathcal{X}^n_t \defeq \{x^n \in \mathcal{X}^n: t(x^n)=t \}.
\end{equation}
Before stating our converse channel coding theorem for DIMC, we again need to
 introduce some definitions from \cite{yang-meng:nep}.
For any  $t \in \mathcal{P}$, define
\begin{equation} \label{eq-def-qtn}
  q_t(y^n) \defeq \prod^n_{i=1} q_t (y_i)
\end{equation}
where
\begin{equation} \label{eq-def-qt}
  q_t (y) \defeq \sum_{x \in \mathcal{X}} t(x) p(y|x),
\end{equation}
\begin{equation} \label{eq3r-3}
  I (t; P) \defeq \sum_{x \in {\cal X}} t(x) \int p(y |x) \ln {p(y|x) \over q_t (y) } d y
\end{equation}
and
\begin{equation} \label{eq-def-lambda*}
   \lambda^*_{-} (t; P) \defeq
   \sup \left \{ \lambda \geq 0:  \sum_{a \in {\cal X}} t(a) \int p(y |a) \left [ {p(y|a)  \over q_t (y) } \right ]^{-\lambda} d y  <\infty \right \}.
\end{equation}
 It is easy to see that $ \lambda^*_{-} (t; P) $ is the same for all $t \in {\cal P}$ with the same support set $\{a\in {\cal X}: t(a) >0 \}$. Suppose that
\begin{equation} \label{eq-def-lambda*+}
\lambda^*_{-} (t; P) > 0.
\end{equation}
Define for any $t \in {\cal P}$ and any $\delta \geq 0$
\begin{equation} \label{eq3rlr}
 r_{-} (t, \delta) \defeq \sup_{\lambda \geq 0} \left [ \lambda (\delta - I(t; P) ) -
 \sum_{x \in {\cal X}} t(x) \ln  \int  p(y |x) \left [ {p (y|x) \over q_t (y) } \right ]^{-\lambda}  d y \right ]
 \end{equation}
and for any $t \in {\cal P}$ and any $\lambda \in [0, \lambda^*_{-} (t; P))$,
random variables $X_{t}$ and $Y_{t,\lambda}$ with joint distribution $t(x)p(y|x)f_{-\lambda}(y|x)$ where
\begin{equation} \label{eq3rlf}
 f_{-\lambda} (y |x) \defeq
  {\left [ {p (y|x) \over q_t (y) } \right ]^{-\lambda} \over    \int p(v |x )  \left [ {p (v|x ) \over q_t (v)} \right ]^{-\lambda} d v  } .
\end{equation}
Then define
\begin{equation} \label{eq3rld-}
 D(t, x, \lambda) \defeq \mathbf{E} \left[ \left. \ln \frac{p(Y_{t,\lambda}|X_t)}{q_t (Y_{t,\lambda})} \right| X_t = x \right]
\end{equation}
 \begin{equation} \label{eq3rld}
 \delta_{-} (t, \lambda)
 \defeq \mathbf{E} \left[ -\ln \frac{p(Y_{t,\lambda}|X_t)}{q_t (Y_{t,\lambda})}  \right]
  +  I(t; P)
\end{equation}
\begin{equation}
  \label{eq3rld-1}
  \Delta^*_{-} (t) \defeq \lim_{\lambda \uparrow \lambda^*_{-} (t;P)} \delta_{-} (t, \lambda)
\end{equation}
\begin{eqnarray} \label{eq3rl-13}
  \sigma^2_{D, -} (t; P, \lambda)  &\defeq&
    \mathbf{E} \left\{ \mathbf{Var} \left[ \left. \ln \frac{p(Y_{t,\lambda}|X_t)}{q_t (Y_{t,\lambda})} \right| X_t \right] \right\}
  \nonumber \\
  &=& \sum_{x \in \mathcal{X}} t(x) \mathbf{Var} \left[ \left. \ln \frac{p(Y_{t,\lambda}|X_t)}{q_t (Y_{t,\lambda})} \right| X_t = x\right]
\end{eqnarray}
\begin{eqnarray} \label{eq3rl-14}
  M_{D, -} (t; P , \lambda)
  &\defeq& \mathbf{E} \left\{ \mathbf{M_3} \left[ \left. \ln \frac{p(Y_{t,\lambda}|X_t)}{q_t (Y_{t,\lambda})} \right| X_t \right] \right\}
  \nonumber \\
  &=& \sum_{x \in \mathcal{X}} t(x)  \mathbf{M_3} \left[ \left. \ln \frac{p(Y_{t,\lambda}|X_t)}{q_t (Y_{t,\lambda})} \right| X_t =x  \right]
  \;
\end{eqnarray}
and
\begin{eqnarray} \label{eq-tm-2}
  \hat{M}_{D, -} (t; P , \lambda)
  &\defeq& \mathbf{E} \left\{ \mathbf{\hat{M}_3} \left[ \left. \ln \frac{p(Y_{t,\lambda}|X_t)}{q_t (Y_{t,\lambda})} \right| X_t \right] \right\}
  \nonumber \\
  &=& \sum_{x \in \mathcal{X}} t(x)  \mathbf{\hat{M}_3} \left[ \left. \ln \frac{p(Y_{t,\lambda}|X_t)}{q_t (Y_{t,\lambda})} \right| X_t =x  \right].
\end{eqnarray}
Note that $  \sigma^2_{D, -} (t; P, \lambda) $,  $M_{D, -} (t; P , \lambda)$, and $ \hat{M}_{D, -} (t; P , \lambda)$ are respectively the conditional variance, conditional third absolute central moment, and conditional third central moment of
$\ln \frac{p(Y_{t,\lambda}|X_t)}{q_t (Y_{t,\lambda})}$ given $X_t$.  Write $\sigma^2_{D, -} (t; P, 0)  $ simply as $\sigma^2_D (t; P)$,
  $M_{D, -} (t; P , 0)$ as $M_{D} (t; P )$, and $\hat{M}_{D, -} (t; P , 0)$ as
  $ \hat{M}_{D} (t; P )$. Assume that
  \begin{equation}
    \label{eq3rl-14-1}
 \sigma^2_D (t; P) >0 \mbox{ and }    M_{D} (t; P ) < \infty.
  \end{equation}
Furthermore $r_{-} (t, \delta)$ has the following parametric expression
 \begin{equation} \label{eq3rp1}
  r_{ -} (t, \delta_{-}(t, \lambda)) =   \lambda ( \delta_{-} (t, \lambda) - I(t; P)) -
  \sum_{x \in {\cal X}} t(x) \ln  \int  p(y |x) \left [ {p (y|x) \over q_t (y) } \right ]^{-\lambda}  d y
   \end{equation}
with $\lambda = {\partial r_{-} (t, \delta) \over \partial \delta }$
satisfying $\delta_{-}(t, \lambda) = \delta$.
In addition, let
  \begin{eqnarray}
    \lefteqn{\bar{\xi}_{D,-} ( t; P, \lambda, n )
      \defeq \frac{2 C_{BE} M_{\mathrm{D},-} (t;P,\lambda)}
                        {\sqrt{n}\sigma^3_{\mathrm{D},-} (t;P,\lambda)}
               } \nonumber \\
    &&{+}\: e^{\frac{n \lambda^2 \sigma^2_{D,-} (t;P,\lambda)}{2}}
                \left[ Q(\sqrt{n} \lambda \sigma_{D,-} (t;P,\lambda))
                       - Q(\rho^*+\sqrt{n} \lambda \sigma_{D,-} (t;P,\lambda)) \right] \\
    \lefteqn{\underline{\xi}_{D,-} (t;P,\lambda,n)
       \defeq e^{\frac{n \lambda^2 \sigma^2_{D,-} (t;P,\lambda)}{2}}
                  Q(\rho_*+\sqrt{n} \lambda \sigma_{D,-} (t;P,\lambda))
               }
  \end{eqnarray}
  with $Q(\rho^*) = \frac{C_{BE} M_{\mathrm{D},-} (t;P,\lambda)}
                                    {\sqrt{n}\sigma^3_{\mathrm{D},-} (t;P,\lambda)}
          $
  and   $Q(\rho_*) = \frac{1}{2} - \frac{2 C_{BE} M_{\mathrm{D},-} (t;P,\lambda)}
                                                       {\sqrt{n}\sigma^3_{\mathrm{D},-} (t;P,\lambda)}
          $.
Similar to the case in Section \ref{sec:non-asympt-conv}, the purpose
of introducing above definitions is to utilize the following results,
proved as Theorem 8 in \cite{yang-meng:nep}, which are valid for any $t \in {\cal P}_n$ satisfying \eqref{eq-def-lambda*+} and \eqref{eq3rl-14-1}.

{\em
\begin{description}

 \item[(a)] There exists a $\delta^* >0$ such that for any $\delta \in (0, \delta^*]$  \begin{equation} \label{eqrl3-4-}
 r_{-} (t, \delta) = {1 \over 2 \sigma^2_D (t; P) } \delta^2 + O(\delta^3).
  \end{equation}

  \item[(b)] For   any $\delta \in (0, \Delta^*_{-} (t))$,  and any $x^n
    \in \mathcal{X}^n_t$,
  \begin{eqnarray} \label{eqrl3-17}
   \underline{\xi}_{D,-} (t; P, \lambda ,n) e^{- n r_{-} (t, \delta) } &\geq&
   { \Pr \left \{ \left. {1 \over n} \ln {p(Y^n |X^n)  \over
          q_t(Y^n) }\leq I(t; P) - \delta \right | X^n = x^n  \right
    \}} \nonumber \\
   &\geq& \underline{\xi}_{D,-} (t; P, \lambda ,n) e^{- n r_{-} (t, \delta) }
  \end{eqnarray}
  where $\lambda = {\partial  r_{-} (t, \delta)  \over \partial \delta} >0$, and $Y^n = Y_1 Y_2 \cdots Y_n$ is the output of the DIMC in response to an independent and identically distributed (IID)  input $X^n = X_1 X_2 \cdots X_n$, the  common distribution of each $X_i$ having $\cal X$ as its support set.
  Moreover, when $\delta = o(1)$ and $\delta = \Omega (1/\sqrt{n})$,
  \begin{eqnarray}
    \label{eql3-17-1}
    \bar{\xi}_{D,-} (t;P,\lambda,n) &=& e^{\frac{n \lambda^2 \sigma^2_{D,-} (t;P,\lambda)}{2}}
                                                   Q \left( \sqrt{n} \lambda \sigma_{D,-} (t;P,\lambda) \right)\left( 1 + o(1) \right) \\
    \label{eql3-17-2}
    \underline{\xi}_{D,-} (t;P,\lambda,n) &=& e^{\frac{n \lambda^2 \sigma^2_{D,-} (t;P,\lambda)}{2}}
                                                   Q \left( \sqrt{n} \lambda \sigma_{D,-} (t;P,\lambda) \right)\left( 1 - o(1) \right)
  \end{eqnarray}
  and
  \begin{equation}
    \label{eql3-17-3}
    e^{\frac{n \lambda^2 \sigma^2_{D,-} (t;P,\lambda)}{2}}
    Q \left( \sqrt{n} \lambda \sigma_{D,-} (t;P,\lambda) \right) = \Theta \left( \frac{1}{\sqrt{n} \lambda} \right)
  \end{equation}
  with $\lambda = r'_X (\delta) = \Theta (\delta)$.

   \item[(c)] For any  $ \delta \leq c \sqrt{\ln n \over n} $, where
     $c < \sigma_D (t; P)$ is a constant, and $x^n
    \in \mathcal{X}^n_t$,
  \begin{eqnarray} \label{eqrl3-17+}
    Q  \left ( {\delta \sqrt{n} \over \sigma_D (t; P)} \right ) - {C_{BE} M_D (t; P) \over \sqrt{n} \sigma^3_D (t; P)}
    & \leq  &   \Pr \left \{ \left. {1 \over n} \ln {p(Y^n |X^n)  \over q_t(Y^n) }\leq I(t; P) - \delta \right | X^n = x^n  \right \}
    \nonumber \\
  &  \leq &  Q  \left ( {\delta \sqrt{n} \over \sigma_D (t; P)} \right ) + {C_{BE} M_D (t; P) \over \sqrt{n} \sigma^3_D (t; P)}.
     \end{eqnarray}
  \end{description}
}

Turn our attention to sequences in  $\mathcal{Y}^n$.
For any $t \in {\cal P}_n$ and any $x^n \in \mathcal{X}^n_t$, define
\begin{equation} \label{eq-def-btd}
  B_t(x^n,\delta) \defeq \left\{ y^n:  -\infty < \frac{1}{n} \ln
  \frac{p(y^n|x^n)}{q_t(y^n)} \leq I(t;P) - \delta \right\}
\end{equation}
and
\begin{eqnarray} \label{eq-def-ptd}
  P_{t,\delta} &\defeq& P_{x^n} (B_t(x^n,\delta))  \nonumber \\
  &=& \Pr \left\{ \left.\frac{1}{n} \ln
  \frac{p(Y^n|X^n)}{q_t(Y^n)} \leq I(t;P) - \delta \right| X^n =
    x^n \right\}
\end{eqnarray}
where $P_{t,\delta}$ only depends on type $t$ and $\delta$. Since for any $y^n \in \mathcal{Y}^n$, the following set
\begin{equation} \label{eq-jar-dimc}
\left\{ x^n \in {\cal X}^n_t: \frac{1}{n} \ln
  \frac{p(y^n|x^n)}{q_t(y^n)} \geq I(t;P) - \delta
 \right\}
 \end{equation}
is referred to as a DIMC jar for $y^n$ based on type $t$ in
\cite{yang-meng:jardecoding}, we shall call $B_t (x^n,\delta) $ the {\em outer mirror image of jar}  corresponding to $x^n$.  Further define
\begin{equation} \label{eq-def-btnd}
  B_{t,n,\delta} \defeq \cup_{x^n \in \mathcal{X}^n_t} B_t (x^n, \delta)
\end{equation}
\begin{equation} \label{eq-def-pbtnd}
  P(B_{t,n,\delta}) \defeq \int_{y^n \in B_{t,n,\delta}} q_t(y^n) d y^n.
\end{equation}

\subsection{Converse Coding Theorem}

For any channel code $\mathcal{C}_n$ of block length $n$ with
  average word error probability $P_e (\mathcal{C}_n) = \epsilon_n$, assume that the message $M$ is uniformly distributed in $\{1,2, \ldots, e^{nR (\mathcal{C}_n)}\}$. Let $x^n(m)$ be the codeword corresponding to the
message $m$, and $\epsilon_{m,n}$  the conditional error probability given message $m$. Then
\begin{equation} \label{eq-proof-dsc--2}
  \epsilon_n = \mathbf{E} [\epsilon_{M,n}].
\end{equation}
Let $\beta_n = \sqrt{-2 \ln \epsilon_n \over n }$ and
\begin{equation} \label{eq-proof-dsc--1}
  \mathcal{M} \defeq \left\{ m: \epsilon_{m,n} \leq \epsilon_n
    (1+\beta_n) \right\}.
\end{equation}
Consider a type $t \in {\cal P}_n$ such that
\begin{equation} \label{eqth2-0}
 | \{ m \in  \mathcal{M}: t(x^n (m)) = t \} | \geq {| \mathcal{M}| \over (n+1)^{|{\cal X}|} }.
 \end{equation}
 Here and throughout the paper, $|S|$ denotes the cardinality of a finite set $S$. Since $|{\cal P}_n | \leq (n+1)^{|{\cal X}|}$, it follows from the pigeonhole principle  that such a type $t \in {\cal P}_n$ exists. In other words, if we classify codewords in $\{ x^n (m): m \in  \mathcal{M} \}$ according to their types, then there is at least one type $ t \in {\cal P}_n$ such that the number of codewords in $\{ x^n (m): m \in  \mathcal{M} \}$ with that type is not less than the average.

 We are now ready to state our converse theorem for DIMC.
\begin{theorem}
  \label{thm-dimc}
  Given a DIMC,
  for any  channel code $\mathcal{C}_n$  of block length $n$ with
  average word error probability $P_e (\mathcal{C}_n) = \epsilon_n$,
  \begin{eqnarray}
    \label{eq-thm-dsc-0}
    R(\mathcal{C}_n) &\leq& I(t;P) - \delta - \frac{\ln \epsilon_n - \ln P(B_{t,n,\delta})}{n} +
    |\mathcal{X}| \frac{\ln (n+1)}{n} \nonumber \\
    &&{-}\: \frac{\ln \frac{-2 \ln \epsilon_n}{ n} - \ln \left( 1 + \sqrt{\frac{-2 \ln \epsilon_n}{n}} \right)}{n}
  \end{eqnarray}
 for any $t \in {\cal P}_n$ satisfying \eqref{eqth2-0},  where $\delta$ is the largest number satisfying
  \begin{equation}
    \label{eq-thm-dsc-0+}
    \left( 1 + 2  \sqrt{\frac{-2 \ln \epsilon_n}{n}} \right) \epsilon_n \leq P_{t,\delta}.
  \end{equation}
  Moreover, if a type $t \in {\cal P}_n$ satisfying \eqref{eqth2-0} also satisfies  \eqref{eq-def-lambda*+} and \eqref{eq3rl-14-1}, then
    the following hold:
  \begin{enumerate}
  \item
  \begin{eqnarray}
    \label{eq-thm-dsc-1}
    R(\mathcal{C}_n) &\leq& I(t;P) - \delta - \frac{\ln \epsilon_n - \ln P(B_{t,n,\delta})}{n} +
    |\mathcal{X}| \frac{\ln (n+1)}{n} \nonumber \\
    &&{-}\: \frac{\ln \frac{-2 \ln \epsilon_n}{ n} - \ln \left( 1 + \sqrt{\frac{-2 \ln \epsilon_n}{n}} \right)}{n}
  \end{eqnarray}
  where $\delta$ is the solution to
  \begin{equation}
    \label{eq-thm-dsc-2}
    \left( 1 +2   \sqrt{\frac{-2 \ln \epsilon_n}{n}} \right)  \epsilon_n = \underline{\xi}_{D,-} (t;P,\lambda,n) e^{-n r_{-} (t,\delta) }
  \end{equation}
  with $\delta_{-} (t,\lambda) = \delta$.
  \item When $\epsilon_n = \frac{e^{- n^{\alpha}}}{2 \sqrt{\pi n^{\alpha}}} \left( 1 - \frac{1}{2 n^{\alpha}} \right)$ for $\alpha \in (0,1)$,
    \begin{equation}
      \label{eq-thm-dsc-2+}
      R(\mathcal{C}_n) \leq I(t;P) - \sqrt{2} \sigma_{D} (t;P) n^{-\frac{1-\alpha}{2}} + O(n^{-(1-\alpha)}) .
    \end{equation}
  \item When $\epsilon_n = \frac{n^{-\alpha}}{2 \sqrt{\pi \alpha \ln n}} \left( 1 - \frac{1}{2 \alpha \ln n} \right)$ for $\alpha > 0$,
    \begin{eqnarray}
      \label{eq-thm-dsc-3}
      R(\mathcal{C}_n) &\leq& I(t;P) - \sigma_D (t;P) \sqrt{\frac{2 \alpha \ln n}{n}} + O \left( {\frac{\ln n}{n} }\right).
    \end{eqnarray}
  \item When $\epsilon_n = \epsilon$ satisfying $\epsilon + \frac{1}{\sqrt{n}} \left( 2\epsilon \sqrt{-2 \ln \epsilon}  + \frac{
            C_{BE} M_{D}
        (t;P)}{\sigma^3_{D} (t;P)} \right) <1$,
    \begin{eqnarray}
      \label{eq-thm-dsc-6}
      R(\mathcal{C}_n)
      &\leq& I(t;P) - \frac{\sigma_{D} (t;P)}{\sqrt{n}} Q^{-1}
      \left( \epsilon + \frac{1}{\sqrt{n}} \left( 2\epsilon \sqrt{-2 \ln \epsilon}  + \frac{
            C_{BE} M_{D}
        (t;P)}{\sigma^3_{D} (t;P)} \right) \right) \nonumber \\
       &&{+}\: (|\mathcal{X}|+1) \frac{\ln
        n}{n} - \frac{\ln \epsilon}{n}\\
      \label{eq-thm-dsc-7}
      &=&  I(t;P) - \frac{\sigma_{D} (t;P)}{\sqrt{n}} Q^{-1}
      \left( \epsilon \right) +  (|\mathcal{X}|+1)\frac{\ln
        n}{n} + O(n^{-1}) .
    \end{eqnarray}
  \end{enumerate}
\end{theorem}

\begin{IEEEproof}
We again apply the outer mirror image of jar converse-proof technique.
By Markov inequality,
\begin{equation}
  \label{eq-proof-dsc-0}
  \Pr \{ M \in {\cal M} \} \geq \frac{\beta_n}{1+\beta_n} \mbox{ and }  |\mathcal{M}| \geq e^{nR(\mathcal{C}_n)
    + \ln \frac{\beta_n}{1+\beta_n} } .
\end{equation}
For any $t \in {\cal P}_n$ satisfying \eqref{eqth2-0}, let
\begin{equation} \label{eq-proof-dsc-0+}
  \mathcal{M}_t \defeq \left\{ m: \epsilon_{m,n} \leq \epsilon_n
    (1+\beta_n), t(x^n(m)) = t \right\} .
\end{equation}
Then
\begin{equation}
  \label{eq-proof-dsc-1}
  |\mathcal{M}_t| \geq \frac{|\mathcal{M}|}{(n+1)^{|\mathcal{X}|}} \geq
 e^{nR(\mathcal{C}_n)
    + \ln \frac{\beta_n}{1+\beta_n} - |\mathcal{X}|
    \ln (n+1)} .
\end{equation}
Denote the decision region for message $m \in \mathcal{M}_t$ as
$D_m$. Now for any $m \in \mathcal{M}_t$,
\begin{eqnarray}
  \label{eq-proof-dsc-2}
  P_{x^n(m)} ( B_t(x^n(m),\delta) \cap D_m ) &=&
  P_{x^n(m)} ( B_t(x^n(m),\delta) ) - P_{x^n(m)} ( B_t(x^n(m),\delta) \cap D^c_m ) \nonumber \\
  &\geq& P_{x^n(m)} ( B_t(x^n(m), \delta) ) - \epsilon_{m,n} \nonumber \\
  &\geq& P_{x^n(m)} ( B_t(x^n(m), \delta) ) - \epsilon_n(1+\beta_n)
\end{eqnarray}
At this point, we select $\delta$ such that for any $x^n \in \mathcal{X}^n_t$,
\begin{equation}
  \label{eq-proof-dsc-4}
  P_{x^n} ( B_t(x^n,\delta) ) = P_{t,\delta} \geq \epsilon_n (1 + 2 \beta_n).
\end{equation}
Substituting \eqref{eq-proof-dsc-4} into
\eqref{eq-proof-dsc-2}, we have
\begin{equation} \label{eq-proof-dsc-4-1}
  P_{x^n(m)}( B_t(x^n(m),\delta) \cap D_m ) \geq \beta_n \epsilon_n.
\end{equation}
By the fact that $D_m$ are disjoint for different $m$ and
\begin{equation} \label{eq-proof-dsc-4-2}
  \cup_{m \in \mathcal{M}_t} (D_m \cap B_t(x^n(m),\delta)) \subseteq B_{t,n,\delta},
\end{equation}
we have
\begin{eqnarray}
  \label{eq-proof-dsc-5-}
  P(B_{t,n,\delta}) &=& \int\limits_{B_{t,n,\delta}} q_t (y^n) dy^n \nonumber \\
     &\geq& \sum_{m \in \mathcal{M}_t } \int\limits_{B(x^n(m),\delta) \cap D_m}
     q_t (y^n) dy^n \nonumber \\
     &\geq& \sum_{m \in \mathcal{M}_t } \int\limits_{B(x^n(m),\delta)
       \cap D_m} p(y^n|x^n(m)) e^{-n(I(t;P)-\delta)} dy^n \nonumber \\
     &=&  \sum_{m \in \mathcal{M}_t } e^{-n(I(t;P) - \delta)} \int\limits_{B(x^n(m),\delta)
       \cap D_m} p(y^n|x^n(m)) dy^n \nonumber \\
      &=&  \sum_{m \in \mathcal{M}_t } e^{-n(I(t;P) - \delta)} P_{x^n(m)}(
      B(x^n(m),\delta) \cap D_m ) \nonumber \\
      &\geq& \sum_{m \in \mathcal{M}_t } e^{-n(I(t;P) - \delta)}
      \beta_n \epsilon_n = |\mathcal{M}_t| e^{-n(I(t;P) - \delta)} \beta_n \epsilon_n
\end{eqnarray}
which implies that
\begin{equation}
  \label{eq-proof-dsc-5}
  |\mathcal{M}_t| \leq e^{n(I(t;P) - \delta) - \ln \beta_n - \ln
    \epsilon_n + \ln P(B_{t,n,\delta})} .
\end{equation}
Then combining \eqref{eq-proof-dsc-1} and \eqref{eq-proof-dsc-5}
yields
\begin{equation}
  \label{eq-proof-dsc-5+}
  R (\mathcal{C}_n) \leq I(t;P) - \delta - \frac{\ln \epsilon_n - \ln P(B_{t,n,\delta})}{n} - \frac{\ln
    \frac{\beta_n}{1+\beta_n} }{n} - \frac{\ln \beta_n}{n} +
  |\mathcal{X}| \frac{\ln (n+1)}{n} .
\end{equation}
Since $\beta_n =  \sqrt{\frac{-2 \ln \epsilon_n}{n}}$ by definition, \eqref{eq-thm-dsc-0} and \eqref{eq-thm-dsc-0+}
directly come from \eqref{eq-proof-dsc-5+} and \eqref{eq-proof-dsc-4}.
\begin{enumerate}
\item According to \eqref{eqrl3-17},
it can be seen that selecting $\delta$ to be the solution to \eqref{eq-thm-dsc-2} will suffice
\eqref{eq-proof-dsc-4}. Consequently, \eqref{eq-thm-dsc-1} is proved.
\item The proof is essentially the same as that for part 2) of Theorem \ref{thm-bimc}, where we can show that
  \begin{equation}
    \label{eq-proof-dsc-5++}
    P_{t,\delta} \geq \left( 1 + 2  \sqrt{\frac{-2 \ln \epsilon_n}{n}}\right) \epsilon_n
  \end{equation}
  when $\epsilon_n = \frac{e^{-n^{\alpha}}}{2 \sqrt{\pi n^{\alpha}}} \left( 1 - \frac{1}{2 n^{\alpha}} \right)$ and
  $\delta = \sqrt{2} \sigma_D (t;P) n^{-\frac{1-\alpha}{2}} - \eta n^{-(1-\alpha)}$ for some constant $\eta$.
\item Apply the trivial bound $P(B_{t,n,\delta}) \leq 1$.
Then similar to the proof for part 3) of Theorem \ref{thm-bimc}, one can verify that
by making $\delta = \sigma_{D,-} (t;P) \sqrt{\frac{2 \alpha \ln n}{n}} - \eta {\frac{\ln n}{n}}$
for some properly chosen constant $\eta$,
\begin{eqnarray} \label{eq-proof-dsc-9+}
  P_{t,\delta} &\geq&
    \underline{\xi}_{D,-}
    \left( t;P, \frac{\partial r_{-} (t,{\delta})}{\partial {\delta}} , n \right)
    e^{- n r_{-} (t,\delta)} \nonumber \\
    &\geq& \left( 1 + 2  \sqrt{\frac{-2 \ln \epsilon_n}{n}}\right) \epsilon
\end{eqnarray}
for $\epsilon_n = \frac{n^{-\alpha}}{2 \sqrt{\pi \alpha \ln n}} \left( 1 - \frac{1}{2 \alpha \ln n} \right)$,
where \eqref{eqrl3-4-}, \eqref{eql3-17-2} and \eqref{eql3-17-3} are utilized.
\item According to
  \eqref{eq-proof-dsc-4}, we should select $\delta$ such that
  \begin{equation}
    \label{eq-proof-dsc-9}
   P_{t,\delta} \geq \left( 1 + 2  \sqrt{\frac{-2 \ln \epsilon}{n}} \right) \epsilon.
  \end{equation}
  Now by \eqref{eqrl3-17+},
  \begin{equation}
    \label{eq-proof-dsc-10}
    \delta = \frac{\sigma_{D} (t;P)}{\sqrt{n}} Q^{-1}
      \left( \epsilon + \frac{1}{\sqrt{n}} \left( 2 \epsilon \sqrt{-2 \ln \epsilon}  + \frac{ C_{BE} M_{D}
        (t;P)}{\sigma^3_{D} (t;P)} \right) \right)
  \end{equation}
  will guarantee \eqref{eq-proof-dsc-9}.
  Consequently,
  \eqref{eq-thm-dsc-6} is proved by substituting
  \eqref{eq-proof-dsc-9} and $\epsilon_n = \epsilon$ into
  \eqref{eq-proof-dsc-5+} and applying the trivial bound
  $P(B_{t,n,\delta}) \leq 1$, and \eqref{eq-thm-dsc-7} is yielded by
  the property of $Q^{-1}$ function shown in the proof of Theorem
  \ref{thm-bimc}.
\end{enumerate}
\end{IEEEproof}

\begin{remark}
  \label{re4}
  Remarks similar to Remarks \ref{re2} and \ref{re3} can be drawn
  here too for Theorem \ref{thm-dimc}.
 \end{remark}

For maximal error probability, we have the following corollary, which can be proved similarly.
\begin{corollary}
  \label{col-dimc}
  Given a DIMC, for any channel code $\mathcal{C}_n$ of block length $n$ with
  maximum error probability $P_m (\mathcal{C}_n) = \epsilon_n$,
  \begin{equation}
    \label{eq-thm-maxdsc-0}
    R(\mathcal{C}_n) \leq I(t;P) - \delta - \frac{\ln \epsilon_n - \ln
    P(B_{t,n,\delta})}{n} + |\mathcal{X}| \frac{\ln (n+1)}{n} - \frac{\ln  \sqrt{\frac{-2 \ln \epsilon_n}{n}}}{n}
  \end{equation}
  for any $t \in {\cal P}_n$ such that there are at least
  $(n+1)^{-|\mathcal{X}|}$ portion of codewords in $\mathcal{C}_n$
  with type $t$, where $\delta$ is the largest number satisfying
  \begin{equation}
    \label{eq-thm-maxdsc-0+}
    \left( 1 +  \sqrt{\frac{-2 \ln \epsilon_n}{n}} \right) \epsilon_n \leq P_{t,\delta}.
  \end{equation}
  Moreover, if  $t \in {\cal P}_n$  satisfies  \eqref{eq-def-lambda*+} and \eqref{eq3rl-14-1}, then the following hold:
  \begin{enumerate}
  \item
  \begin{equation}
    \label{eq-thm-maxdsc-1}
    R(\mathcal{C}_n) \leq I(t;P) - \delta - \frac{\ln \epsilon_n - \ln
    P(B_{t,n,\delta})}{n} + |\mathcal{X}| \frac{\ln (n+1)}{n}
    - \frac{\ln  \sqrt{\frac{-2 \ln \epsilon_n}{n}}}{n}
  \end{equation}
  where $\delta$ is the solution to
  \begin{equation}
    \label{eq-thm-maxdsc-2}
    \left( 1 +  \sqrt{\frac{-2 \ln \epsilon_n}{n}} \right) \epsilon_n = \underline{\xi}_{D,-} (t;P,\lambda,n) e^{-n r_{-} (t,\delta) }
  \end{equation}
  with $\delta_{-} (t,\lambda) = \delta$.
  \item When $\epsilon_n = \epsilon$ satisfying $\epsilon + \frac{1}{\sqrt{n}} \left( \epsilon \sqrt{-2 \ln \epsilon}  + \frac{
           C_{BE} M_{D} (t;P)}{\sigma^3_{D} (t;P)} \right) <1$,
    \begin{eqnarray}
      \label{eq-thm-maxdsc-6}
      R(\mathcal{C}_n) &\leq& I(t;P) - \frac{\sigma_{D} (t;P)}{\sqrt{n}} Q^{-1}
     \left( \epsilon + \frac{1}{\sqrt{n}} \left( \epsilon \sqrt{-2 \ln \epsilon}  + \frac{
           C_{BE} M_{D} (t;P)}{\sigma^3_{D} (t;P)} \right) \right) \nonumber
         \\
      &&{+}\:  (|\mathcal{X}|+0.5) \frac{\ln
        n}{n} - \frac{\ln \epsilon}{n} \\
    &=& I(t;P) - \frac{\sigma_{D} (t;P)}{\sqrt{n}} Q^{-1}
     \left( \epsilon \right) + (|\mathcal{X}|+0.5) \frac{\ln
        (n+1)}{n} + O(n^{-1}).
    \end{eqnarray}
  \end{enumerate}
\end{corollary}

  \subsection{Taylor-Type Expansion}

  Fix a DIMC $ P = \{p(y|x), x \in \mathcal{X}, y \in \mathcal{Y}\}$ with its capacity $C_{\mathrm{DIMC}} >0$. For any block length $n$ and average error probability $\epsilon$, let $R_n (\epsilon)$ be the best coding rate achievable with block length $n$ and average error probability $\leq \epsilon$, as defined in \eqref{eq-bt-1}. In this subsection, we extend Theorem~\ref{thm-BIMSC-second-order}  to establish a Taylor-type expansion of $R_n (\epsilon)$ in the case of DIMC.

We begin with reviewing the non-asymptotic achievability of jar decoding established in \cite{yang-meng:jardecoding}.   It has been proved
  in \cite{yang-meng:jardecoding} that under jar decoding, Shannon random codes  $\mathcal{C}_n$ of
  block length $n$ based on any type $t \in {\cal P}_n $ satisfying \eqref{eq-def-lambda*+} and \eqref{eq3rl-14-1} have the following performance:
  \begin{enumerate}
  \item
    \begin{equation}
      \label{eq-thm-dimc-2}
      {R} (\mathcal{C}_{n}) \geq {I} (t;P) - \delta - r_{-} (t,\delta) -
                                  \frac{( 0.5+ |\mathcal{X}|) \ln (n+1) - \ln \frac{2(1- C_{BE}) M_{\mathrm{D},-} (t;P,\lambda)}
                        {\sqrt{n}\sigma^3_{\mathrm{D},-} (t;P,\lambda)}  }{n}
    \end{equation}
    while maintaining
    \begin{equation}
      \label{eq-thm-dimc-1}
      P_e (\mathcal{C}_{n}) \leq \left( \bar{\xi}_{D,-} (t;P,\lambda,n) + \frac{2(1- C_{BE}) M_{\mathrm{D},-} (t;P,\lambda)}
                        {\sqrt{n}\sigma^3_{\mathrm{D},-} (t;P,\lambda)}  \right) e^{-n r_{-} (t,\delta) }
    \end{equation}
    for any $\delta \in (0, \Delta^*_{-} (t))$,
    where $ \lambda = {\partial r_{-} (t, \delta) \over \partial \delta }$
    satisfying $\delta_{-} (t,\lambda) = \delta$.
  \item
    \begin{equation}
      \label{eq-thm-dimc-4}
      {R} (\mathcal{C}_{n}) \geq I(t;P) -
      \sigma_{\mathrm{D}} (t; P) \sqrt{\frac{2 \alpha
          \ln n}{n}} - \frac{ (0.5+\alpha  + |\mathcal{X}|) \ln (n+1)}{n}
      - O \left( \frac{\ln \ln n}{n} \right)
    \end{equation}
    while maintaining
    \begin{equation}
      \label{eq-thm-dimc-3}
      P_e (\mathcal{C}_{n}) \leq
      \frac{n^{-\alpha}}{2 \sqrt{ \pi \alpha \ln n}} + O \left( n^{-\alpha} \frac{\ln n}{\sqrt{n}} \right)
      = \Theta \left( \frac{n^{-\alpha}}{\sqrt{\ln n}}\right)
    \end{equation}
    for any $\alpha \geq 0$.
  \item
    \begin{eqnarray}
      \label{eq-thm-dimc-6}
      {R} (\mathcal{C}_{n}) &\geq&
      {I}(t;P) -
      \frac{c}{\sqrt{n}} - \left( \frac{1}{2} + |\mathcal{X}| \right) \frac{\ln (n+1)}{n}
       - \frac{1}{n} \ln \frac{(1-C_{BE}) M_{\mathrm{D}}
      (t;P)}{\sigma^3_{\mathrm{D}} (t;P)}
       \end{eqnarray}
    while maintaining
    \begin{eqnarray}
      \label{eq-thm-dimc-5}
      P_e (\mathcal{C}_{n}) &\leq&
      Q \left(  \frac{c}{\sigma_{\mathrm{D}} (t;P)} \right) +
      \frac{M_{\mathrm{D}} (t; P)}{\sigma^3_{\mathrm{D}} (t; P)}
      \frac{1}{\sqrt{n}}
    \end{eqnarray}
    for any real number $c$.
  \end{enumerate}

By combining \eqref{eq-thm-dimc-2} and \eqref{eq-thm-dimc-1} with \eqref{eq-thm-dsc-0} and \eqref{eq-thm-dsc-0+} or with \eqref{eq-thm-dsc-1} and \eqref{eq-thm-dsc-2}, it is expected that $R_n (\epsilon)$ would be expanded as
\begin{equation} \label{eq-so-dimc-0-}
  R_n (\epsilon) = {I} (t;P) - \delta + o(\delta)
  \end{equation}
 for some $t \in {\cal P}$, where $\delta$ is defined according to \eqref{eq-thm-dimc-1},  \eqref{eq-thm-dsc-0+}, or \eqref{eq-thm-dsc-2}. In the rest of this subsection, we shall demonstrate with mathematic rigor that this is indeed the case. To simplify our argument, we impose the following conditions\footnote{Some of these conditions, for example, Condition C3, can be relaxed. Here we choose not to do so in order not to make our subsequent argument unnecessary complicated.} on the channel:
 \begin{description}
 \item[(C1)] For any $t \in {\cal P}$,  $M_{D} (t; P ) < \infty $.

 \item[(C2)]  $\sigma^2_D (t; P) =0$  implies $I(t; P) =0$.

 \item[(C3)] For any $t \in {\cal P}$,  $ \lambda^*_{-} (t; P) = +\infty$.

 \item[(C4)] There exists $\lambda^* >0$ such that $\delta_{-} (t, \lambda)$,  $\sigma^2_{D, -} (t; P, \lambda) $, $  M_{D, -} (t; P , \lambda)$,  $  \hat{M}_{D, -} (t; P , \lambda)$,  and $ r_{ -} (t, \delta_{-}(t, \lambda)) $ are continuous functions of $t $ and $\lambda$ over $(t, \lambda) \in {\cal P} \times [0, \lambda^*]$.

 \item[(C5)] There exists $s^* >0$ such that $r_{-}^{-1} (t, s) $ is a continuous  function of $t$ and $s$ over  $(t, s) \in {\cal P} \times [0, s^*]$, where $ r_{-}^{-1} (t, \cdot)  $ is an inverse function of $r_{-} (t, \cdot)$.
  \end{description}
  Since $r_{-} (t, \delta)$ is a continuous and strictly increasing function of $\delta$ before it reaches $+\infty$---which may or may not happen---it can be easily verified that for any $s \geq 0$
   \begin{eqnarray} \label{eq-so-dimc-0}
   r_{-}^{-1} (t, s) & = & \max \{ \delta: r_{-} (t, \delta) \leq s \} \nonumber \\
              & = & \inf \{ \delta:  r_{-} (t, \delta) >  s   \}.
      \end{eqnarray}
 In view of the definitions and properties of  $\delta_{-} (t, \lambda)$,  $\sigma^2_{D, -} (t; P, \lambda) $, $  M_{D, -} (t; P , \lambda)$,  $  \hat{M}_{D, -} (t; P , \lambda)$,  and $ r_{ -} (t, \delta) $  (see
 \cite{yang-meng:nep} for details and examples), Conditions (C1) to (C5) are generally met by most channels, particularly by channels with discrete output alphabets, and discrete input additive white Gaussian channels.

 To characterize $\delta$ in \eqref{eq-so-dimc-0-} analytically, we need a counterpart of Lemma~\ref{le1}. To this end, define for any $ 0< c < C_{\mathrm{DIMC}}$
  \begin{equation} \label{eq-so-dimc-0+}
  {\cal P} (c) \defeq \{ t \in {\cal P}: I(t; P) \geq c \}
  \end{equation}
  \begin{equation} \label{eq-so-dimc-0++}
  {\cal P}_n (c) \defeq \{ t \in {\cal P}_n: I(t; P) \geq c \}
  \end{equation}
  and  for any type $t \in {\cal P}$ satisfying $\sigma^2_D (t; P) >0$
  \begin{equation}
  \label{eq-so-dimc-1}
  g_{t;P,n} (\delta) \defeq e^{\frac{n \lambda^2 \sigma^2_{D, -} (t;P,\lambda)}{2}} Q(\sqrt{n} \lambda \sigma_{D, -} (t;P,\lambda)) e^{- n r_{-} (t,\delta)}
\end{equation}
where $\lambda = {\partial r_{-} (t, \delta) \over \partial \delta }$. Note that ${\cal P}(c)$ is a closed set, and it follows from Condition (C2) that $\sigma^2_D (t; P) >0$ for any $t \in {\cal P} (c) $. Interpret $ g_{t;P,n} (\delta)$ as a function of $\lambda$ through $\delta = \delta_{-} (t, \lambda)$. Then we have the following lemma.
\begin{lemma} \label{le2}
There exists $\lambda^+ >0$ such that for any $n >0$ and $ t \in {\cal P}(c) $, $ g_{t;P,n} ( \delta_{-} (t, \lambda) )$ is a strictly decreasing function of $\lambda$ over $\lambda \in [0, \lambda^+]$.
\end{lemma}

\begin{IEEEproof}
The proof is in parallel with that of Lemma~\ref{le1}. As such, we point out only places where differences occur. In the place of \eqref{eq-so-3-}, we now have
 \begin{equation} \label{eq-tm-3}
 \frac{d \sigma^2_{D, -}(t;P ,\lambda)}{d \lambda} = - \hat{M}_{D, -}(t;P ,\lambda) \;.
 \end{equation}
In parallel with \eqref{eq-so-14+} and \eqref{eq-so-15+}, we now have for any $t \in {\cal P}(c)$
\begin{eqnarray}
   \lefteqn{\frac{d g_{t;P,n} (\delta_{-} (t, \lambda))}{d \lambda}} \nonumber \\
     &\leq & e^{- n r_{-} (t, \delta_{-}(t, \lambda))}
           \frac{\sqrt{n} \sigma_{D, -} (t;P, \lambda)}{\sqrt{2 \pi}}
           \left (
    \left| - \frac{  \lambda \frac{d \sigma^2_{D, -}(t;P ,\lambda)}{d \lambda} }
                  {2 \sigma^2_{D, -}(t;P ,\lambda) \left( 1+ n \lambda^2 \sigma^2_{D,-}(t;P,\lambda) \right)} \right| -1 \right ) \nonumber \\
    & = & e^{- n r_{-} (t, \delta_{-}(t, \lambda))}
           \frac{\sqrt{n} \sigma_{D, -} (t;P, \lambda)}{\sqrt{2 \pi}}
           \left (
    \left|  \frac{  \lambda \hat{M}_{D, -}(t;P ,\lambda)  }
                  {2 \sigma^2_{D, -}(t;P ,\lambda) \left( 1+ n \lambda^2 \sigma^2_{D,-}(t;P,\lambda) \right)} \right| -1 \right )
                  \label{eq-so-le2-1} \\
  &  \leq &  e^{- n r_{-} (t, \delta_{-}(t, \lambda))}
           \frac{\sqrt{n} \sigma_{D, -}(t;P ,\lambda)}{\sqrt{2 \pi}}
           \left (
 \left|  \frac{  \lambda \hat{M}_{D, -}(t;P ,\lambda)  }
                  {2 \sigma^2_{D, -}(t;P,\lambda) } \right| -1 \right) .\label{eq-so-le2-2}
       \end{eqnarray}
 Since ${\cal P}(c)$ is closed, it then follows from Condition (C4) that  there is a $\lambda^+ >0$ such that for any $\lambda \in [0, \lambda^+] $ and any $t \in {\cal P}(c)$
 \[         \left|  \frac{  \lambda \hat{M}_{D, -}(t;P ,\lambda) }
                  {2 \sigma^2_{D, -}(t;P,\lambda) } \right| -1  <0 \]
 and hence
 \[       \frac{d g_{t;P,n} (\delta_{-} (t, \lambda))}{d \lambda} <0 \]
 for any $n >0$.  This completes the proof of Lemma~\ref{le2}.
 \end{IEEEproof}

 \begin{remark}
 In view of \eqref{eq-so-le2-1}, it is clear that when $n$ is large, $  g_{t;P,n} (\delta_{-} (t, \lambda)) $ is a strictly decreasing function of $\lambda$ over an interval even larger than $[0, \lambda^+]$ for each and every $t \in {\cal P}(c)$.
  \end{remark}

Now let
 \[ \epsilon_n^+ \defeq \max \{ g_{t;P,n} (\delta_{-} (t, \lambda^+ /2 )): t \in {\cal P}(c) \} \]
 which, in view of Condition (C4) and the fact that ${\cal P} (c)$ is closed, is well defined and also an exponential function of $n$. For any $ \epsilon_n^+ \leq \epsilon \leq 1/2$ and $t \in {\cal P}(c)$, let $\delta_{t,n} (\epsilon)$ be the unique solution to
\begin{equation}
  \label{eq-so-dimc-2}
  g_{t;P,n} (\delta) = \epsilon\;.
\end{equation}
Further define
\begin{equation} \label{eq-so-sc}
 s(c) \defeq \max \left\{ s: 0< s \leq s^*, r_{-}^{-1} (t, s) \leq { C_{\mathrm{DIMC}} - c \over 2} \;\; \forall t \in {\cal P}  \right\}
 \end{equation}
 and let $\epsilon_n (c)$ be the unique solution $\epsilon$ to
 \begin{equation} \label{eq-so-sc1}
    {- \ln \epsilon \left( 1 +2\sqrt{-2 \ln \epsilon \over n} \right) \over n} = s(c) .
    \end{equation}
 It is easy to see that in view of Condition (C5), $s(c) >0$ is well defined and once again $\epsilon_n (c)$ is also an exponential function of $n$. Let $\epsilon_n^u <1$ be the unique solution $\epsilon$ to
 \begin{equation} \label{eq-so-sc2}
  \epsilon \left( 1 +2\sqrt{-2 \ln \epsilon \over n} \right) =1 .
  \end{equation}
  Note that
  \[ \max\{ I (t; P): t \in {\cal P}_n \} = C_{\mathrm{DIMC}} - O \left( {1 \over n^2 } \right). \]
  Let $N(c)$ be the smallest integer $N >0$ such that
  \begin{equation} \label{eq-so-sc3}
      \max\{ I (t; P): t \in {\cal P}_n \} \geq C_{\mathrm{DIMC}} - {C_{\mathrm{DIMC}} -c \over 2}
      \end{equation}
  for all $n \geq N$. Then we have the following Taylor-type expansion of $R_n (\epsilon)$.

\begin{theorem}
  \label{thm-DIMC-second-order}
  For any $n \geq N(c)$ and any $\max\{\epsilon_n^+, \epsilon_n (c) \} \leq   \epsilon<  \epsilon_n^u $, let
  \begin{eqnarray}
    \label{eq-so-dimc-3+}
    t^* &\defeq& \argmax_{t \in \mathcal{P}_n (c) } \left[ I(t;P) - \delta_{t,n} (\epsilon) \right] \\
    \label{eq-so-dimc-3-}
    t^{\#} &\defeq& \argmax_{t \in \mathcal{P}_n (c)} \left[ I(t;P) - \frac{\sigma_D (t;P)}{\sqrt{n}} Q^{-1} (\epsilon) \right].
  \end{eqnarray}
  Then
  \begin{equation}
    \label{eq-so-dimc-3}
    \left| R_n (\epsilon) - \left( I(t^*;P) - \delta_{t^*,n} (\epsilon) \right) \right|
    \leq o \left( \delta_{t^*,n} (\epsilon) \right)
  \end{equation}
  where
  \begin{eqnarray}
    \label{eq-so-dimc-4}
    o \left( \delta_{t^*,n} (\epsilon) \right)
      &=& r_{-} (t^*, \delta_{t^*,n} (\epsilon)) + \frac{ (|\mathcal{X}|+1.5) \ln (n+1) + d_1}{n}
  \end{eqnarray}
  if $\epsilon \leq \frac{1}{3} $, and
  \begin{equation}
    \label{eq-so-dimc-5}
    \left| R_n (\epsilon) - \left( I(t^{\#};P) - \frac{\sigma_D (t^{\#};P)}{\sqrt{n}} Q^{-1} (\epsilon) \right) \right|
    \leq \frac{(|\mathcal{X}|+1) \ln (n+1) + d_2}{n}
  \end{equation}
   otherwise, where $d_1$ and $d_2$ are constants depending on the channel, but independent of $n$ and  $\epsilon$.
\end{theorem}

\begin{IEEEproof}
For any $t \in {\cal P}_n $ and $0 < \epsilon < 1$, let
 \[ \delta_{t, n}^P (\epsilon) =\sup \left \{ \delta >0: P_{t,\delta} \geq  \left( 1 + 2  \sqrt{\frac{-2 \ln \epsilon}{n}} \right) \epsilon \right \} .\]
 By Theorem~\ref{thm-dimc} and the trivial bound $P(B_{t,n,\delta}) \leq 1$, it is not hard to verify that
\begin{equation}
  \label{eq-so-dimc-6}
  R_n (\epsilon) \leq \max_{t \in {\cal P}_n } [ I(t;P) - \delta_{t, n}^P ]  - \frac{\ln \epsilon + \ln \frac{-2 \ln \epsilon}{n}}{n} +
  \frac{\ln \left( 1 + \sqrt{\frac{-2 \ln \epsilon}{n}} \right) + |\mathcal{X}| \ln (n+1)}{n}.
\end{equation}
Let us now examine
 \[\max_{t \in {\cal P}_n } [ I(t;P) - \delta_{t, n}^P ]  .\]
 In view of the Chernoff bound (see Theorem 8 in \cite{yang-meng:nep}),
   \[ P_{t,\delta} \leq e^{-n r_{-} (t, \delta)} \]
 for any $t \in {\cal P}_n $ and $\delta >0$, which, together with  \eqref{eq-so-dimc-0}, implies
  \begin{eqnarray} \label{eq-so-dt-1}
    \delta_{t, n}^P & \leq   & r_{-}^{-1} \left (t, { -\ln  \left( 1 + 2  \sqrt{\frac{-2 \ln \epsilon}{n}} \right) \epsilon \over n }\right )  \\
     & \leq  &  r_{-}^{-1} (t, s(c))  \label{eq-so-dt-2} \\
     & \leq & {C_{\mathrm{DIMC}} -c \over 2} \label{eq-so-dt-3}
        \end{eqnarray}
 whenever $\max\{\epsilon_n^+, \epsilon_n (c) \} \leq   \epsilon<  \epsilon_n^u $. In the above derivation, \eqref{eq-so-dt-1} is due to \eqref{eq-so-dimc-0}; and \eqref{eq-so-dt-2}  and \eqref{eq-so-dt-3}   follow from  \eqref{eq-so-sc}, \eqref{eq-so-sc1}, and \eqref{eq-so-sc2}.  Therefore,
 \begin{eqnarray} \label{eq-so-dt-4}
  \max_{t \in {\cal P}_n } [ I(t;P) - \delta_{t, n}^P ]
  & \geq &    \max_{t \in {\cal P}_n }  I(t;P) -  {C_{\mathrm{DIMC}} -c \over 2} \nonumber \\
  & \geq & c
  \end{eqnarray}
  where the last inequality is due to \eqref{eq-so-sc3}. In view of  \eqref{eq-so-dt-4}, it is not hard to see that for any $t \in {\cal P}_n$ achieving $\max_{t \in {\cal P}_n } [ I(t;P) - \delta_{t, n}^P ] $,
  \[  I(t; P) \geq c  + \delta_{t, n}^P \geq c \]
  and hence
    \[ \max_{t \in {\cal P}_n } [ I(t;P) - \delta_{t, n}^P ]  =  \max_{t \in {\cal P}_n (c) } [ I(t;P) - \delta_{t, n}^P ]  \]
 which, together with \eqref{eq-so-dimc-6}, implies
    \begin{equation}
  \label{eq-so-dt-5}
  R_n (\epsilon) \leq \max_{t \in {\cal P}_n (c) } [ I(t;P) - \delta_{t, n}^P ]  - \frac{\ln \epsilon + \ln \frac{-2 \ln \epsilon}{n}}{n} +
  \frac{\ln \left( 1 + \sqrt{\frac{-2 \ln \epsilon}{n}} \right) + |\mathcal{X}| \ln (n+1)}{n}.
\end{equation}

When $\epsilon > \frac{1}{3}$, it follows from \eqref{eqrl3-17+} and \eqref{eq-proof-dsc-10} that for any $t \in {\cal P}_n (c)$,
\begin{eqnarray} \label{eq-so-dt-6}
  \delta_{t, n}^P & \geq  & \frac{\sigma_{D} (t;P)}{\sqrt{n}} Q^{-1}
      \left( \epsilon + \frac{1}{\sqrt{n}} \left( 2 \epsilon \sqrt{-2 \ln \epsilon}  + \frac{ C_{BE} M_{D}
        (t;P)}{\sigma^3_{D} (t;P)} \right) \right)  \nonumber \\
        & \geq & \frac{\sigma_{D} (t;P)}{\sqrt{n}} Q^{-1}
      ( \epsilon ) - \sqrt{2\pi} e^{{[Q^{-1} (\epsilon)]^2 \over 2} } \frac{\sigma_{D} (t;P)}{n} \left( 2 \epsilon \sqrt{-2 \ln \epsilon}  + \frac{ C_{BE} M_{D}
        (t;P)}{\sigma^3_{D} (t;P)} \right)
      \end{eqnarray}
   Since ${\cal P} (c)$ is closed, it follows Condition (C4) that  $ \sigma_{D} (t;P)$ and   $\frac{  M_{D} (t;P)}{\sigma^3_{D} (t;P)}$ are bounded over ${\cal P} (c)$. Plugging \eqref{eq-so-dt-6} into \eqref{eq-so-dt-5} yields
   \[  R_n (\epsilon) \leq \max_{t \in {\cal P}_n (c) } \left [ I(t;P) - \frac{\sigma_{D} (t;P)}{\sqrt{n}} Q^{-1}
      ( \epsilon ) \right ]   + \frac{( |\mathcal{X}| +1) \ln (n+1) +d }{n}
    \]
  for some constant $d$,  which, together with the achievability in \eqref{eq-thm-dimc-6} and \eqref{eq-thm-dimc-5}, implies
      \eqref{eq-so-dimc-5}.

 Now let us focus on the case when $\epsilon \leq \frac{1}{3}$. For any $t \in {\cal P}(c)$, let $\underline{\delta}_{t, n} (\epsilon) $ be the unique solution to
\begin{equation}
  \label{eq-so-dimc-7}
  \left( 1 + 2 \sqrt{\frac{-2 \ln \epsilon}{n}} \right) \epsilon  =  \xi_{D,-} (t;P,\lambda,n) e^{-n r_{-} (t,\delta)}
\end{equation}
where $\lambda = {\partial r_{-} (t, \delta) \over \partial \delta}$.
By following the argument in the proof of Theorem \ref{thm-BIMSC-second-order}, it is not hard to verify that for any $t \in {\cal P}_n (c)$
\begin{equation} \label{eq-so-dt-7}
 \delta_{t, n}^P (\epsilon) \geq \underline{\delta}_{t, n} (\epsilon) \geq \delta_{t, n} (\epsilon) - { d \over n}
 \end{equation}
 for some constant $d$ independent of $n$, $\epsilon$, and $t$.  Plugging \eqref{eq-so-dt-7} into \eqref{eq-so-dt-5} then yields
\begin{equation}
  \label{eq-so-dimc-7-}
  R_n (\epsilon) \leq I(t^*;P) - \delta_{t^*,n} (\epsilon)
  - \frac{\ln \epsilon + \ln \frac{-2 \ln \epsilon}{n}}{n} +
  \frac{ \sqrt{\frac{-2 \ln \epsilon}{n}}+ |\mathcal{X}| \ln (n+1) +  d}{n} .
\end{equation}
In the meantime,
\begin{eqnarray}
  \label{eq-so-dimc-8}
  \epsilon &=& g_{t^*;P,n} (\delta_{t^*,n}) \nonumber \\
               &\geq&
  \frac{1}{\sqrt{2 \pi} \left( \sqrt{n} \lambda_{t^*,n} \sigma_{D,-} (t^*;P,\lambda_{t^*,n})
      + \frac{1}{ \sqrt{n} \lambda_{t^*,n} \sigma_{D, -} (t^*;P,\lambda_{t^*,n}) } \right)}
  e^{- n r_{-} (t^*,\delta_{t^*,n} (\epsilon))}
\end{eqnarray}
where $\lambda_{t^*,n} = \left. \frac{\partial r_{-} (t^*, \delta)}{\partial \delta} \right|_{\delta = \delta_{t^*,n} (\epsilon)}$. Consequently,
\begin{eqnarray}
  \label{eq-so-dimc-9}
  \frac{- \ln \epsilon}{n} &\leq& r_{-} (t^*, \delta_{t^*,n} (\epsilon))
  + \frac{\ln \left[ \sqrt{2 \pi} \left( \sqrt{n} \lambda_{t^*,n} \sigma_{D,-} (t^*;P,\lambda_{t^*,n})
      + \frac{1}{ \sqrt{n} \lambda_{t^*,n} \sigma_{D,-} (t^*;P,\lambda_{t^*,n}) } \right) \right]}{n}
  \nonumber \\
  &\leq& r_{-} (t^*, \delta_{t^*,n} (\epsilon)) + \frac{\ln n}{2 n} + \frac{\eta_1}{n}
\end{eqnarray}
where $\eta_1$ is a constant independent of $n$, $\epsilon$, and $t^*$. Now substituting
\eqref{eq-so-dimc-9} and $\epsilon \leq \frac{1}{3}$ into \eqref{eq-so-dimc-7-}
yields
\begin{eqnarray}
  \label{eq-so-dimc-11}
  R_n (\epsilon) &\leq&  I(t^*;P) - \delta_{t^*,n} (\epsilon) +
  r_{-} (t^*, \delta_{t^*,n} (\epsilon)) \nonumber \\
  &&{+}\:
  \frac{ - \ln \frac{2 \ln 3}{n} + \eta_1+\sqrt{r_{-} (t^*, \delta_{t^*,n} (\epsilon)) + \frac{1}{2 e} + \frac{\eta_1}{n}}
  + \frac{1}{2}\ln n + |\mathcal{X}| \ln (n+1) + d }{n} \nonumber \\
  &\leq& I(t^*;P) - \delta_{t^*,n} (\epsilon) +
  r_{-} (t^*, \delta_{t^*,n} (\epsilon)) + \frac{\underline{d}_1 + \left( |\mathcal{X}| + \frac{3}{2} \right) \ln (n+1)}{n}
\end{eqnarray}
for some constant $\underline{d}_1$  independent of $n$, $\epsilon$, and $t^*$, where the last inequality is due to the fact that in view of Condition (4), $ r_{-} (t^*, \delta_{t^*,n} (\epsilon)) $ is bounded over $t\in {\cal P}(c)$ and $\epsilon \geq \max \{ \epsilon_n^+, \epsilon_n (c) \}$.

To complete the proof, let us go back to the achievability given in \eqref{eq-thm-dimc-2} and \eqref{eq-thm-dimc-1}. Now choose $t$  to be $t^*$,  and fellow the argument in the proof of Theorem \ref{thm-BIMSC-second-order}. Then it is not hard to show that
\begin{equation}
  \label{eq-so-dimc-12}
  R_n (\epsilon) \geq I(t^*;P) - \delta_{t^*,n} (\epsilon)
  - r_{-} (t^*,\delta_n (\epsilon)) - \frac{(|\mathcal{X}|+1) \ln (n+1) + \bar{d}_1}{n}
\end{equation}
where $\bar{d}_1$ is a constant independent of $n$, $\epsilon$, and $t^*$.  Combining \eqref{eq-so-dimc-12} with \eqref{eq-so-dimc-11} completes the proof of Theorem~\ref{thm-DIMC-second-order}.
\end{IEEEproof}

Remarks similar to those immediately after Theorem \ref{thm-BIMSC-second-order} also apply here. In particular, Theorem~\ref{thm-DIMC-second-order} and the achievability of jar decoding given in \eqref{eq-thm-dimc-2}and  \eqref{eq-thm-dimc-1} to \eqref{eq-thm-dimc-6} and \eqref{eq-thm-dimc-5} once again imply that jar decoding is indeed optimal up to the second order coding performance in the non-asymptotical regime for any DIMC. In addition, the following remarks are helpful to the computation of the Taylor-type expansion of $R_n (\epsilon)$ as expressed in \eqref{eq-so-dimc-3+} to \eqref{eq-so-dimc-5}.

\begin{remark} \label{re_c}
When $I(t; P)$, $\delta_{-} (t, \lambda)$,  $\sigma^2_{D, -} (t; P, \lambda) $, $  M_{D, -} (t; P , \lambda)$,  $  \hat{M}_{D, -} (t; P , \lambda)$,  and $ r_{ -} (t, \delta_{-}(t, \lambda)) $ are all continuously differentiable with respect to $t$ over $ t \in {\cal P}(c)$ and $ \lambda \in [0, \lambda^*]$, which is true for most channels including particularly channels with discrete output alphabets, and discrete input additive white Gaussian channels,  $\mathcal{P}_n (c) $ in the definitions of $t^*$ and $t^{\#}$ can be replaced by  $\mathcal{P}(c)$. Thus, in this case,
\begin{eqnarray}
    \label{eq-so-dimc-3++}
    t^* &\defeq& \argmax_{t \in \mathcal{P} (c) } \left[ I(t;P) - \delta_{t,n} (\epsilon) \right] \\
    \label{eq-so-dimc-3--}
    t^{\#} &\defeq& \argmax_{t \in \mathcal{P} (c)} \left[ I(t;P) - \frac{\sigma_D (t;P)}{\sqrt{n}} Q^{-1} (\epsilon) \right].
  \end{eqnarray}
 Hereafter, we shall assume that the channel satisfies this continuously differentiable condition, and use \eqref{eq-so-dimc-3++} and \eqref{eq-so-dimc-3+}, or  \eqref{eq-so-dimc-3--} and \eqref{eq-so-dimc-3-}  interchangeably.

 \end{remark}

 \begin{remark}  \label{re_ci1}
It is worth pointing out the impact of $c$ on the maximization problems given in \eqref{eq-so-dimc-3++}, \eqref{eq-so-dimc-3+}, \eqref{eq-so-dimc-3--}, and \eqref{eq-so-dimc-3-}. In view of the definitions of $s(c)$ and $\epsilon_n (c)$ in \eqref{eq-so-sc} and \eqref{eq-so-sc1}, it is not hard to see that when $\epsilon $ is relatively large with respect to $n$ (in the sense that ${-\ln \epsilon \over n}$ is small), one can select $c$ to be close to $C_{\mathrm{DIMC}}$. In this case, it suffices to search a small range ${\cal P} (c)$ for optimal $t^*$.  On the other hand,
when $\epsilon $ is relatively small with respect to $n$, e.g., a exponential function of $n$, $c$ should be selected to be far below $C_{\mathrm{DIMC}}$ and hence one has to search a large range ${\cal P} (c)$ for optimal $t^*$.
\end{remark}

\begin{remark} \label{re_bt_dt}
  When the Taylor-type expansion of $R_n (\epsilon)$ in Theorem \ref{thm-DIMC-second-order} is applied to the case of BIMSC, it yields essentially the same result as in Theorem \ref{thm-BIMSC-second-order}, with explanation as follows. For any BIMSC, $t(0)$ fully charaterizes the type $t$. Then by symmetry, $\frac{\partial \delta_{t,n} (\epsilon)}{\partial t(0)} = 0$ at $t(0)=0.5$ for any $n$ and $\epsilon$. Note that $\delta_{t,n} (\epsilon) = \delta_n (\epsilon)$ when $t(0) = 0.5$, the capacity achieving input distribution. Therefore,
  \begin{eqnarray}
    \label{eq-so-dimc-13}
    \max_{t \in \mathcal{P} (c)} [ I(t;P) - \delta_{t,n} (\epsilon) ]
    &=& \max_{t \in \mathcal{P} (C_{\mathrm {BIMSC}} - O \left( \delta_n (\epsilon) \right) )} [ I(t;P) - \delta_{t,n} (\epsilon) ]
    \nonumber \\
    &=& C_{\mathrm{BIMSC}} - \delta_n (\epsilon) + O \left( \delta^2_n (\epsilon) \right) .
  \end{eqnarray}
  Consequently, by observing that the high order term $o(\delta_n (\epsilon))$ in Theorem \ref{thm-BIMSC-second-order} is also in the order of $\delta^2_n (\epsilon)$, the Taylor-type expansion of $R_n (\epsilon)$ for BIMSC in Theorem \ref{thm-DIMC-second-order} is shown to be the same as that in Theorem \ref{thm-BIMSC-second-order}.
\end{remark}

\subsection{Comparison with Asymptotic Analysis and Implication}

It is instructive to compare Theorem~\ref{thm-DIMC-second-order} with the second order asymptotic performance analysis as $n$ goes to $\infty$.

{\em Asymptotic analysis with constant $0< \epsilon <1$ and $n \to \infty$}: Fix $0 < \epsilon < 1$. It was shown  in \cite{strassen-1962}, \cite{Hayashi-2009}, \cite{Yury-Poor-Verdu-2010} that for a DIMC with a discrete output alphabet and $C_{\mathrm{DIMC}} >0$,
   \begin{equation} \label{eq-comp-dt1}
   R_n (\epsilon) =  C_{\mathrm{DIMC}} - \frac{\sigma_D (P)}{\sqrt{n}} Q^{-1} (\epsilon) + O \left ({\ln n \over n} \right )
   \end{equation}
 for sufficiently large $n$, where
    \[ \sigma_D (P) = \left \{ \begin{array}{cc}
      \min \{ \sigma_D (t; P): t \in {\cal P} \& I(t; P) = C_{\mathrm{DIMC}} \} & \mbox{ if } \epsilon < {1\over  2} \\
       \max \{ \sigma_D (t; P): t \in {\cal P} \& I(t; P) = C_{\mathrm{DIMC}} \} & \mbox{ if } \epsilon > {1\over  2} .
     \end{array}  \right. \]
Once again, the expression $  C_{\mathrm{DIMC}} - \frac{\sigma_D (P)}{\sqrt{n}} Q^{-1} (\epsilon) $ was referred to as the normal approximation for $R_n (\epsilon)$ in \cite{Yury-Poor-Verdu-2010}. It is not hard to verify that for sufficiently large $n$,
 \begin{eqnarray} \label{eq-comp-dt2}
 C_{\mathrm{DIMC}} - \frac{\sigma_D (P)}{\sqrt{n}} Q^{-1} (\epsilon)
 &\leq &  \max_{t \in \mathcal{P} (c)} \left[ I(t;P) - \frac{\sigma_D (t;P)}{\sqrt{n}} Q^{-1} (\epsilon) \right] \nonumber \\
  & = &  \max_{t : \exists p_X,  |t-p_X| = O \left( {1 \over n^{1/2} } \right) } \left[ I(t;P) - \frac{\sigma_D (t;P)}{\sqrt{n}} Q^{-1} (\epsilon) \right] \nonumber \\
  & = & C_{\mathrm{DIMC}} - \frac{\sigma_D (P)}{\sqrt{n}} Q^{-1} (\epsilon)  + O \left( {1 \over n} \right)
  \end{eqnarray}
 where the first equality is due to the fact that for any $p_X$ satisfying $I(p_X; P) = C_{\mathrm{DIMC}}$ and $t $ satisfying $|t -p_X | = \omega ( 1/n^{1/2} )$,
 \begin{displaymath}
   I(t;P) - \frac{\sigma_D (t;P)}{\sqrt{n}} Q^{-1} (\epsilon) \leq C_{\mathrm{DIMC}} - \frac{\sigma_D (p_X;P)}{\sqrt{n}} Q^{-1} (\epsilon)
 \end{displaymath}
  as
 \begin{displaymath}
   \frac{Q^{-1} (\epsilon)}{\sqrt{n}} |\sigma_D (t; P) - \sigma_D (p_X; P) |
   = O \left( \frac{|t -p_X|}{\sqrt{n}} \right) = o(|t -p_X |^2) = o(C_{\mathrm{DIMC}} - I(t;P)) .
 \end{displaymath}
 Therefore, when $\epsilon > 1/3$, \eqref{eq-comp-dt1} and \eqref{eq-so-dimc-5} are essentially the same for sufficiently large $n$.

 Let us now look at the case $\epsilon \leq 1/3$. Again, $ 0< \epsilon \leq 1/3$ is fixed. In parallel with \eqref{eq-comp-2} and \eqref{eq-comp-3}, we have for each $t \in {\cal P} (c)$
 \begin{eqnarray} \label{eq-comp-dt2+}
r_{-} (t, \delta) & = & {1 \over 2 \sigma^2_D (t; p) } \delta^2  +  \frac{ - \hat{M}_D (t; P) }{6 \sigma^6_D (t; P)}
 \delta^3 + O(\delta^4)
 \end{eqnarray}
 and
 \begin{equation} \label{eq-comp-dt3}
  \delta_{t, n} (\epsilon) = \frac{\sigma_D (t; P)}{\sqrt{n}} Q^{-1} (\epsilon)  + O \left ( {1 \over n} \right ) .
  \end{equation}
 Combining \eqref{eq-comp-dt3} with \eqref{eq-comp-dt2} yields
 \begin{eqnarray} \label{eq-comp-dt4}
 C_{\mathrm{DIMC}} - \frac{\sigma_D (P)}{\sqrt{n}} Q^{-1} (\epsilon) + O(1/n)
 &\leq &  \max_{t \in \mathcal{P} (c)} \left[ I(t;P) - \delta_{t; n} (\epsilon) \right] \nonumber \\
  & \leq & C_{\mathrm{DIMC}} - \frac{\sigma_D (P)}{\sqrt{n}} Q^{-1} (\epsilon)  + O \left( {1 \over n} \right) .
  \end{eqnarray}
 Thus the Taylor-type expansion of $R_n (\epsilon )$ in Theorem~\ref{thm-DIMC-second-order}  implies the second order asymptotic analysis with constant $0< \epsilon <1$ and $n \to \infty$ shown in \eqref{eq-comp-dt1}.

 {\em Asymptotic analysis with $n \to \infty$ and non-exponentially decaying $\epsilon$}: Suppose now $\epsilon$ is a function of $n$ and goes to $0$ as $n \to \infty$, but at a non-exponential speed. Using arguments similar to those made above and in Subsection~\ref{sec2-d}, one can show that  the Taylor-type expansion of $R_n (\epsilon )$ in Theorem~\ref{thm-DIMC-second-order} implies that in this case,
 $C_{\mathrm{DIMC}}$ and $ - \frac{\sigma_D (P)}{\sqrt{n}} Q^{-1} (\epsilon) $  are still respectively the first order and second order terms of of $R_n (\epsilon)$ in the asymptotic analysis with $n \to \infty$. Once again, to the best of our knowledge, the second order asymptotic analysis with $n \to \infty$ and non-exponentially decaying $\epsilon$ has not been addressed before in the literature.

 {\em Divergence from the normal approximation}: In the non-asymptotic regime where $n$ is finite and $\epsilon$ is generally relatively small with respect to $n$, the first two terms
 \[ \max_{t \in \mathcal{P} (c) } \left[ I(t;P) - \delta_{t,n} (\epsilon) \right]  \]
  in  the Taylor-type expansion of $R_n (\epsilon )$ in Theorem~\ref{thm-DIMC-second-order} differ from the normal approximation in a strong way. In particular, the optimal distribution $t^*$ defined in \eqref{eq-so-dimc-3++}  is not necessarily a capacity achieving distribution. In this case, the normal approximation would fail to provide a reasonable estimate for $R_n (\epsilon)$.

\begin{figure}[h]
  \centering
  \begin{tikzpicture}[rounded corners,ultra thick]
    \path (0,0) node(X0) {$0$} (5.5,0) node(Y0) {$0$}
             (0,1) node(X)   {$X$} (5.5,1) node(Y)   {$Y$}
             (0,2) node(X1) {$1$} (5.5,2) node(Y1) {$1$};

    \draw[->] (X0) -- (Y0) node[above,pos=0.5] {$1-p$};
    \draw[->] (X0) -- (Y1) node[above,pos=0.5] {$p$};
    \draw[->] (X1) -- (Y1) node[above,pos=0.5] {$1$};
  \end{tikzpicture}
  \caption{Z Channel}
  \label{fig:zch}
\end{figure}
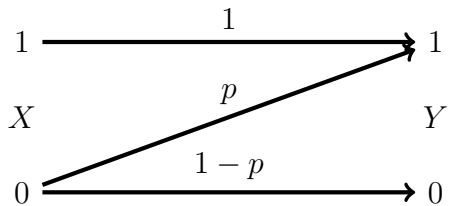

\begin{figure}[h]
  \centering
  \subfloat[$t(0)$ vs. $p$]{\includegraphics[scale=0.4]{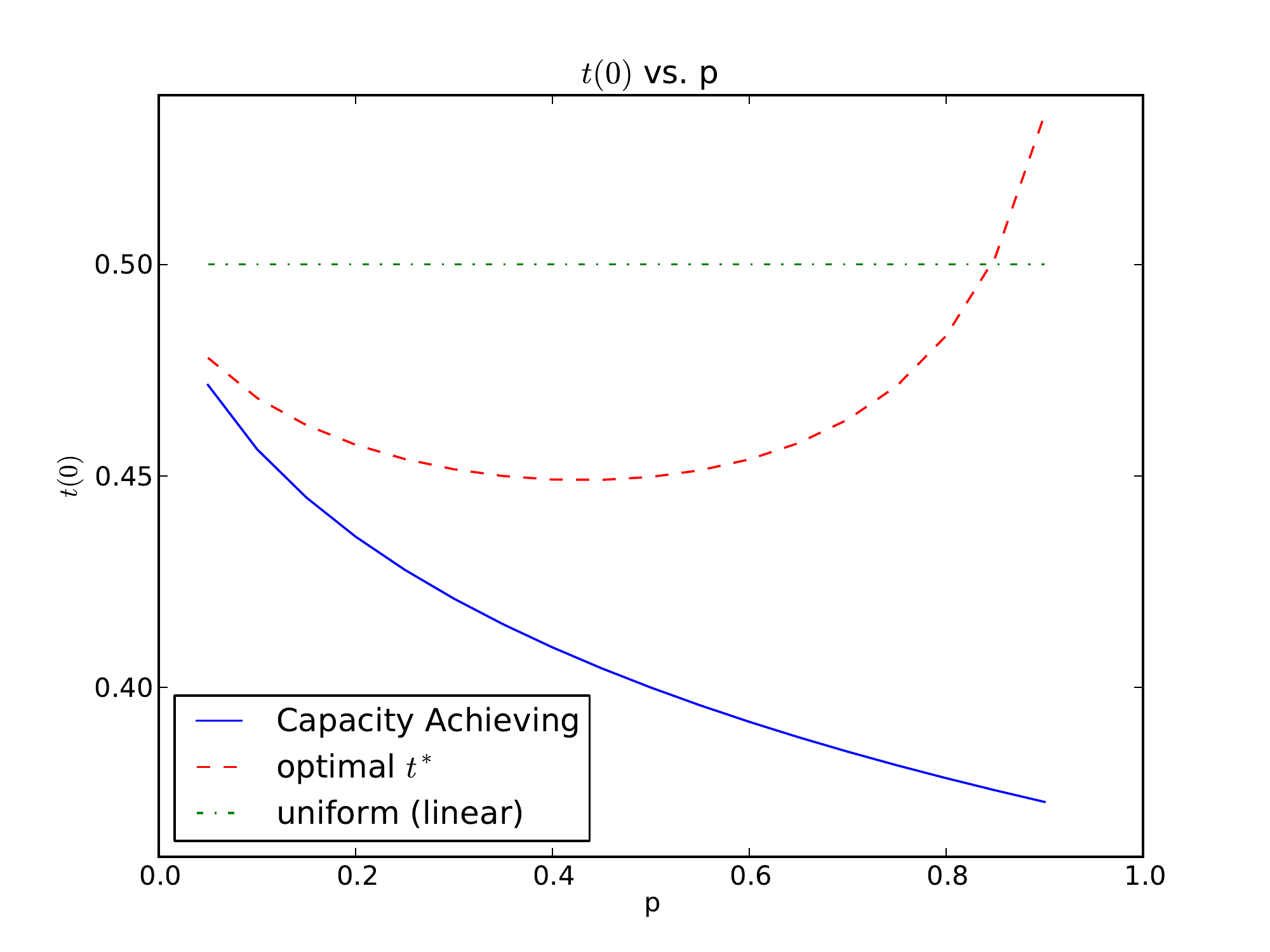}}
  \subfloat[$\frac{I(t;P) - \delta_{t,n} (\epsilon)}{I(t^*;P) - \delta_{t^*,n} (\epsilon)}$ for different $t$]{\includegraphics[scale=0.4]{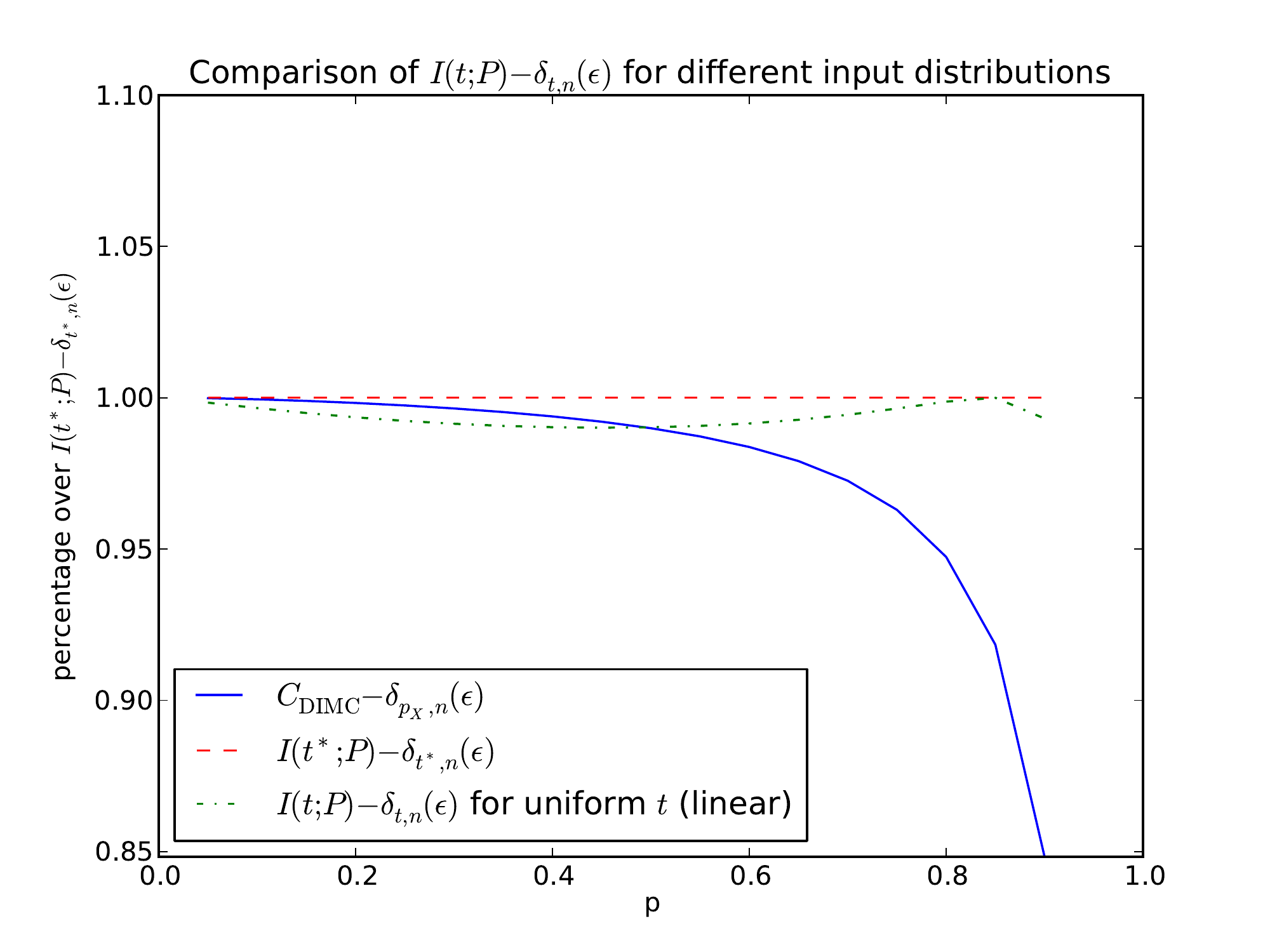}}
  \caption{Illustration for the Z channel with $n=1000$ and $\epsilon = 10^{-6}$: (a) comparison of $t^*$ with the capacity achieving distribution;  and (b) comparison of  $I(t;P) - \delta_{t,n} (\epsilon)$ among different distributions $t$.}
  \label{fig-zp}
\end{figure}

{\em Example:} Consider the Z channel shown in Figure~\ref{fig:zch}.
In this example, we  show that the optimal distribution $t^*$ defined in \eqref{eq-so-dimc-3++} is not a capacity achieving distribution. In the numerical calculation shown in Figure \ref{fig-zp}, the transition probability $p$ (i.e. $\Pr \{Y=1|X=0\}$) ranges from $0.05$ to $0.95$ with block length $n=1000$ and error probability $\epsilon = 10^{-6}$. As can be seen from Figure \ref{fig-zp}(a), $t^* (0)$ is always different from the capacity achieving $t (0)$. Moreover, Figure \ref{fig-zp}(b) shows the percentage of $I(t;P) - \delta_{t,n} (\epsilon)$ over $I(t^*;P) - \delta_{t^*,n} (\epsilon)$ when $t$ is capacity achieving, $t^*$,  and uniform respectively. It is clear  that $C_{\mathrm{DIMC}} - \delta_{p_X,n} (\epsilon)$ is apart from $I(t^*;P) - \delta_{t^*,n} (\epsilon)$ further and further when $p$ gets larger and larger, where $p_X$ is the capacity achieving distribution, indicating that under the practical block length and error probability requirement, Shannon random coding based on the capacity achieving distribution is not optimal.  It is also interesting to note that  for uniform $t$, $I(t;P) - \delta_{t,n} (\epsilon)$  is quite close to $I(t^*;P) - \delta_{t^*,n} (\epsilon)$ within the whole range,  implying  that linear block coding is quit suitable for the Z channel even under the practical block length and error probability requirement.

{\em Implication on code design}: An important implication arising from the Taylor-type expansion of $R_n (\epsilon )$ in Theorem~\ref{thm-DIMC-second-order} in the non-asymptotic regime is that for values of $n$ and $\epsilon$ with practical interest, the optimal marginal codeword symbol distribution is not necessarily a capacity achieving distribution. This is illustrated above for the Z channel. Indeed, other than for symmetric channels like BIMSC, it would expect that the optimal distribution $t^*$ defined in \eqref{eq-so-dimc-3++}  is in general not  a capacity achieving distribution for values of $n$ and $\epsilon$ for which $\delta_{t^*,n} (\epsilon)$ is not relatively small. As such, to design efficient channel codes under the practical block length and error probability requirement, one approach is to solve the maximization problem in \eqref{eq-so-dimc-3++}, get $t^*$, and then design codes so that the marginal codeword symbol distribution is approximately  $t^*$.

\section{Approximation and Evaluation}
\label{sec:appr-eval}
\setcounter{equation}{0}

Based on our converse theorems and Taylor-type expansion of $R_n (\epsilon)$, in this section, we first derive two approximation formulas for  $R_n (\epsilon)$. We then compare them numerically with the normal approximation and some tight (achievable and converse) non-asymptotic bounds, for the  BSC, BEC, BIAGC,  and Z Channel. In all Figures \ref{fig-bsc} to \ref{fig-z-2},  rates are expressed in bits.

\subsection{Approximation Formulas}

In view of the Taylor-type expansion of $R_n (\epsilon )$ in Theorem~\ref{thm-DIMC-second-order}, one reasonable approximation formula is to use the first two terms in Taylor-type expansion of $R_n (\epsilon )$ as an estimate for $R_n (\epsilon )$. We refer to this formula as the second order (SO) formula:
\begin{eqnarray}
  \label{eq-nep-z-2}
  R^{\mathrm{SO}}_n  (\epsilon) &=&   \max_{t \in \mathcal{P} (c) } \left[ I(t;P) - \delta_{t,n} (\epsilon) \right]  \nonumber \\
 & = &  I(t^*;P) - \delta_{t^*; P} (\epsilon)
\end{eqnarray}
where $c$ is selected according to Remark~\ref{re_ci1}.

To derive the other approximation formula for $R_n (\epsilon)$, let us put Theorem~\ref{thm-dimc}, Theorem~\ref{thm-DIMC-second-order}, and the achievability given in \eqref{eq-thm-dimc-2} and \eqref{eq-thm-dimc-1} together. It would make sense for  an optimal code of block length $n$ to draw all its codewords from the same type
$t$ with $|t - t^*| = O(1/n)$. In this case, it is not hard to see that the term $|\mathcal{X}|\frac{\ln (n+1)}{n}$ in the bounds of Theorems~\ref{thm-dimc} and  \ref{thm-DIMC-second-order} (i.e. \eqref{eq-thm-dsc-0}, \eqref{eq-thm-dsc-1},  \eqref{eq-so-dimc-4},  and \eqref{eq-so-dimc-5}) can be dropped. By ignoring the higher order term $ \frac{\ln \frac{-2 \ln \epsilon_n}{ n} - \ln \left( 1 + \sqrt{\frac{-2 \ln \epsilon_n}{n}} \right)}{n}$ in  \eqref{eq-thm-dsc-0} and \eqref{eq-thm-dsc-1}, we get the following approximation  formula (dubbed ``NEP'') :
\begin{equation}
  \label{eq-nep-z-1}
  R^{\mathrm{NEP}}_n (\epsilon) = I(t^*;P) - \delta_{t^*; P} (\epsilon) - \frac{\ln \epsilon}{n} +
  \frac{1}{n} \ln P(B_{t^*,n,\delta_{t^*; P} (\epsilon)})
  \end{equation}

  Rewrite the normal approximation as
  \begin{equation}
  \label{eq-nep-z-3}
  R^{\mathrm{Normal}}_n (\epsilon) = C_{\mathrm{DIMC}} - \frac{\sigma_D (P)}{\sqrt{n}} Q^{-1} (\epsilon).
\end{equation}

\subsection{BIMSC}
\label{sec:bimsc}

In the case of BIMSC, it follows from  Theorem \ref{thm-BIMSC-second-order} and Remark~\ref{re_bt_dt} that  $
 R^{\mathrm{SO}}_n  (\epsilon)$, $ R^{\mathrm{NEP}}_n (\epsilon)$, and $ R^{\mathrm{Normal}}_n (\epsilon)$ become respectively
 \begin{displaymath}
  R^{\mathrm{SO}}_n (\epsilon) =  C_{\mathrm{BIMSC}} - \delta_n (\epsilon)
\end{displaymath}
\begin{equation}
  \label{eq-app-1}
  R^{\mathrm{NEP}}_n (\epsilon) =  C_{\mathrm{BIMSC}} - \delta_n (\epsilon) - \frac{\ln \epsilon}{n} +
  \frac{1}{n} \ln P(B_{n,\delta_n (\epsilon)})
\end{equation}
and
\begin{equation}
  \label{eq-app-3}
  R^{\mathrm{Normal}}_n (\epsilon) = C_{\mathrm{BIMSC}} - \frac{\sigma_H
    (X|Y)}{\sqrt{n}} Q^{-1} (\epsilon) .
\end{equation}
From Theorem \ref{thm-BIMSC-second-order} and its comparison with asymptotic analysis, we can expect that when $\delta_n (\epsilon)$ is extremely small,  $
 R^{\mathrm{SO}}_n  (\epsilon)$ and $ R^{\mathrm{Normal}}_n (\epsilon)$ are close, and both can provide a good approximation for $R_n (\epsilon)$. However, as $\delta_n (\epsilon)$ increases,
the relative position of  $
 R^{\mathrm{SO}}_n  (\epsilon)$ and $ R^{\mathrm{Normal}}_n (\epsilon)$ depends on
\begin{displaymath}
 \zeta_{X|Y} = - \frac{\hat{M}_H (X|Y)}{6 \sigma^6_H (X|Y)}.
\end{displaymath}
Specifically, given a channel with large magnitude of $\zeta_{X|Y}$, $ R^{\mathrm{Normal}}_n (\epsilon)$  is not reliable, as it can be much below achievable bounds or  above converse bounds. On the other hand, as shown later on, $
 R^{\mathrm{SO}}_n  (\epsilon)$  is much more reliable. Moreover, $ R^{\mathrm{NEP}}_n (\epsilon)$, which has  some terms beyond second order on top of $ R^{\mathrm{SO}}_n (\epsilon)$, always provides a good approximation for $R_n (\epsilon)$ even if $\delta_n (\epsilon)$ is relatively large.

\subsubsection{BSC}

For this channel, the trivial bound $P(B_{n,\delta_n (\epsilon)}) \leq 1$ is applied in the evaluation of $ R^{\mathrm{NEP}}_n (\epsilon)$,.
Before jumping into the comparison of those approximations, let us first get some insight by
investigating $\zeta_{X|Y}$. It can be easily verified that for BSC with cross-over probability $p$,
\begin{equation}
  \label{eq-app-4}
  \zeta_{X|Y} = - \frac{1}{6 \ln^5 \frac{1-p}{p}} \frac{1-2p}{p^3(1-p)^3} .
\end{equation}
As can be seen, $\zeta_{X|Y}$ is always negative for any $p \in (0,1)$ and $\zeta_{X|Y} \rightarrow -\infty$
as $p \rightarrow 0$. Therefore, in the case of a very small $p$,  $ R^{\mathrm{Normal}}_n (\epsilon)$ will be larger than $ R^{\mathrm{SO}}_n (\epsilon)$ by a relatively large margin, and even larger than the converse bound.

\begin{figure}[h]
  \centering
  \subfloat[Bounds with $P_m = 10^{-3}$]{\includegraphics[scale=0.4]{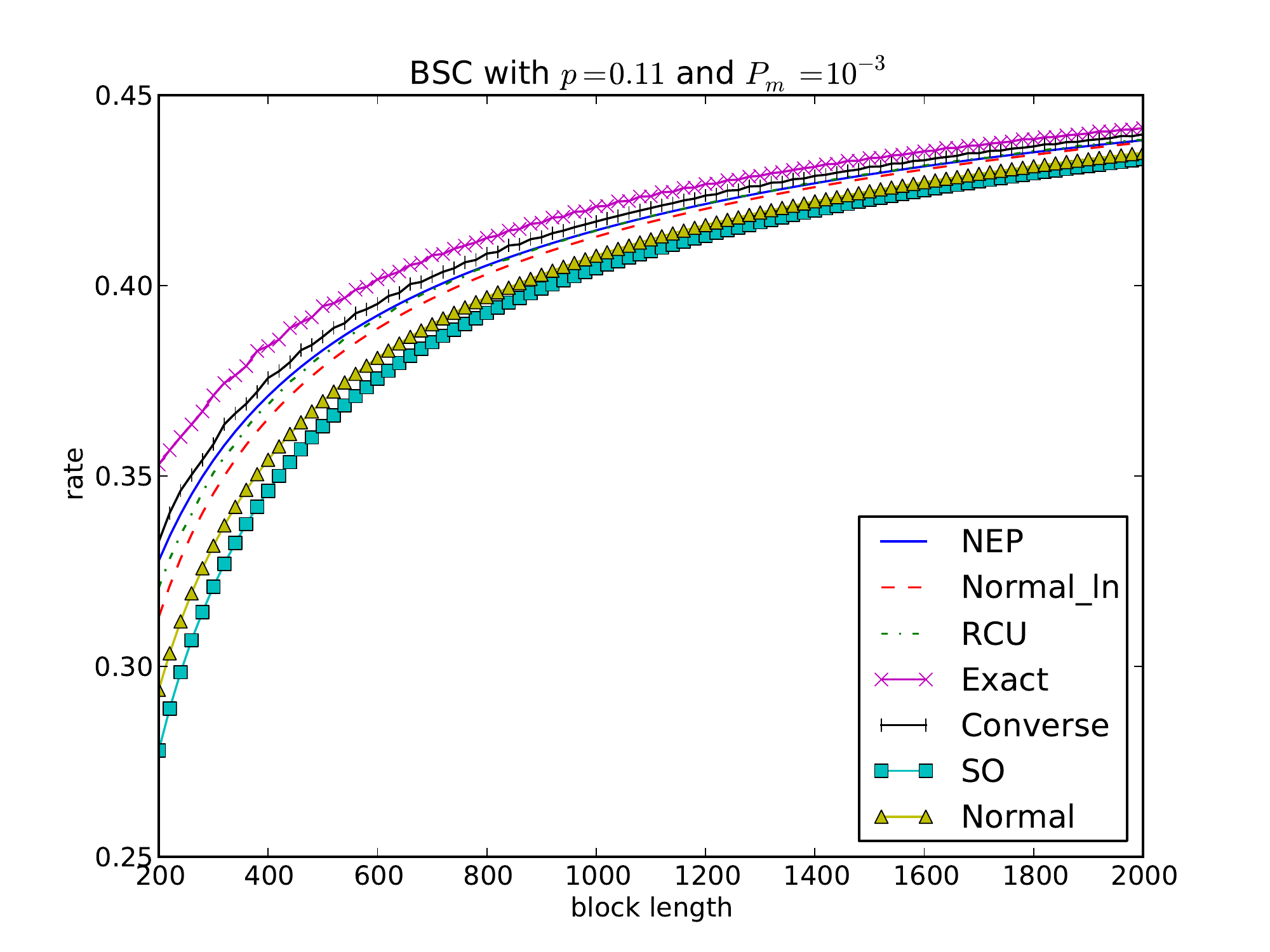}}
  \subfloat[$\delta_n(\epsilon)$ with $P_m = 10^{-3}$]{\includegraphics[scale=0.4]{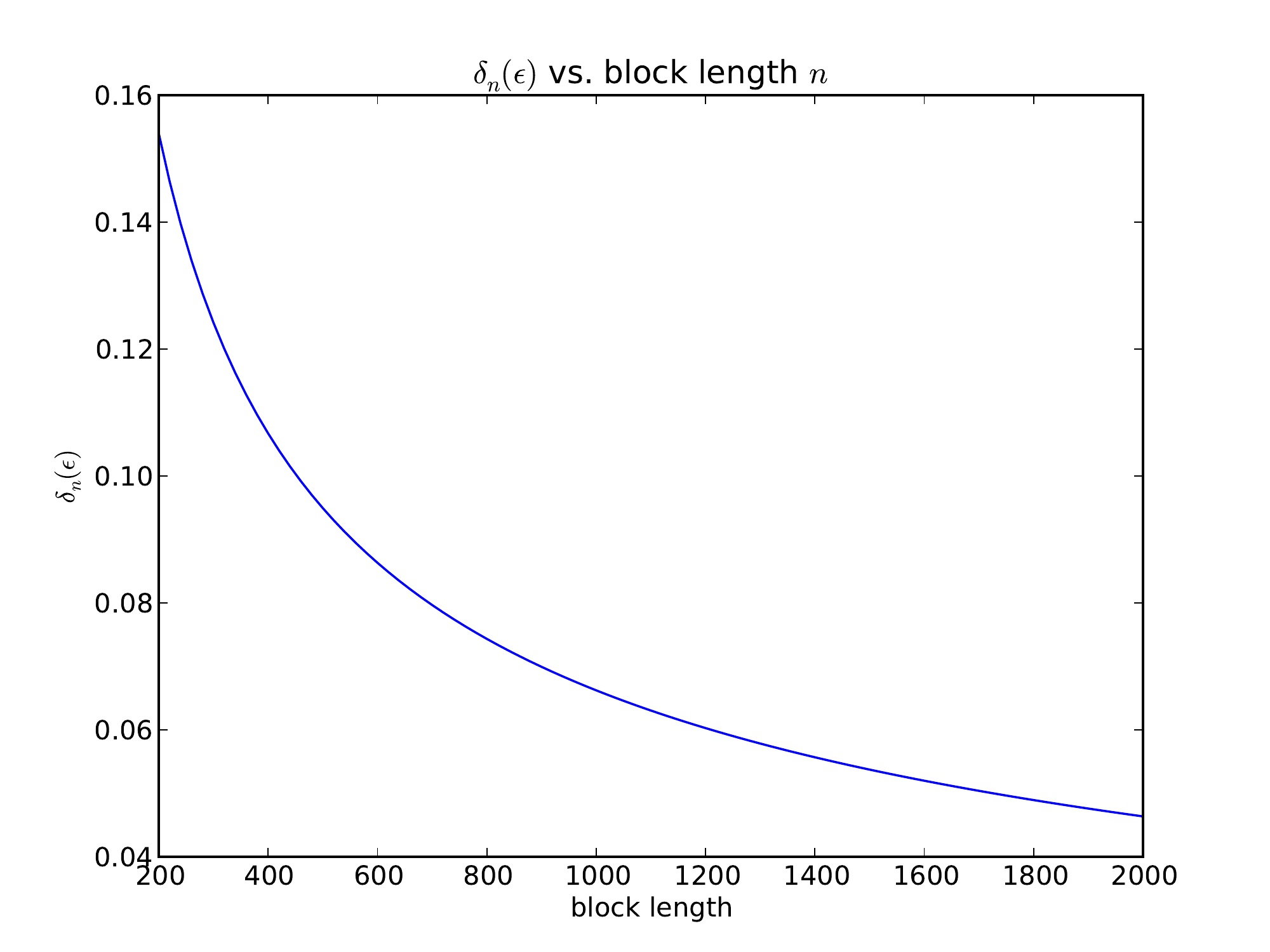}} \\
  \subfloat[Bounds with $P_m = g_{X|Y,n} (\delta)$ and $\delta=0.06$]{\includegraphics[scale=0.4]{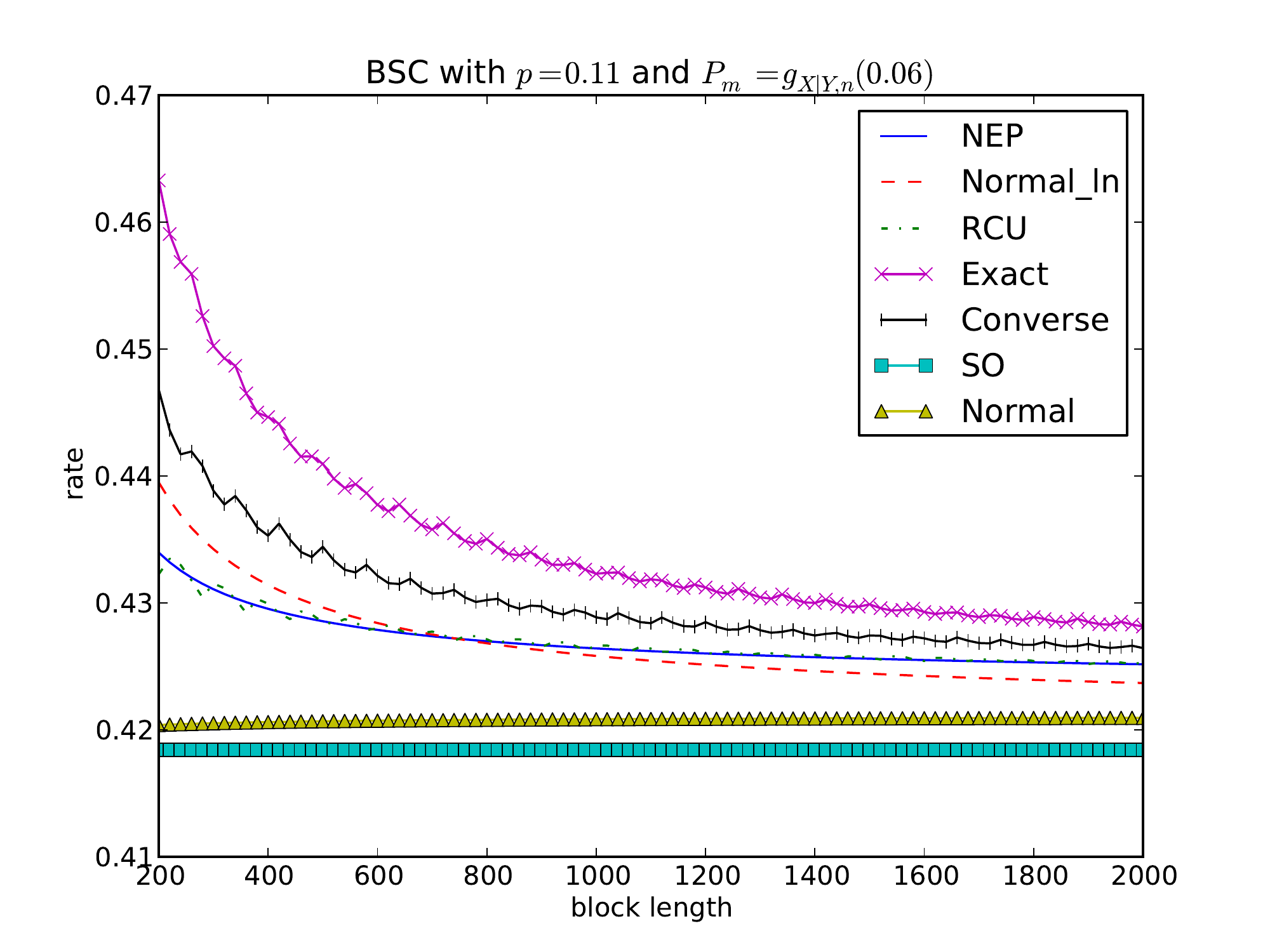}}
  \subfloat[$\log_{10} P_m$ with $P_m = g_{X|Y,n} (\delta)$ and $\delta=0.06$]{\includegraphics[scale=0.4]{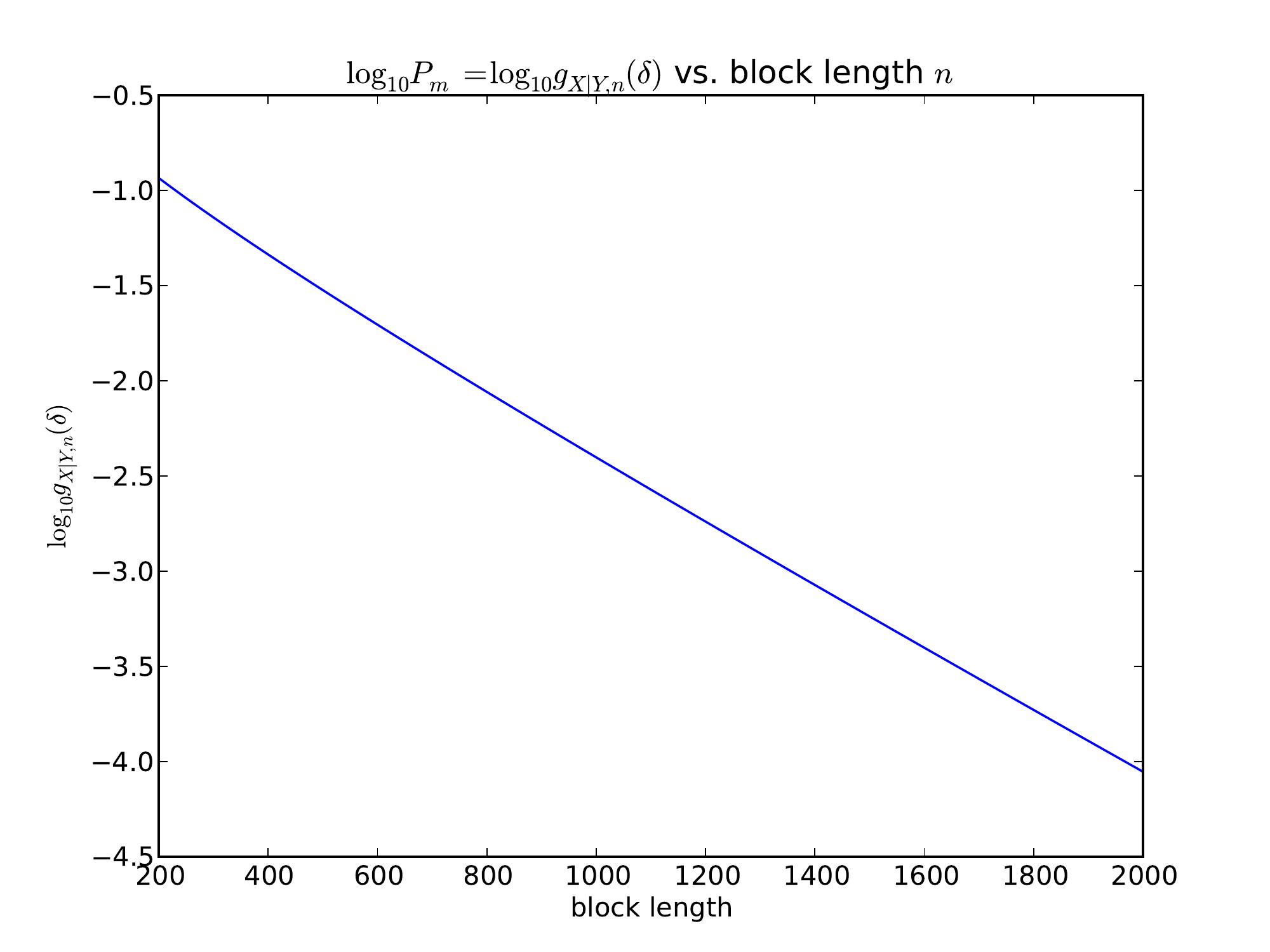}}
  \caption{Comparison of different bounds for BSC with $p=0.11$.}
  \label{fig-bsc}
\end{figure}

\begin{figure}[h]
  \centering
  \subfloat[Bounds with $P_m = 10^{-6}$]{\includegraphics[scale=0.4]{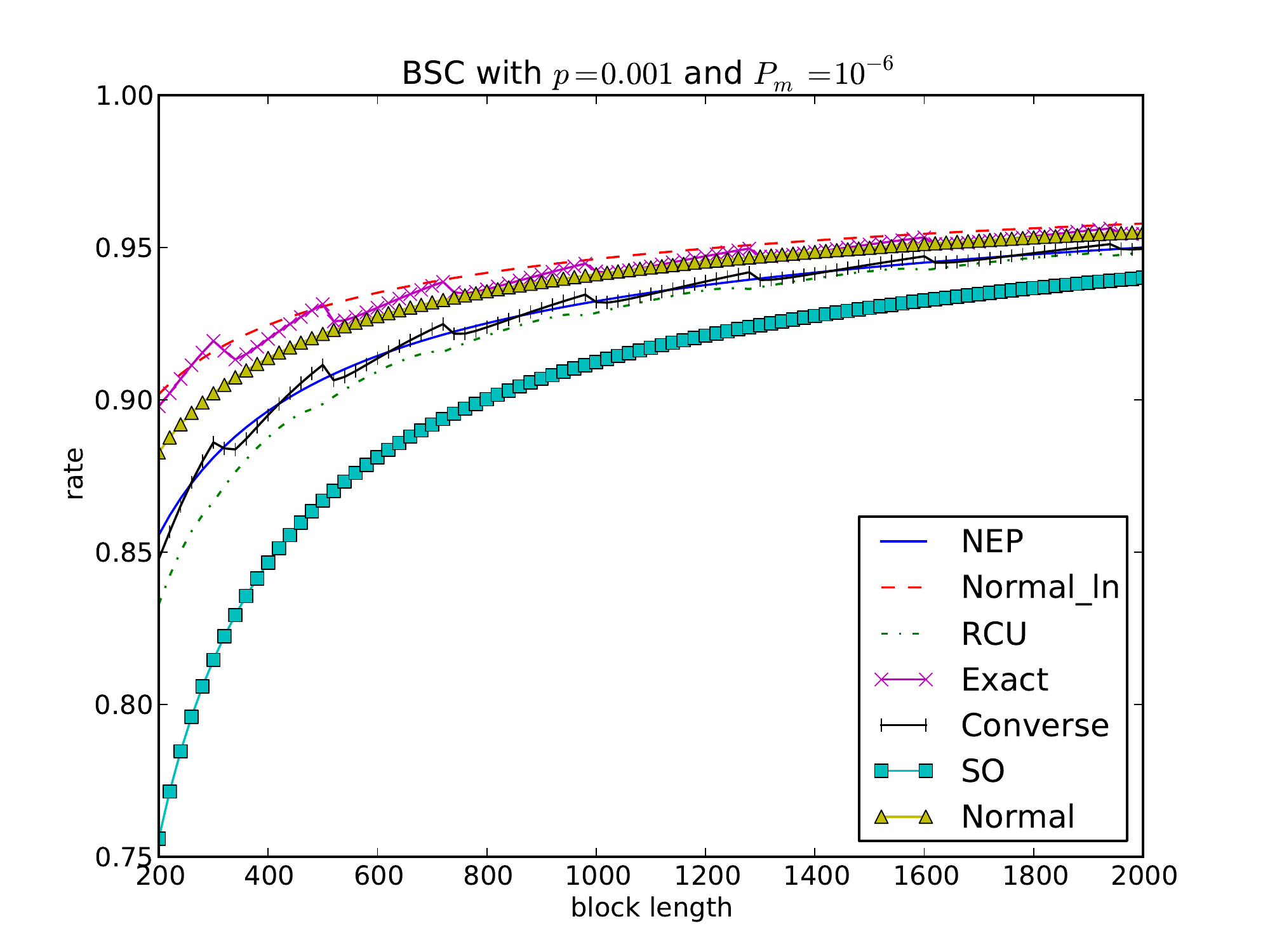}}
  \subfloat[$\delta_n(\epsilon)$ with $P_m = 10^{-6}$]{\includegraphics[scale=0.4]{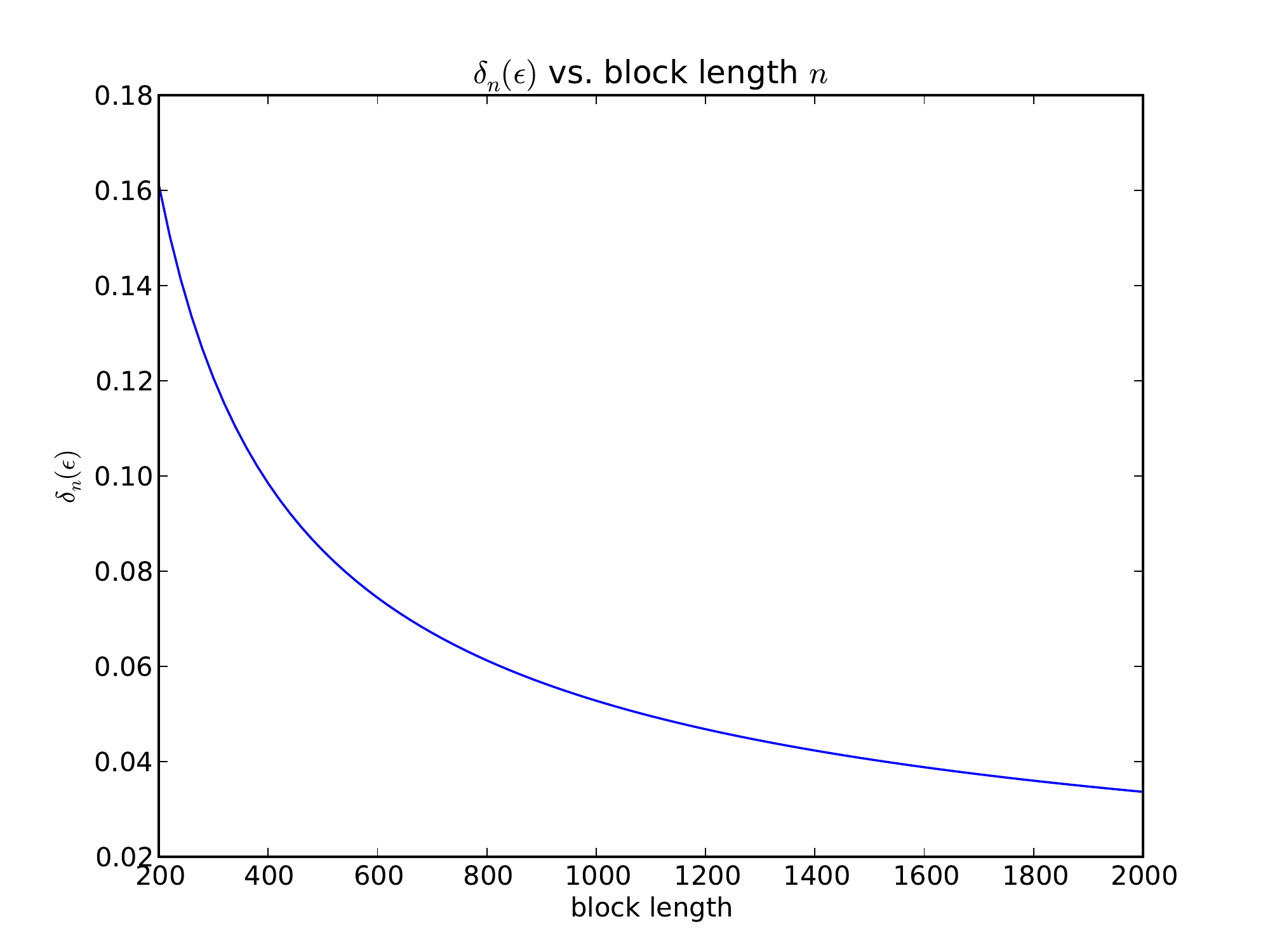}} \\
  \subfloat[Bounds with $P_m = g_{X|Y,n} (\delta)$ and $\delta=0.04$]{\includegraphics[scale=0.4]{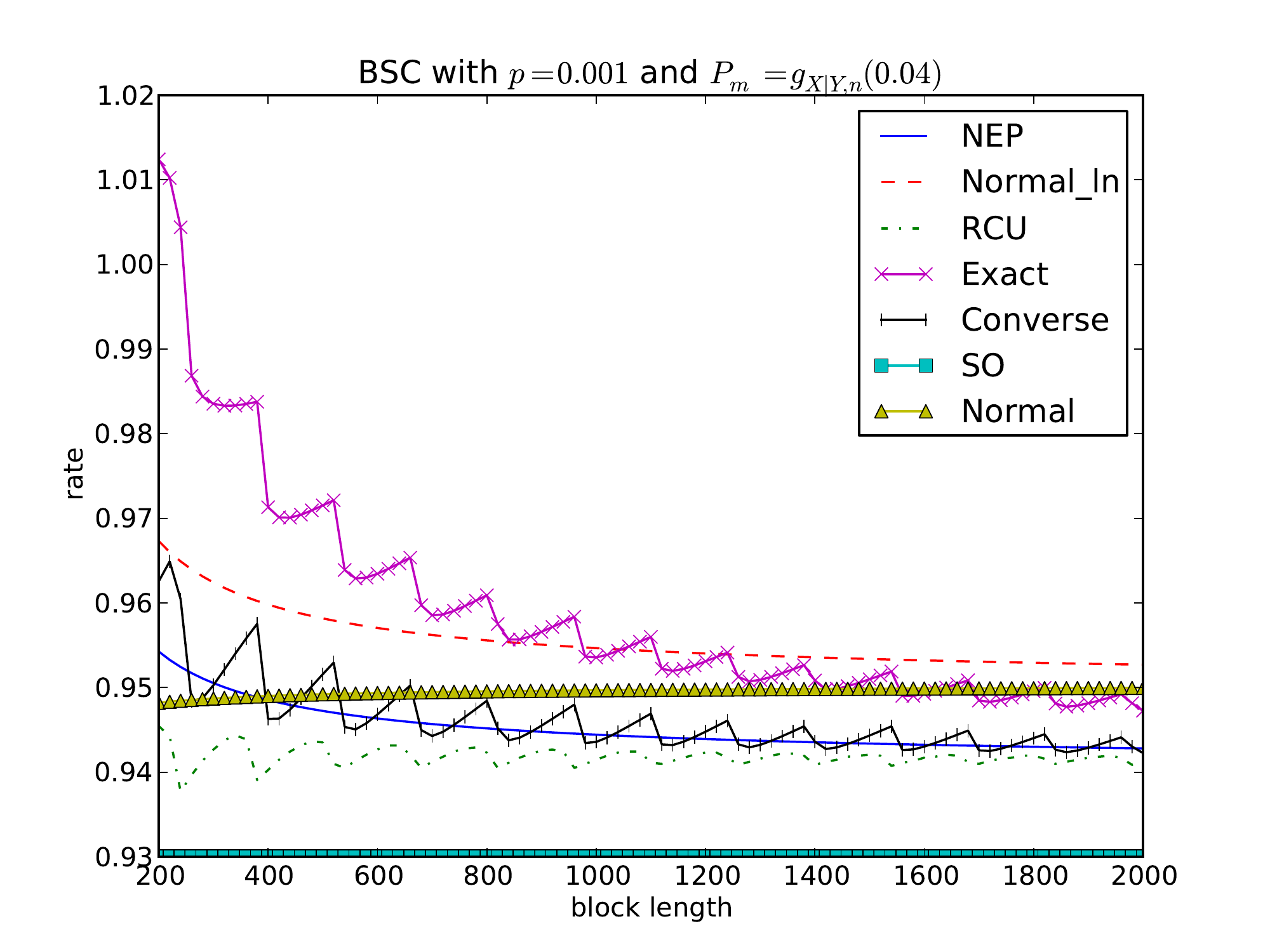}}
  \subfloat[$\log_{10} P_m$ with $P_m = g_{X|Y,n} (\delta)$ and $\delta=0.04$]{\includegraphics[scale=0.4]{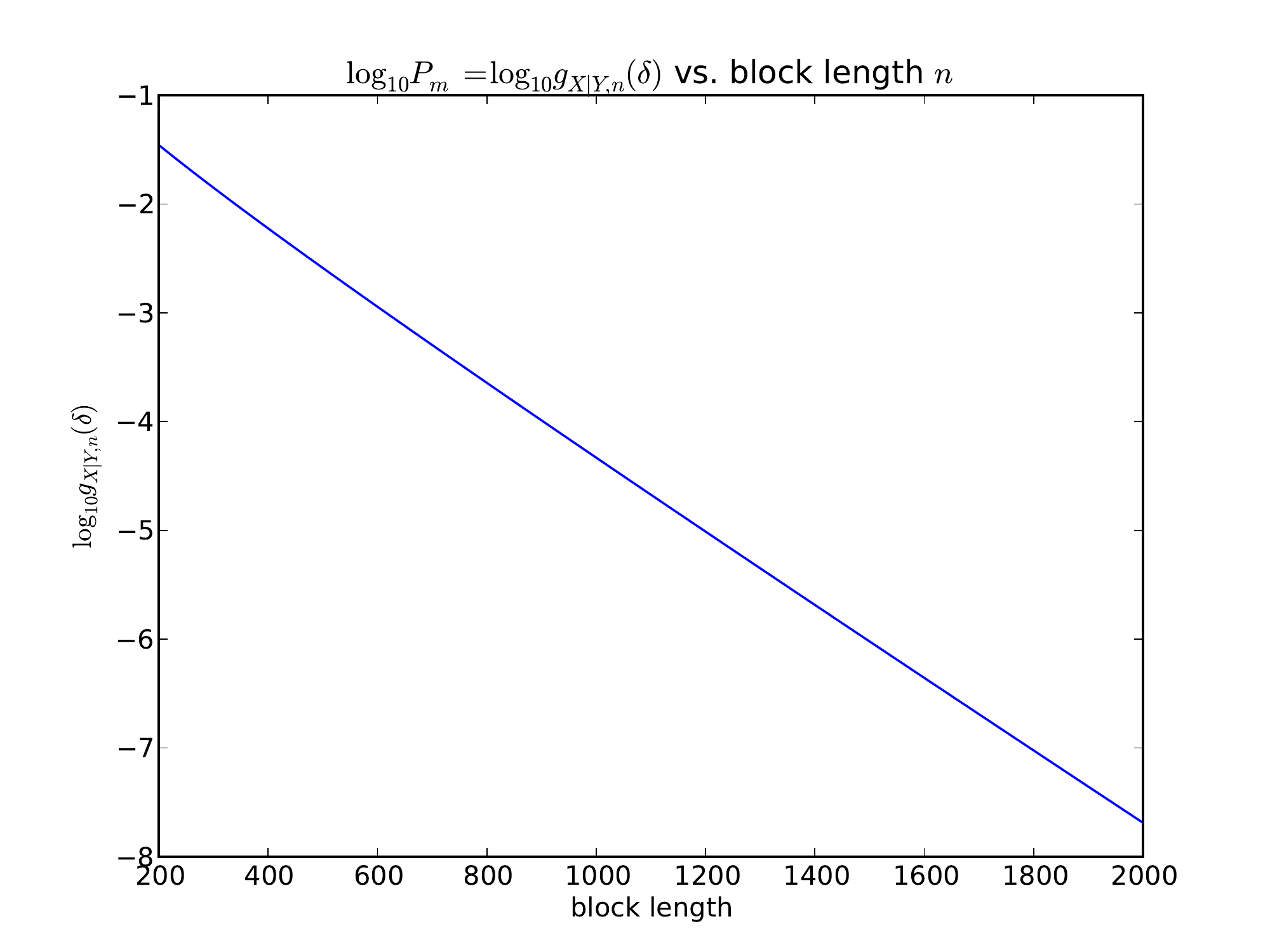}}
  \caption{Comparison of different bounds for BSC with $p=0.001$.}
  \label{fig-bsc-2}
\end{figure}

Now in order to compare those approximations, we invoke Theorem 33 (dubbed ``RCU'') and Theorem 35 (dubbed ``Converse'') in \cite{Yury-Poor-Verdu-2010}, which serve as an achievable bound and a converse bound, respectively. In addition, another converse bound is provided by the exact calculation of \eqref{eq-thm-maxbimc-3} and \eqref{eq-thm-maxbimc-4} in Corollary \ref{col-bimc} (dubbed ``Exact''). Moreover, by Theorem 52 in \cite{Yury-Poor-Verdu-2010}, $\frac{\ln n}{2 n}$ is the third order in the asymptotic analysis of $R_n (\epsilon)$ as $ n \to \infty$ for BSC, and therefore, another approximation is yielded by adding $\frac{\ln n}{2 n}$ to the normal approximation (dubbed ``Normal\_ln''). Then these four approximation formulas (NEP, Normal\_ln,  Normal, SO), two converse bounds (Converse, Exact), and one achievable bound (RCU) are compared against each other with block length $n$ ranging from 200 to 2000; their respective performance is  shown in Figures \ref{fig-bsc} and \ref{fig-bsc-2}.

 In Figure \ref{fig-bsc}, the target channel is the BSC with cross-over probability 0.11, where $\zeta_{X|Y}$ is relatively small. In Figure~\ref{fig-bsc}(a),  bounds are compared with fixed maximum error probability $P_m=10^{-3}$, while $\delta_n (\epsilon)$ changes with respect to block length $n$, shown in Figure~\ref{fig-bsc}(b). In the meantime, Figure~\ref{fig-bsc}(c) shows comparison of these bounds when $\delta_n (\epsilon)$ is fixed to be $0.06$, while $P_m = g_{X|Y,n} (0.06)$ is shown in Figure~\ref{fig-bsc}(d). As can be seen, when $\delta_n (\epsilon)$ gets smaller, the SO and Normal curves tend to coincide with each other. Moreover, since the SO and Normal approximation formulas  are quite close in this case,  both the NEP and Normal\_ln provide quite accurate approximations for $R_n (\epsilon)$ with the NEP slightly better.

 Figure \ref{fig-bsc-2} shows the same curves as those in Figure \ref{fig-bsc}, but for the BSC with cross-over probability $0.001$. In this case, the magnitude of $\zeta_{X|Y}$ is large, and therefore,  the SO and Normal curves are well apart. In fact, the Normal curve is even above those two converse bounds, and so does the Normal\_ln curve, thus confirming our analysis based on $\zeta_{X|Y}$ made at the beginning of this discussion for BSC. On the other hand, the SO curve stays at the same relative position to achievable and converse bounds, and  the NEP  still provides  an accurate approximation for $R_n (\epsilon)$.

\subsubsection{BEC}

 This special channel serves as another interesting example to illustrate the difference
between the SO and Normal approximations. On one hand, it can be easily verified that
\begin{equation}
  \label{eq-app-5}
  P(B_{n,\delta}) = \Pr \left\{ -\frac{1}{n} \ln p(X^n|Y^n) > H(X|Y) + \delta \right\}  \approx g_{X|Y,n} (\delta)
\end{equation}
and therefore, $- \frac{\ln \epsilon}{n}$ and $ \frac{1}{n} \ln P(B_{n,\delta_n (\epsilon)})$ are cancelled out in $ R^{\mathrm{NEP}}_n (\epsilon)$, which is then identical to $ R^{\mathrm{SO}}_n (\epsilon)$. On the other hand,
\begin{equation}
  \label{eq-app-6}
  \zeta_{X|Y} = - \frac{(1-2p)}{6 p^2 (1-p)^2 \ln^3 2}
  \left\{
  \begin{array}{cc}
    < 0 & \mbox{if $p<0.5$} \\
    = 0 & \mbox{if $p=0.5$}\\
    > 0 & \mbox{if $p>0.5$}
  \end{array}
  \right. .
\end{equation}
Therefore, the Normal curve can be all over the map, i.e. it can be above some converse when $p < 0.5$, and  below an achievable bound when $p > 0.5$. When $p=0.5$, the Normal curve happens to be close to the SO curve, hereby explaining why it provides an accurate  approximation for $R_n (\epsilon)$ in this particular case, as shown in \cite{Yury-Poor-Verdu-2010}.

\begin{figure}[h]
  \centering
  \subfloat[Bounds with $P_m = 10^{-6}$]{\includegraphics[scale=0.4]{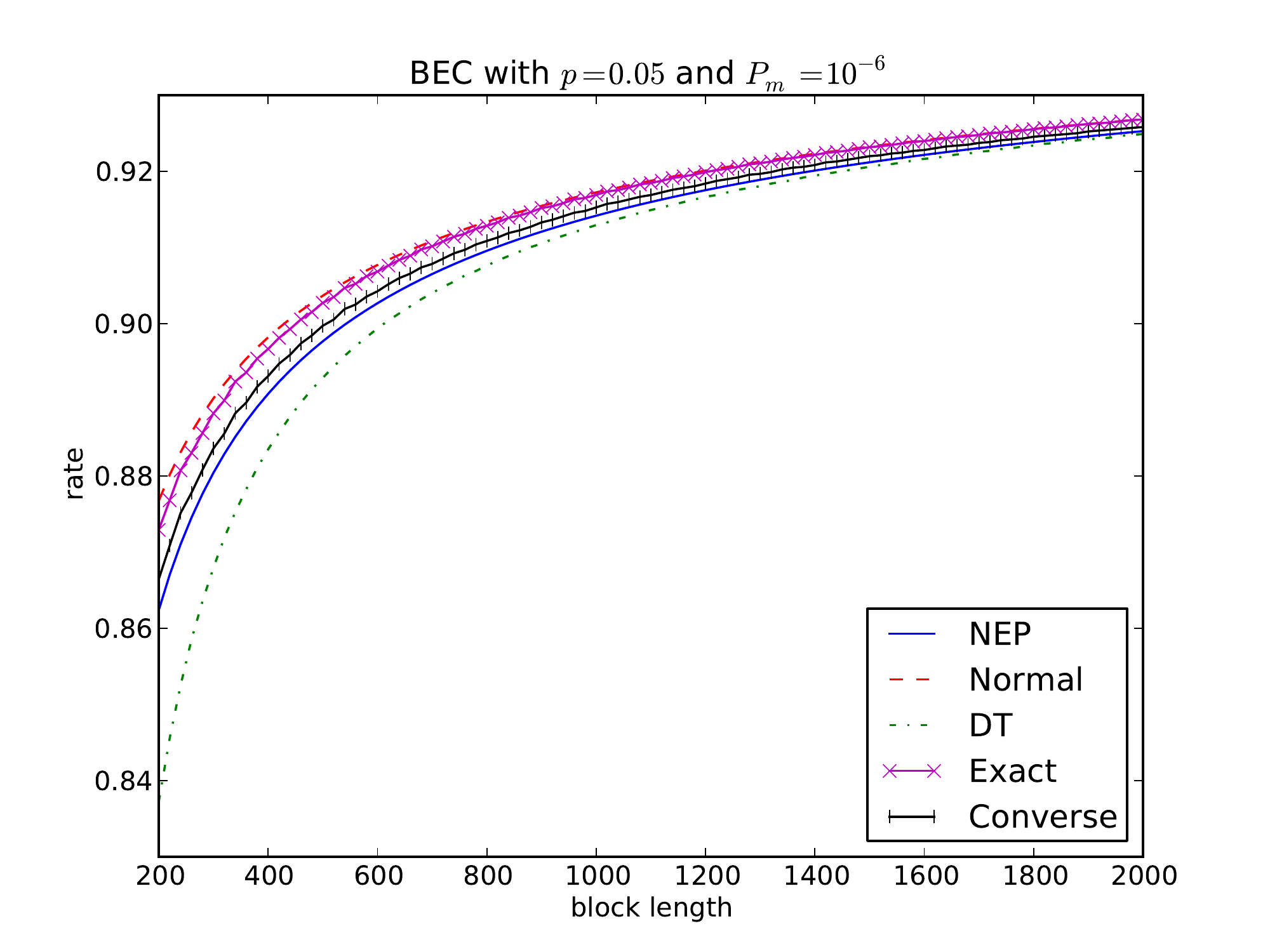}}
  \subfloat[$\delta_n (\epsilon)$ with $P_m = 10^{-6}$]{\includegraphics[scale=0.4]{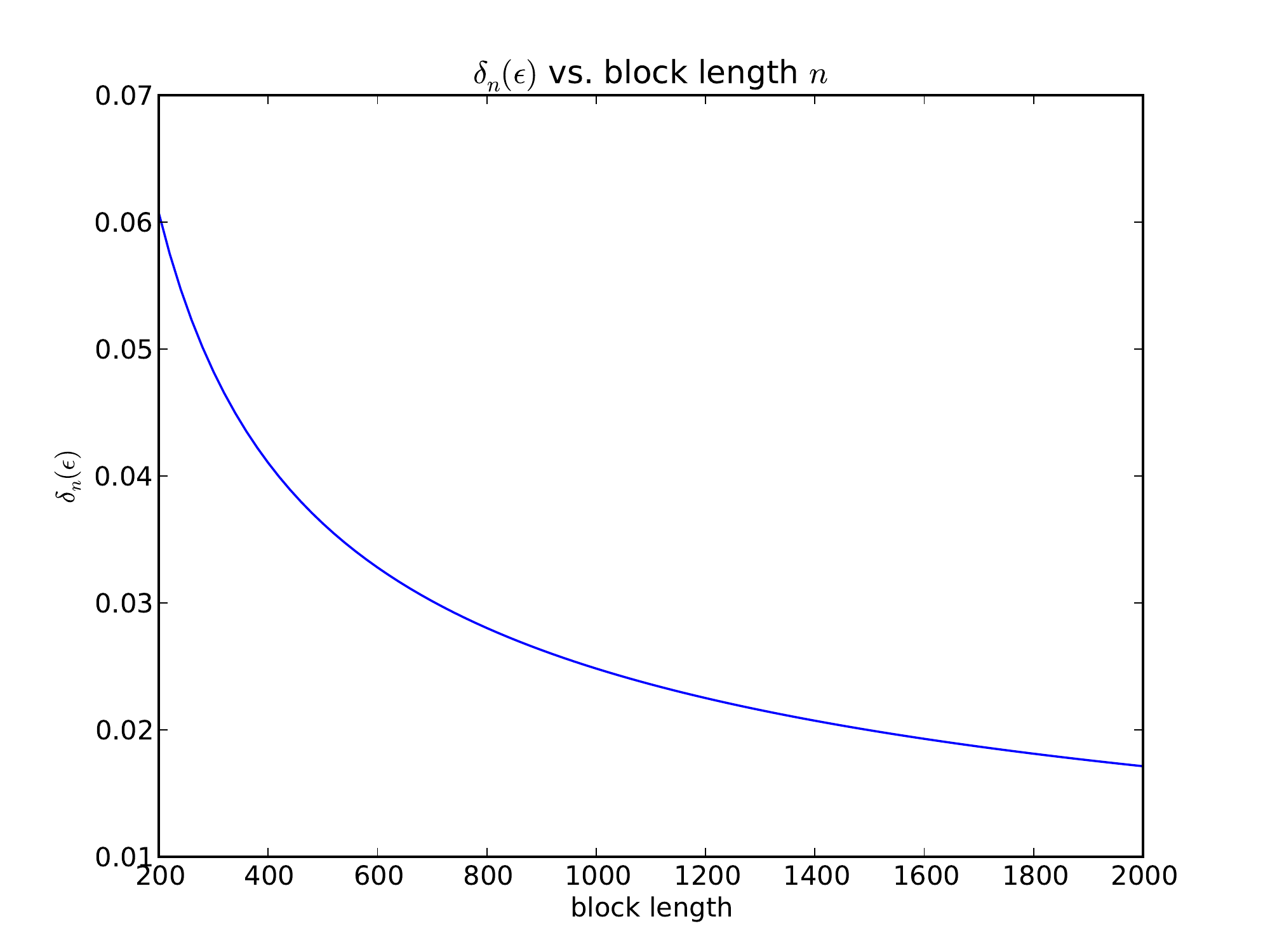}}
  \\
  \subfloat[Bounds with $P_m = g_{X|Y,n} (\delta)$ and $\delta=0.0199$]{\includegraphics[scale=0.4]{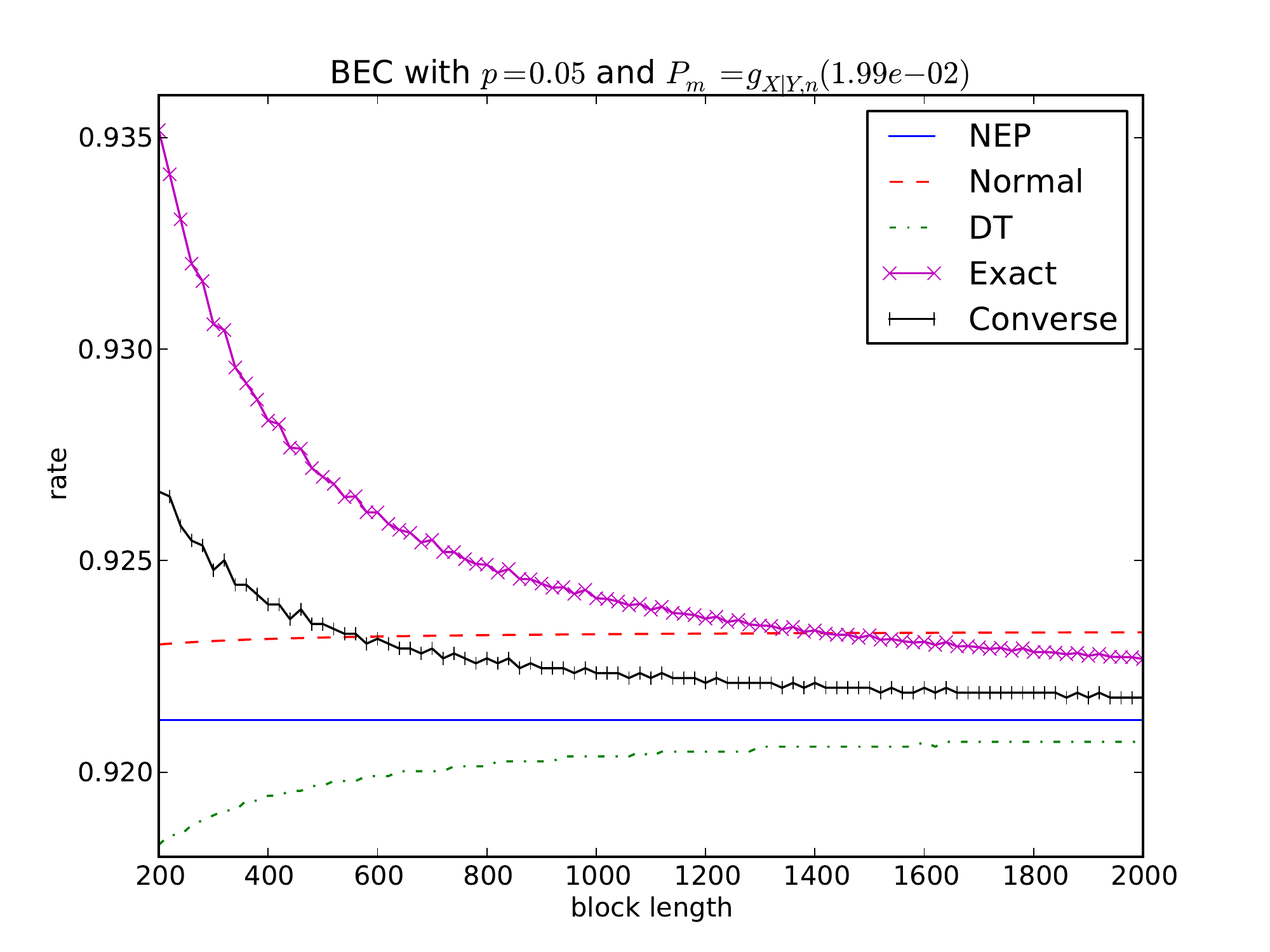}}
  \subfloat[$\log_{10} P_m$ with $P_m = g_{X|Y,n} (\delta)$ and $\delta=0.0199$]{\includegraphics[scale=0.4]{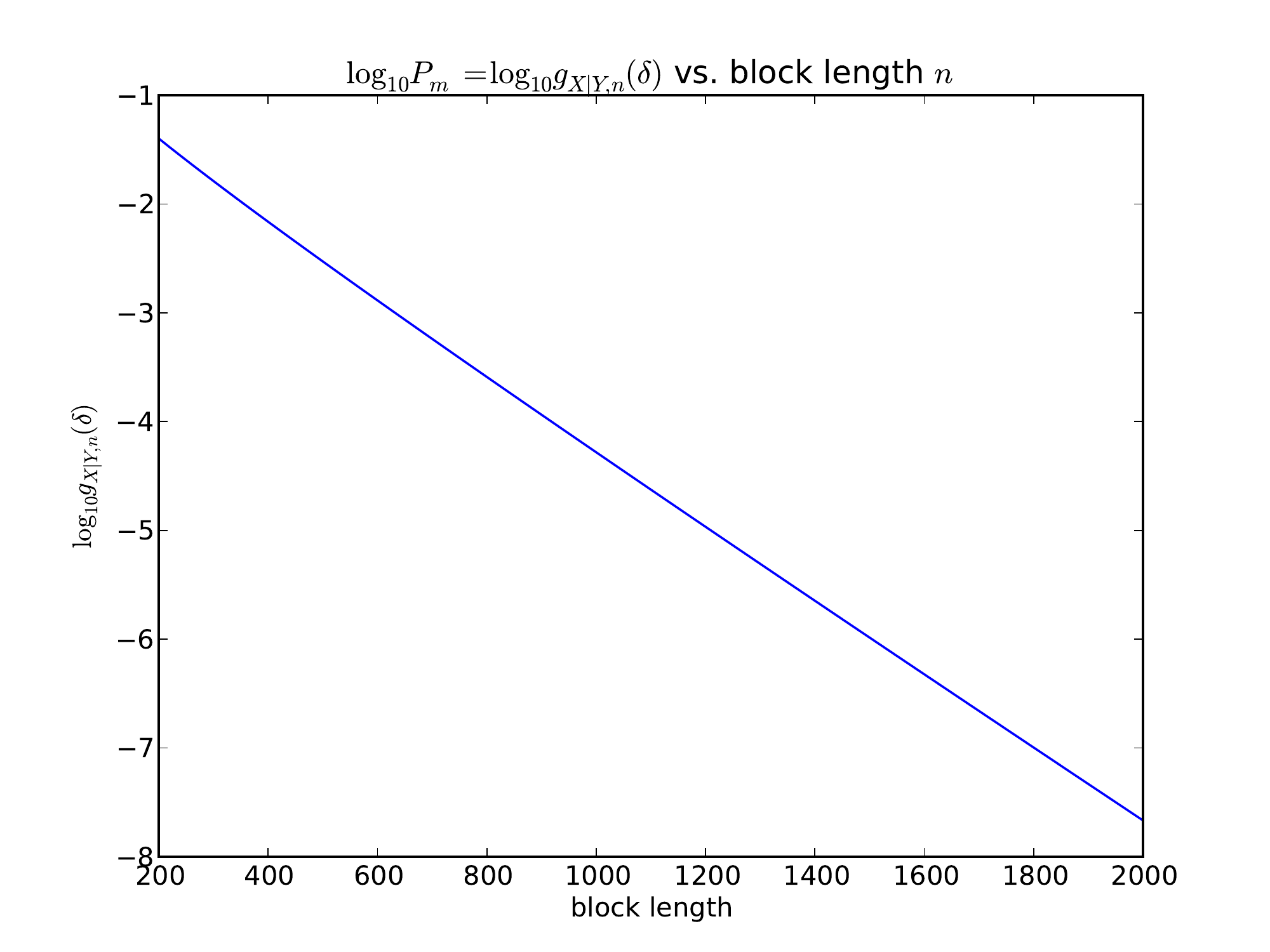}}
  \caption{Comparison of different bounds for BEC with $p=0.05$.}
  \label{fig-bec}
\end{figure}

\begin{figure}[h]
  \centering
  \subfloat[Bounds with $P_m = 10^{-6}$]{\includegraphics[scale=0.4]{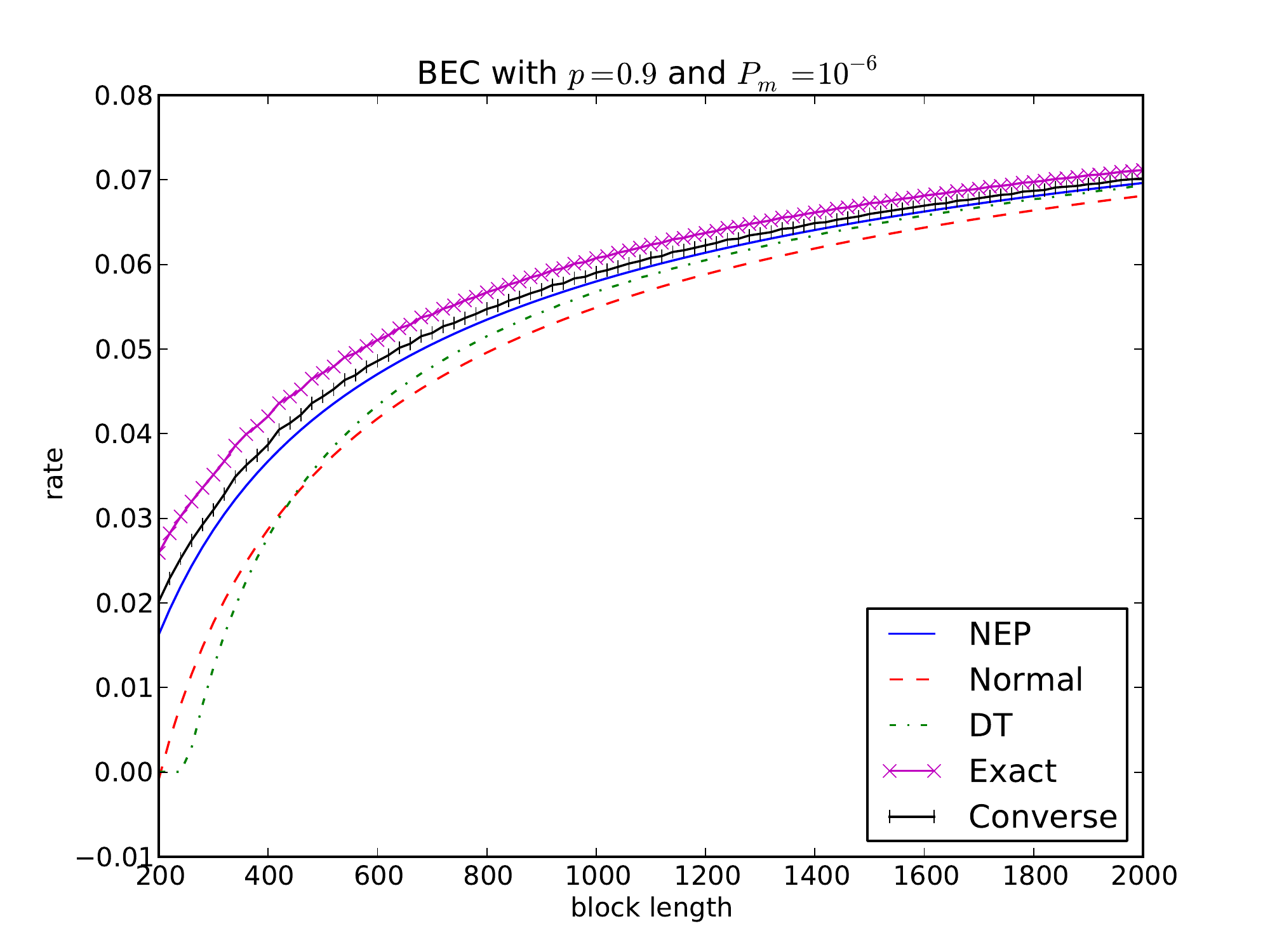}}
  \subfloat[$\delta_n (\epsilon)$ with $P_m = 10^{-6}$]{\includegraphics[scale=0.4]{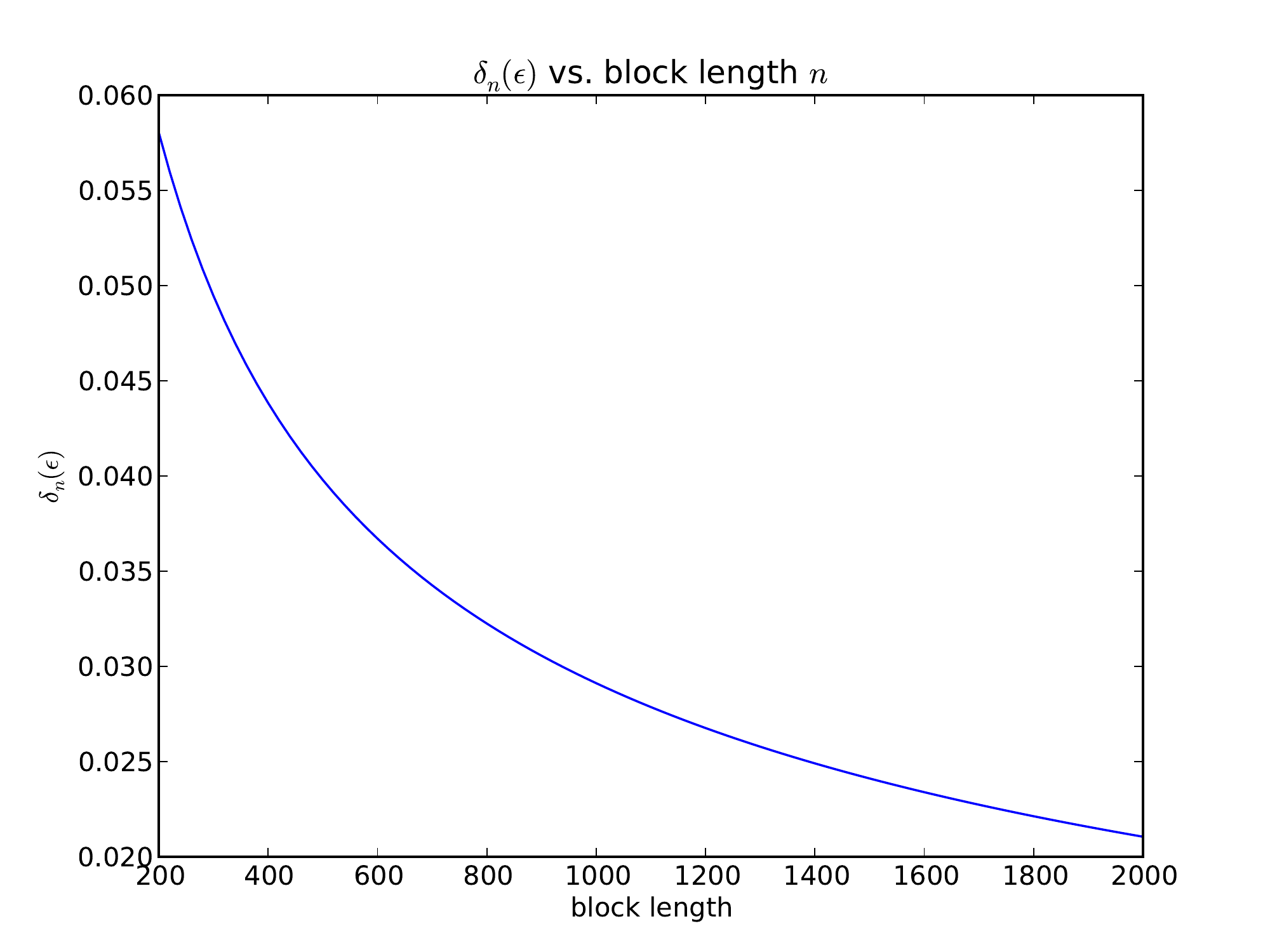}}
  \\
  \subfloat[Bounds with $P_m = g_{X|Y,n} (\delta)$ and $\delta=0.022$]{\includegraphics[scale=0.4]{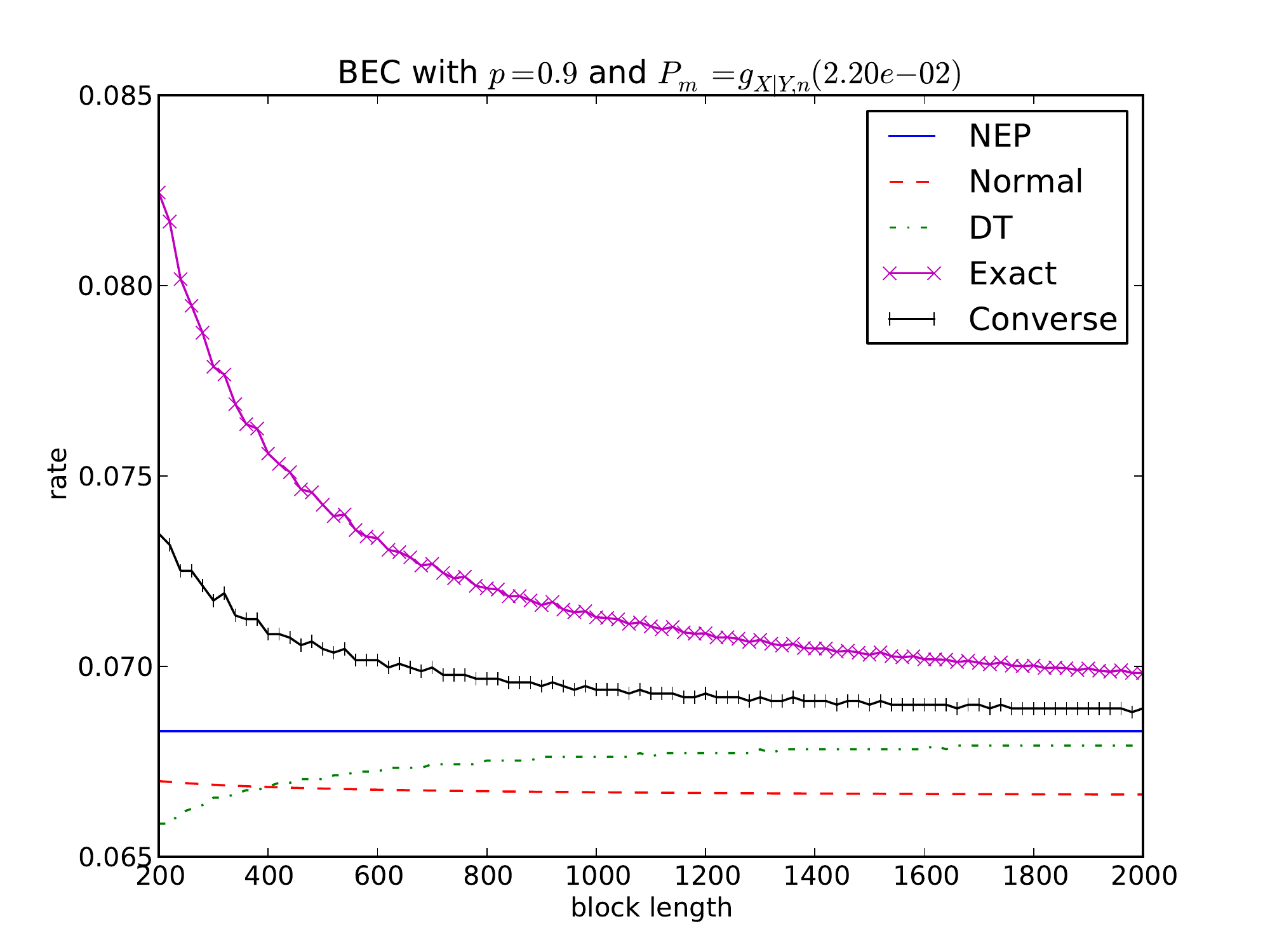}}
  \subfloat[$\log_{10} P_m$ with $P_m = g_{X|Y,n} (\delta)$ and $\delta=0.022$]{\includegraphics[scale=0.4]{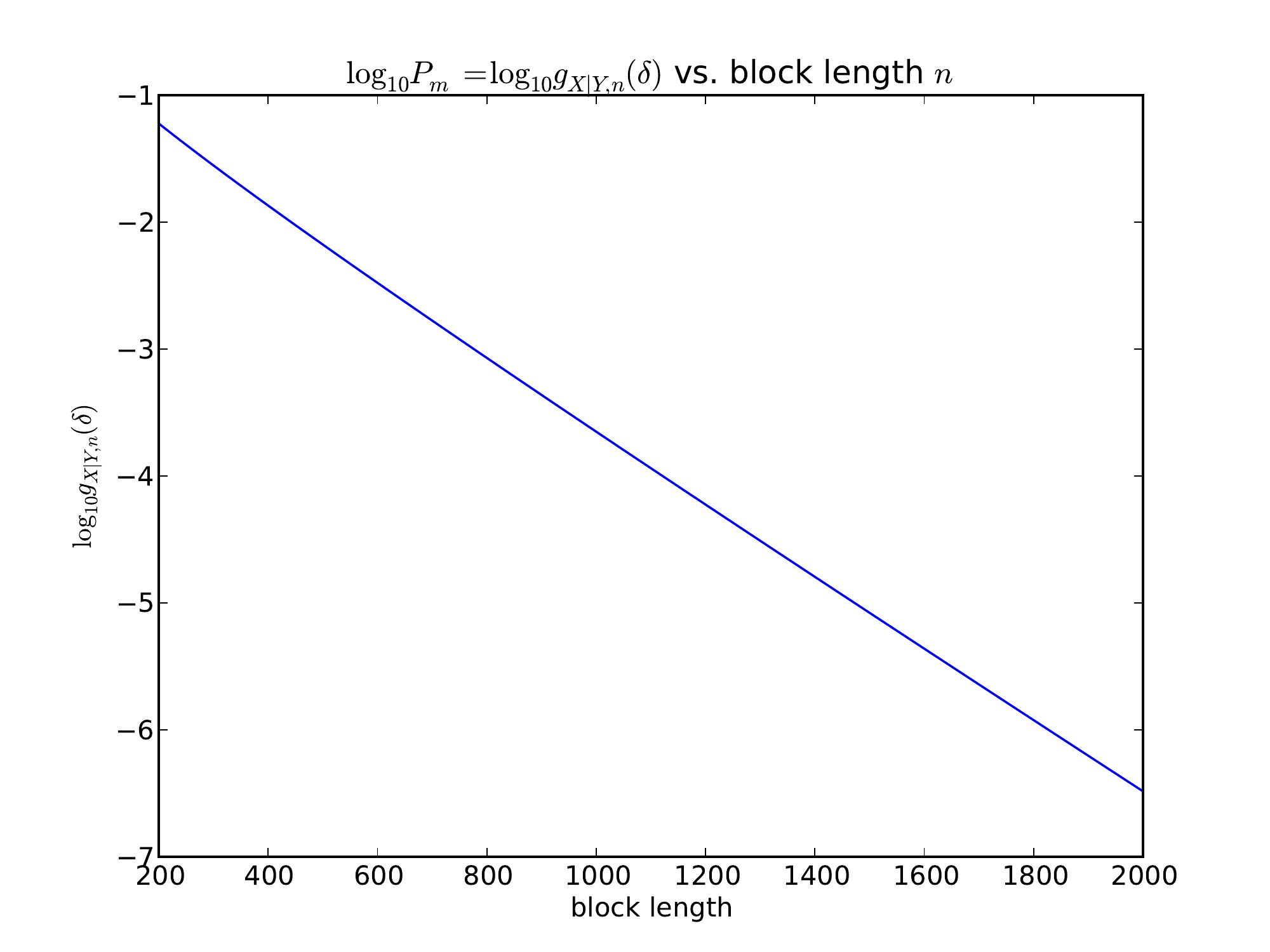}}
  \caption{Comparison of different bounds for BEC with $p=0.9$.}
  \label{fig-bec-2}
\end{figure}

To provide benchmarks for the comparison of approximation formulas, Theorem 37 and 38 in \cite{Yury-Poor-Verdu-2010} are used here, dubbed ``DT'' and ``Converse'' respectively. The exact calculation of
\eqref{eq-thm-maxbimc-3} and \eqref{eq-thm-maxbimc-4} in Corollary \ref{col-bimc} (dubbed ``Exact'')
again serves as an additional converse bound. Then those bounds are drawn in Figures~\ref{fig-bec} and \ref{fig-bec-2} in the same way as those in figure~\ref{fig-bsc}, where erasure probabilities are selected to be $0.05$ and $0.9$,  respectively. Once again, numeric results confirm our analysis and discussion above.

\subsubsection{BIAGC}
\begin{figure}[h]
  \centering
  \includegraphics[scale=0.4]{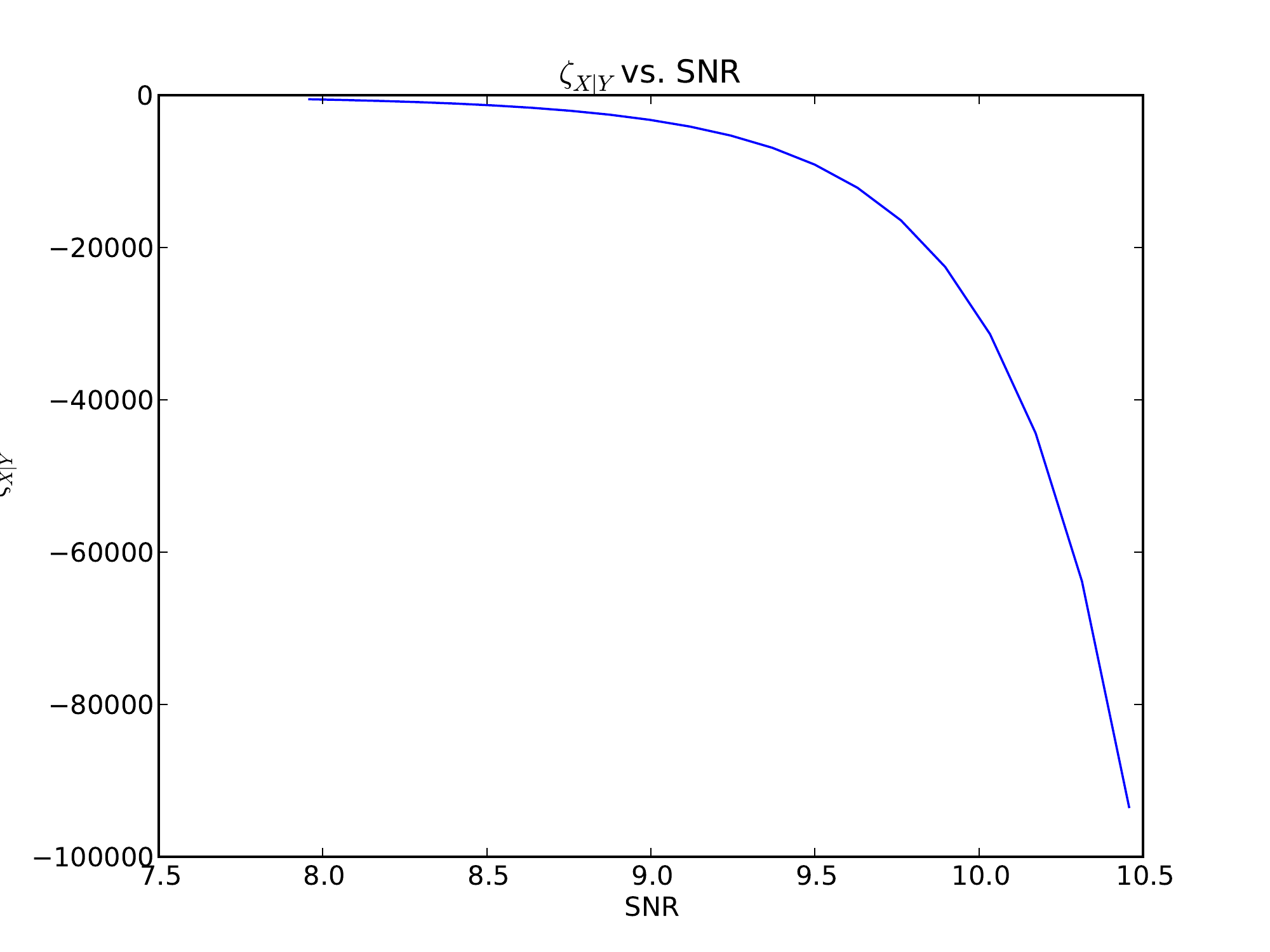}
  \caption{$\zeta_{X|Y}$ of BIAGC}
  \label{fig:biagc-zeta}
\end{figure}

\begin{figure}[h]
  \centering
  \subfloat[Bounds with $P_m = 10^{-3}$]{\includegraphics[scale=0.4]{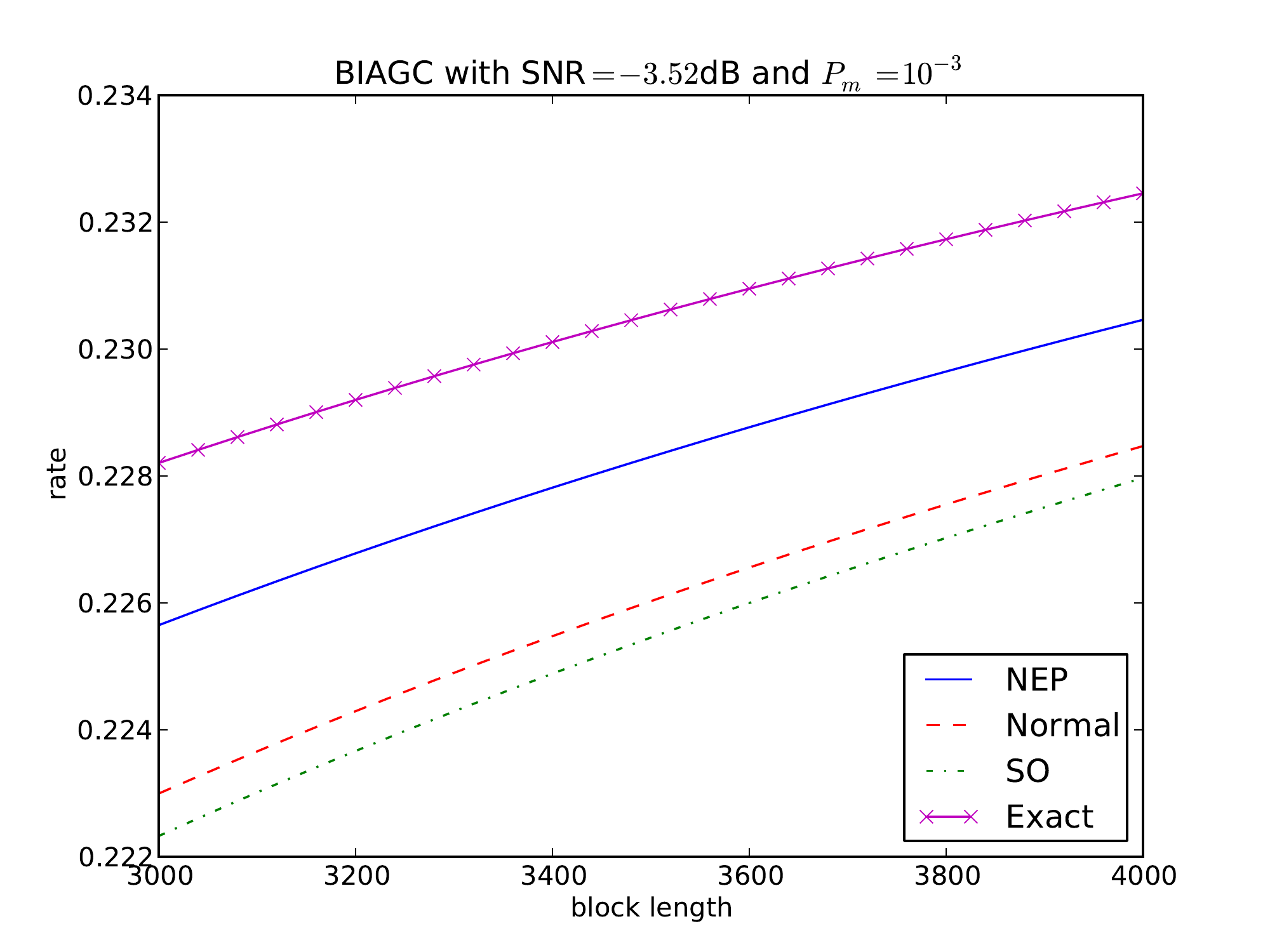}}
  \subfloat[$\delta_n (\epsilon)$ with $P_m = 10^{-3}$]{\includegraphics[scale=0.4]{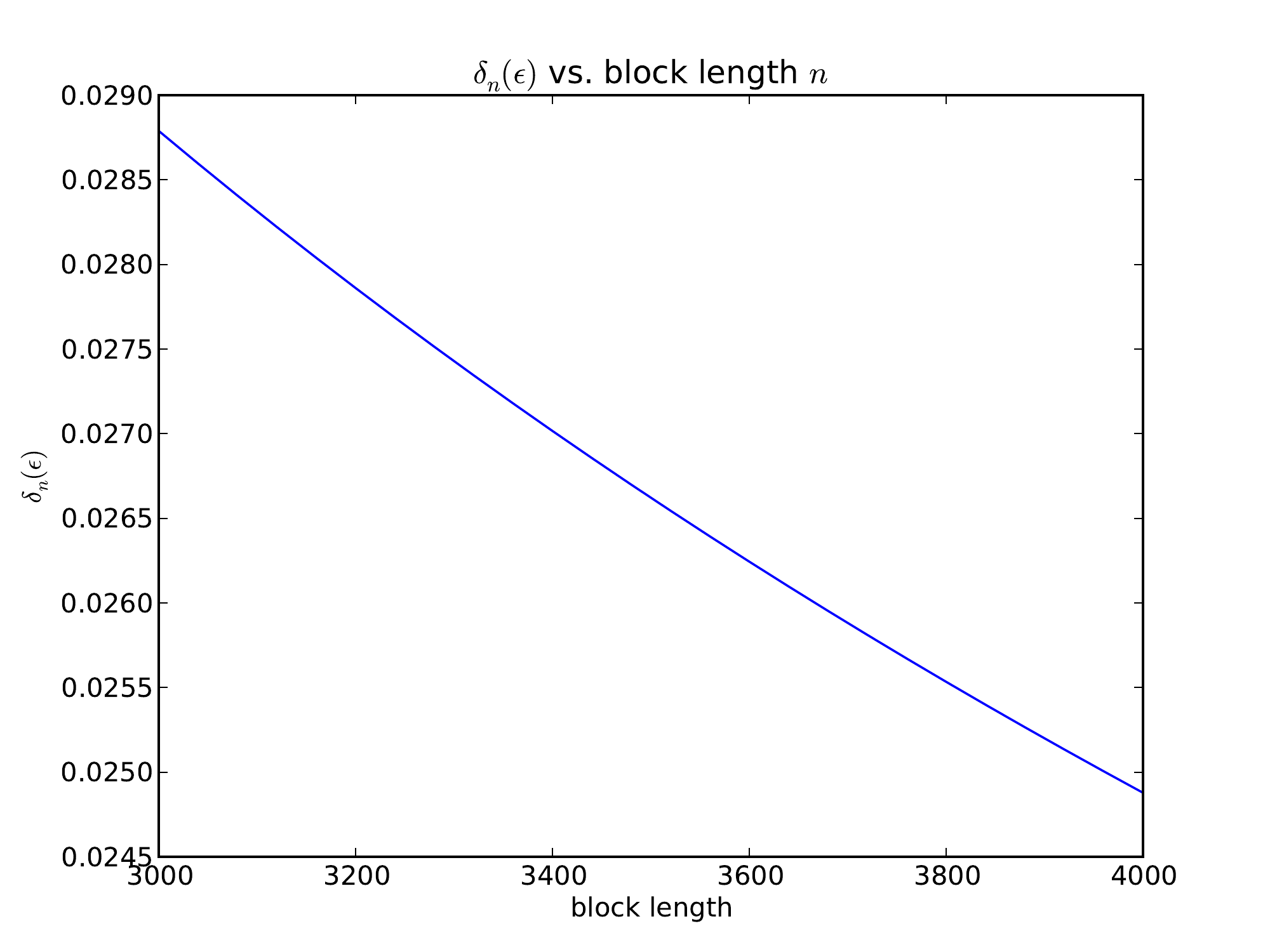}}
  \\
  \subfloat[Bounds with $P_m = g_{X|Y,n} (\delta)$ and $\delta=0.0265$]{\includegraphics[scale=0.4]{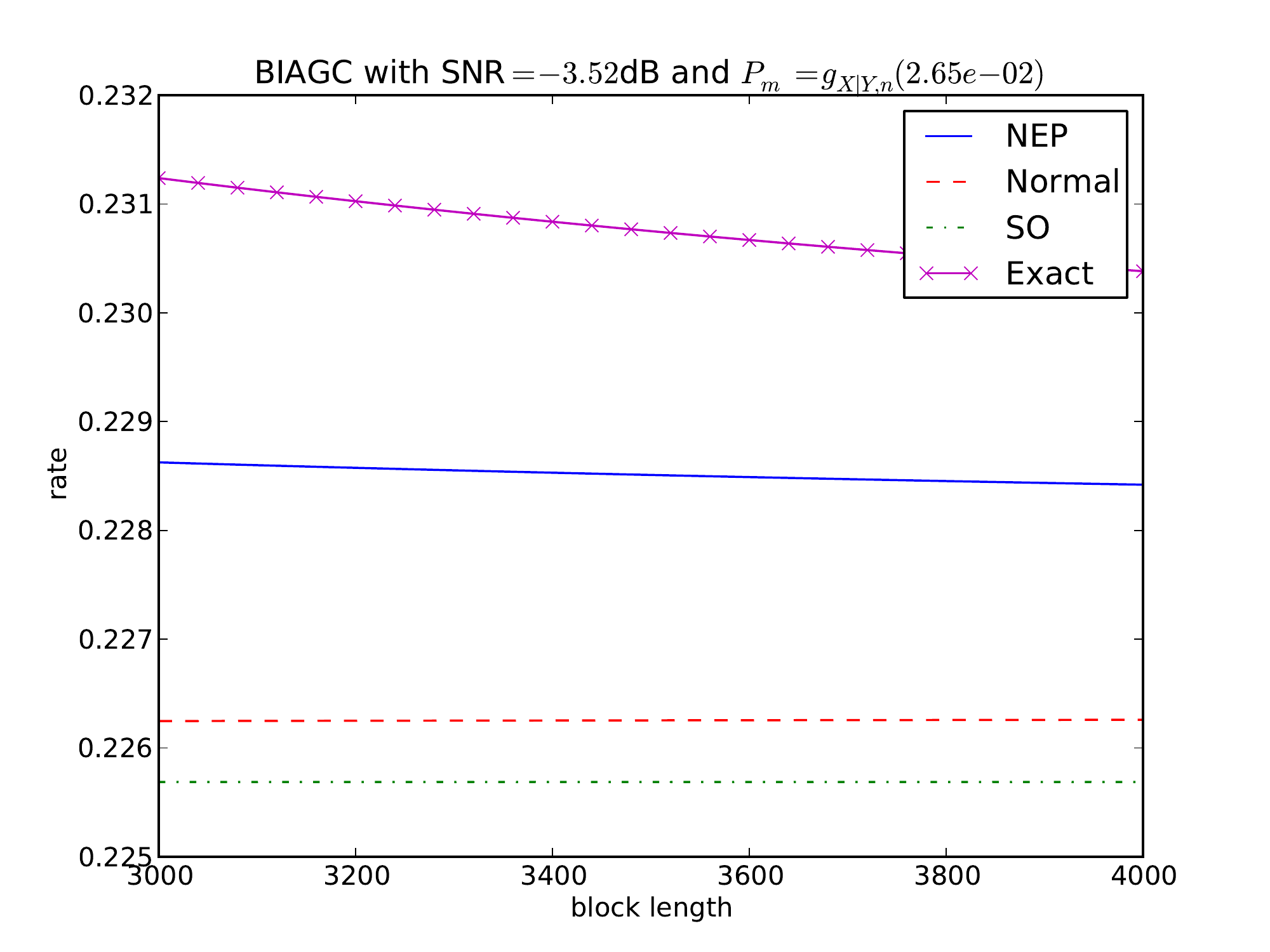}}
  \subfloat[$\log_{10} P_m$ with $P_m = g_{X|Y,n} (\delta)$ and $\delta=0.0265$]{\includegraphics[scale=0.4]{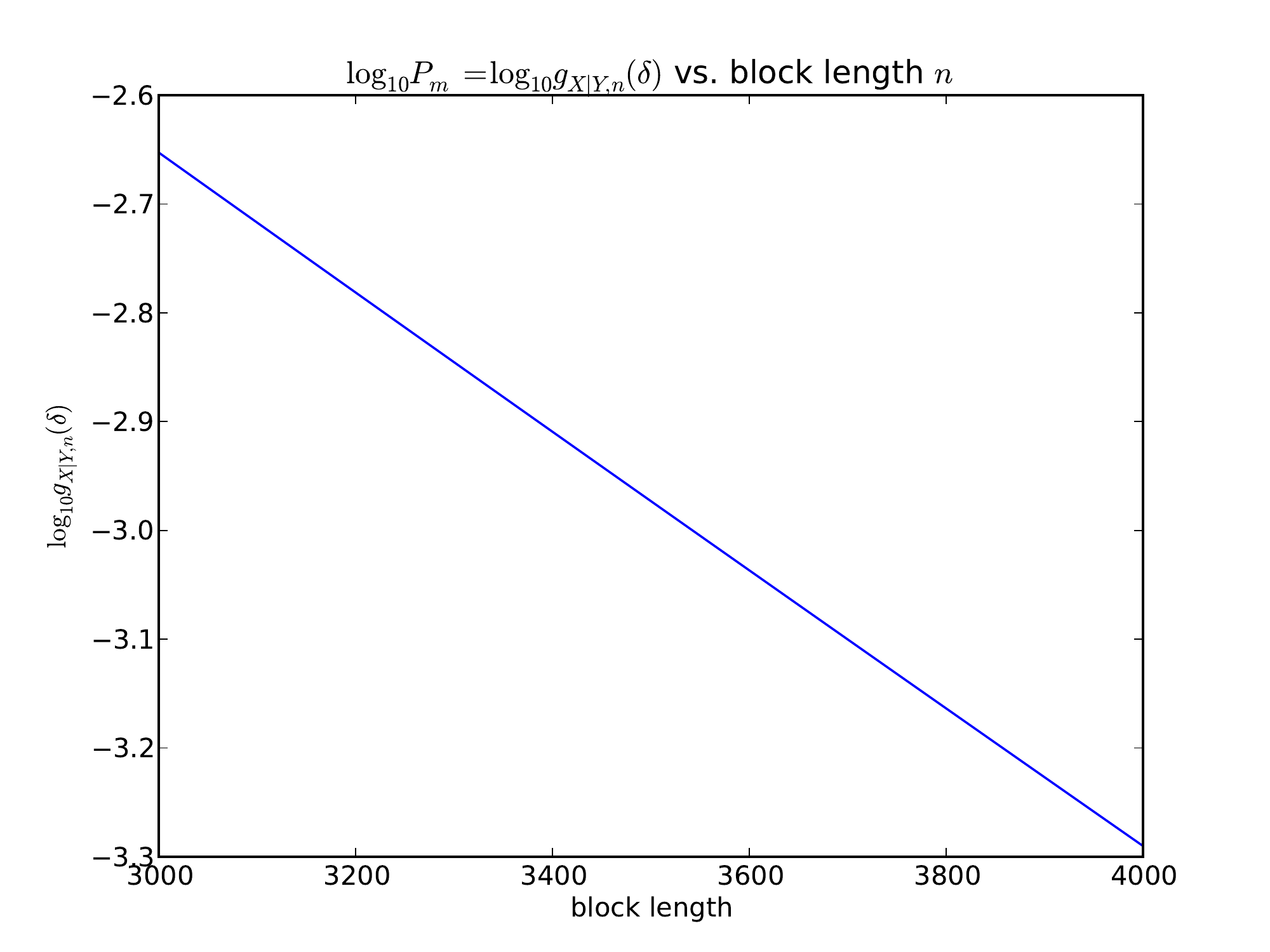}}
  \caption{Comparison of different bounds for BIAGC with SNR $=-3.52$ dB.}
  \label{fig-biagc}
\end{figure}

\begin{figure}[h]
  \centering
  \subfloat[Bounds with $P_m = 10^{-9}$]{\includegraphics[scale=0.4]{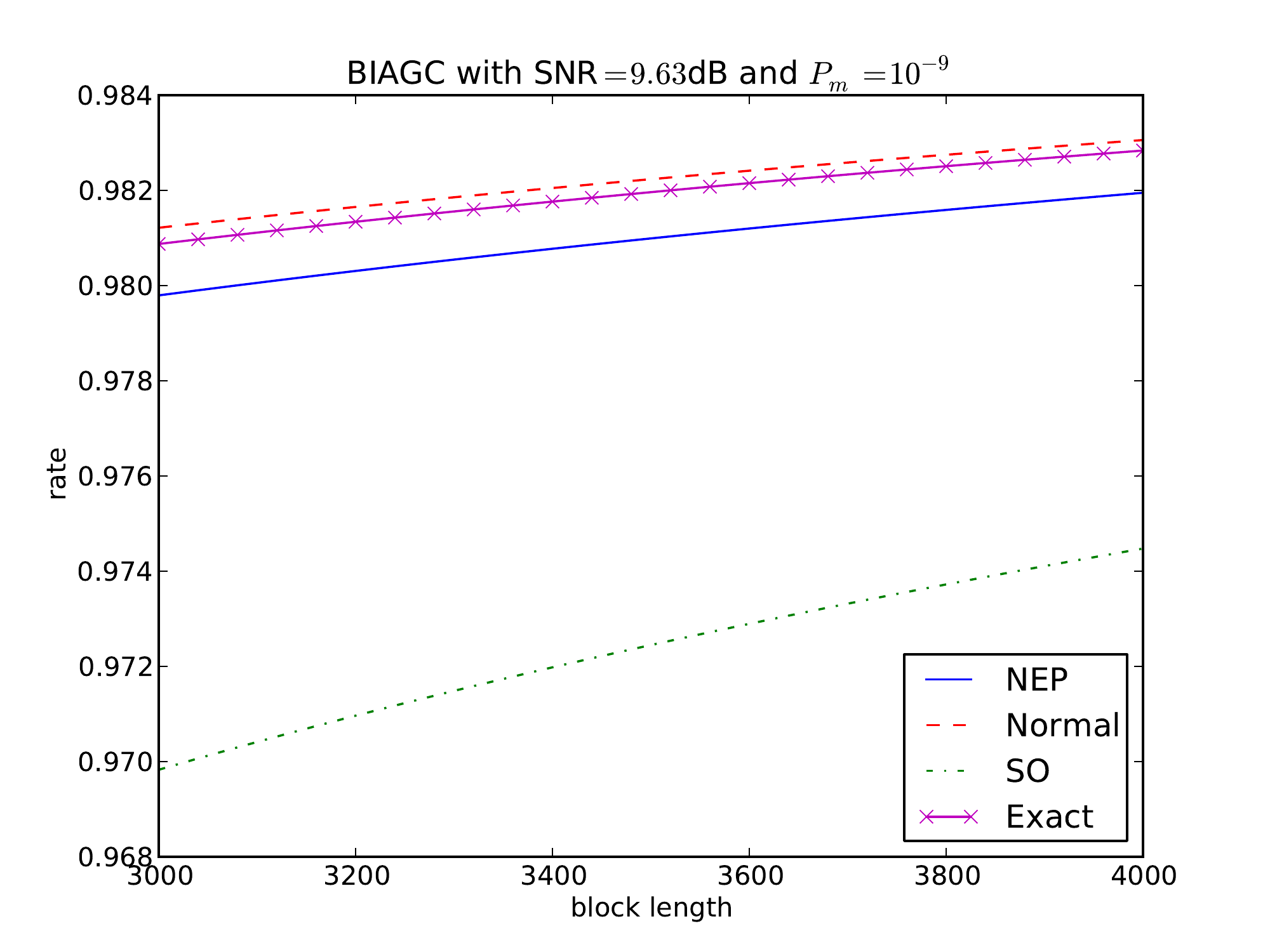}}
  \subfloat[$\delta_n (\epsilon)$ with $P_m = 10^{-9}$]{\includegraphics[scale=0.4]{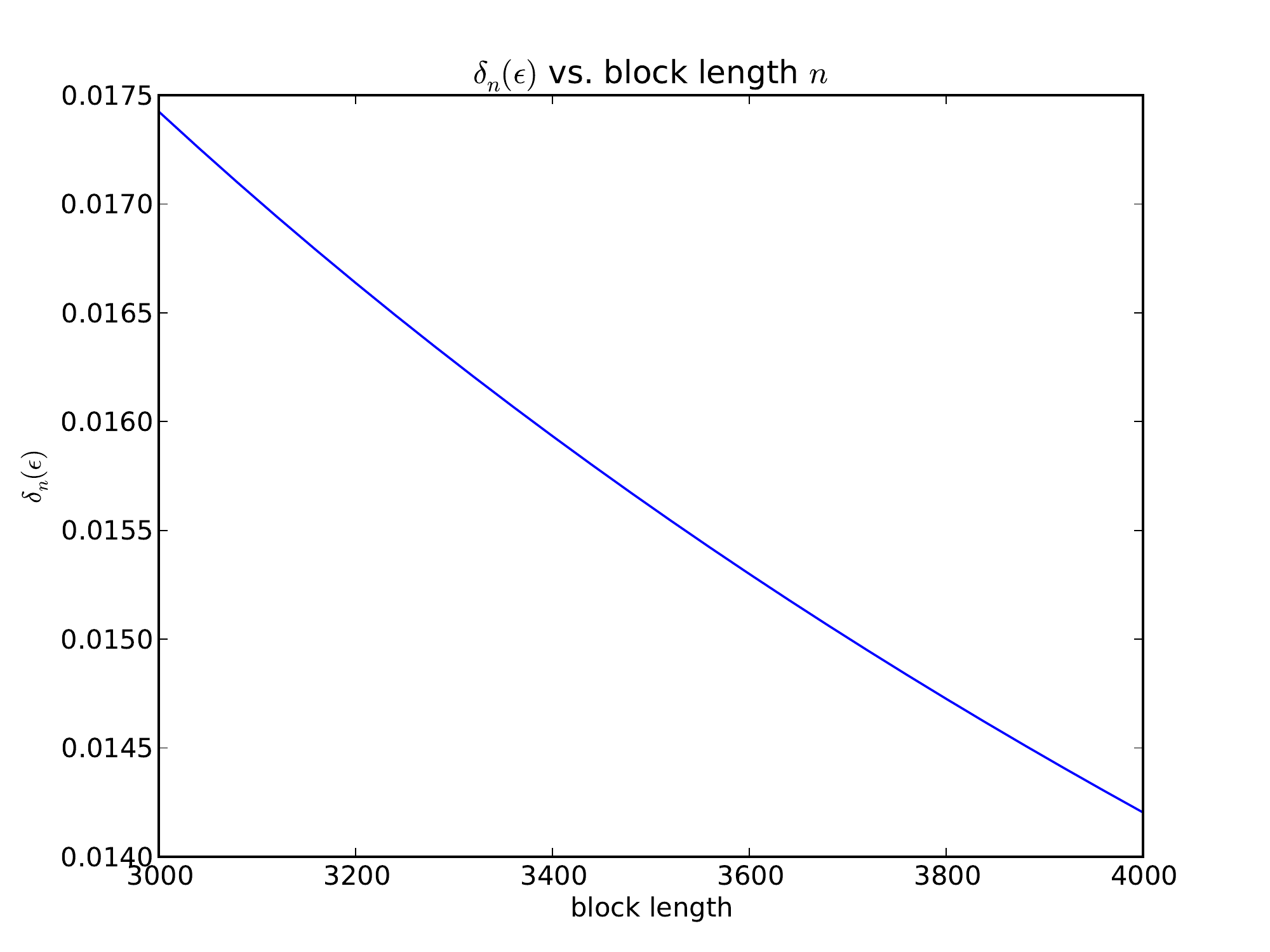}}
  \\
  \subfloat[Bounds with $P_m = g_{X|Y,n} (\delta)$ and $\delta=0.0175$]{\includegraphics[scale=0.4]{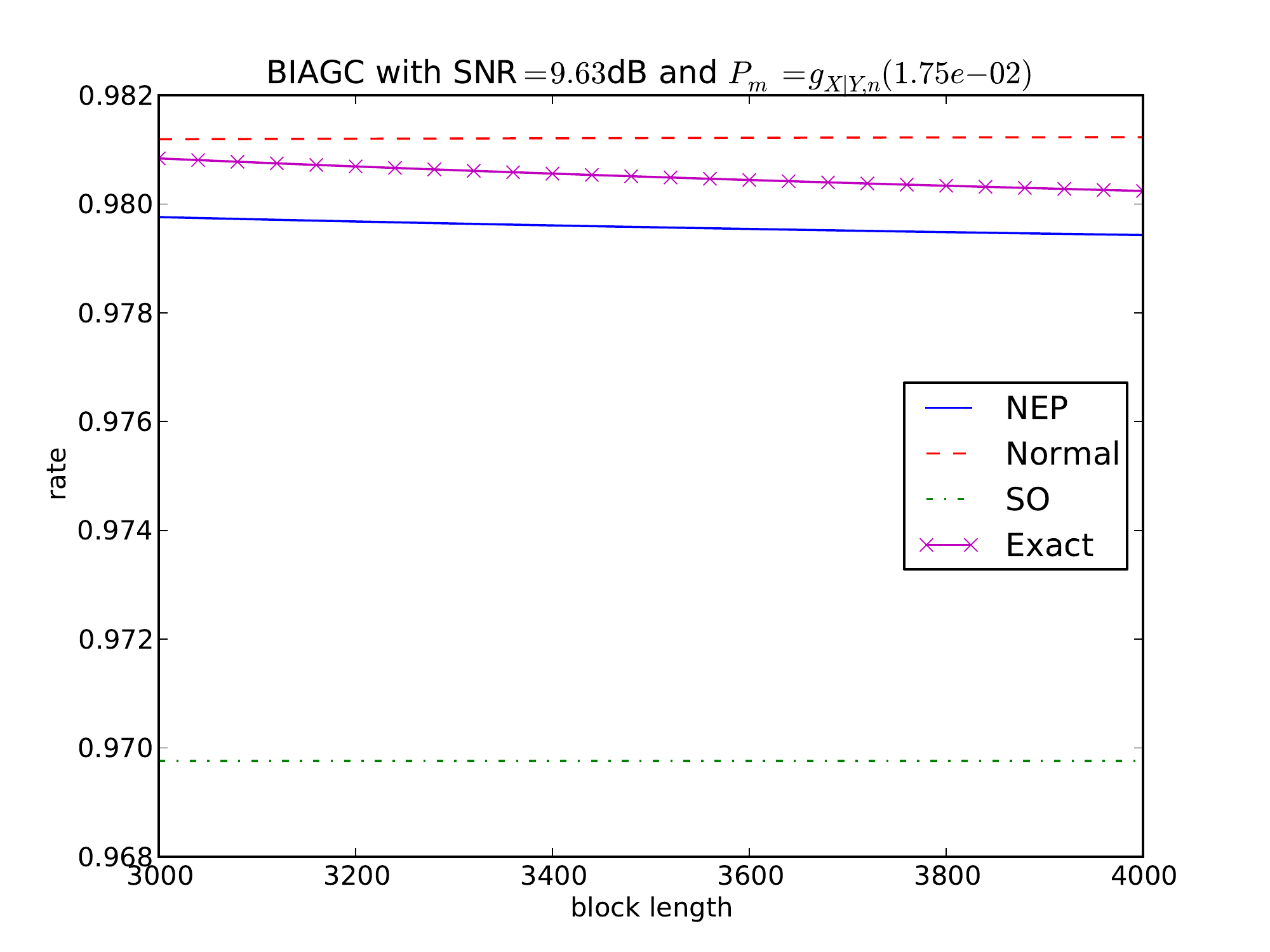}}
  \subfloat[$\log_{10} P_m$ with $P_m = g_{X|Y,n} (\delta)$ and $\delta=0.0175$]{\includegraphics[scale=0.4]{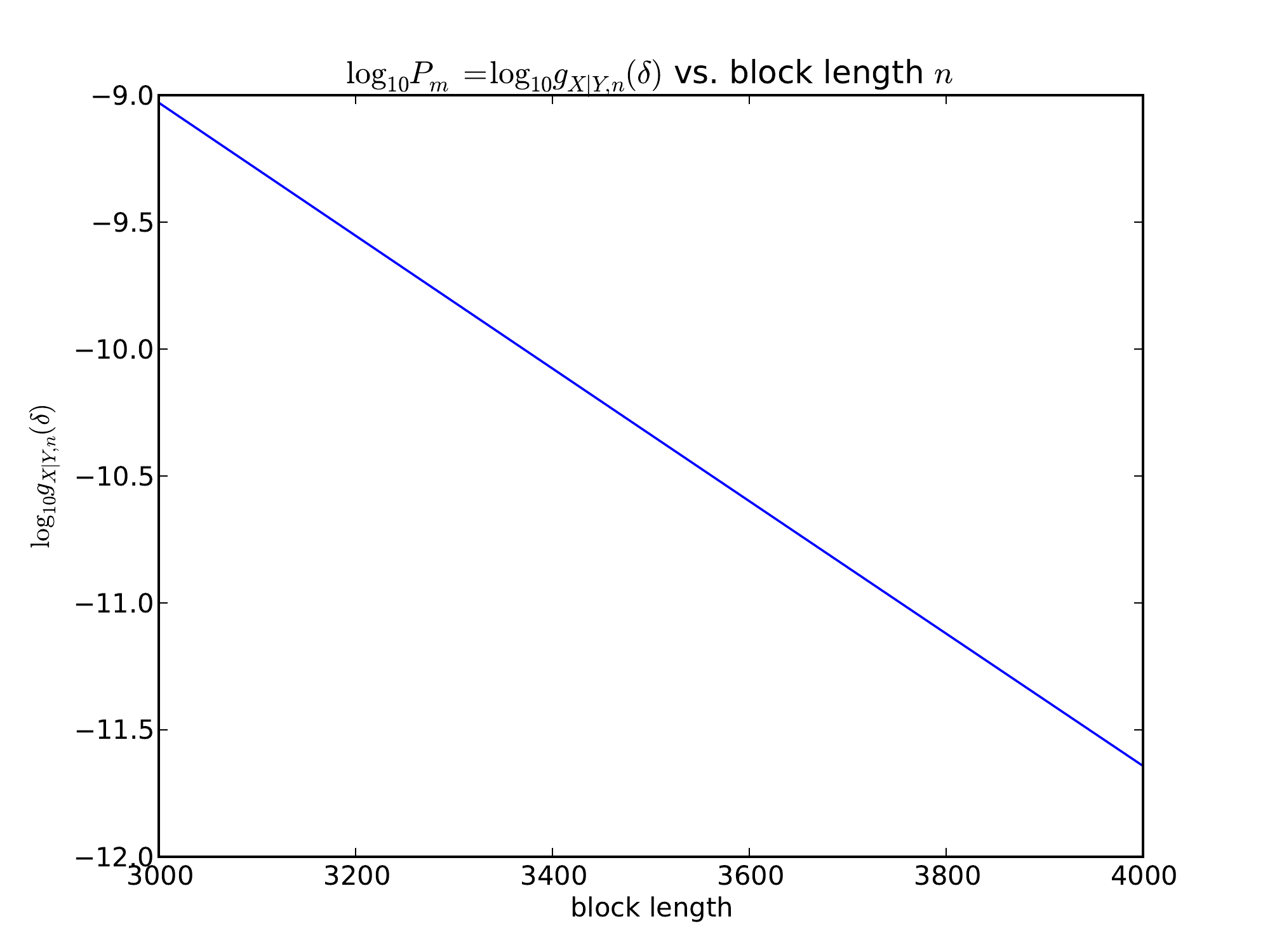}}
  \caption{Comparison of different bounds for BIAGC with SNR $=9.63$ dB. }
  \label{fig-biagc-2}
\end{figure}

 Here we assume that codewords are modulated to $\{+1,-1\}$ before going through an AWGN channel, and apply the trivial bound $P(B_{n,\delta_n (\epsilon)}) \leq 1$ in the NEP formula. Similarly to BSC and BEC, we would like to get some insight by investigating $\zeta_{X|Y}$. Since in this case,  $\zeta_{X|Y}$ does not seem to have a simple close form expression which can be easily computed, numerical calculation of $\zeta_{X|Y}$ is shown in Figure~\ref{fig:biagc-zeta}, where SNR ranges from 8dB to 10.5dB. As can be seen, BIAGC is similar to BSC, i.e. $\zeta_{X|Y}$ is always negative and its magnitude increases with SNR. Therefore,  $ R^{\mathrm{Normal}}_n (\epsilon)$  is close to $ R^{\mathrm{SO}}_n (\epsilon)$  when SNR is low, but can be above some converse bounds when SNR is high. This  is confirmed in Figures \ref{fig-biagc} and \ref{fig-biagc-2}, where exact evaluation of \eqref{eq-thm-maxbimc-1} and \eqref{eq-thm-maxbimc-2} in Corollary \ref{col-bimc} (dubbed ``Exact'') serves as a converse bound.


\subsection{DIMC: Z Channel}
\label{sec:z-channel}

To show an example of DIMC which is not a BIMSC, we consider again the Z channel shown in Figure \ref{fig:zch}. The capacity of Z channel is well known and given by
\begin{equation} \label{eq-z-cap}
  C_Z = \ln \left( 1 + (1-p) p^{\frac{p}{1-p}} \right)
\end{equation}
with the capacity-achieving distribution
\begin{equation} \label{eq-z-px}
  p_X(x) =
    \left\{
      \begin{array}{cc}
        \frac{1}{1- p + p^{-\frac{p}{1-p}}} & \mbox{for $x=0$} \\
          \\
        \frac{p^{-\frac{p}{1-p}}-p}{1- p + p^{-\frac{p}{1-p}}} & \mbox{for $x=1$}
      \end{array}
    \right.
\end{equation}
and the corresponding output distribution
\begin{equation} \label{eq-z-py}
  p_Y(y) =
    \left\{
      \begin{array}{cc}
        \frac{1-p}{1- p + p^{-\frac{p}{1-p}}} & \mbox{for $y=0$} \\
        \\
        \frac{p^{-\frac{p}{1-p}}}{1- p + p^{-\frac{p}{1-p}}} & \mbox{for $y=1$ .}
      \end{array}
    \right.
\end{equation}

To calculate $R^{\mathrm{NEP}}_n (\epsilon) $, $P(B_{t,n,\delta})$ needs to be further investigated, where an interesting observation is that given $x^n$ with type $t$, $\frac{1}{n} \ln \frac{p(y^n | x^n)}{q_t(y^n)} > -\infty$ if and only if $y_i=1$ when $x_i=1$, and the value of $\frac{1}{n} \ln \frac{p(y^n | x^n)}{q_t(y^n)}$ only depends on the number of $y_i$ being $1$ for $i \in \{j: x_j=0 \}$. One can then verify that
\begin{equation}
  B_{t,n,\delta} = \left\{ y^n: \frac{1}{n} |\{i : y_i=0\}| \leq q_t (0) - \frac{\delta}{\ln \frac{1 - t(0) + p t(0)}{p t(0)}} \right\} .
\end{equation}
When $q_t (0) \neq 0.5$,
\begin{equation}
  \label{eq-nep-z-4}
  P (B_{t,n,\delta}) =
  \left\{
    \begin{array}{ll}
      \Pr \left\{ - \frac{1}{n} \ln q_t (Y^n_t) \leq H(Y_t) - \frac{\delta}{\ln \frac{1-t(0)+pt(0)}{pt(0)}} \ln \frac{1-q_t(0)}{q_t(0)} \right\} & \mbox{if $q_t(0) < 0.5$} \\
      \Pr \left\{ - \frac{1}{n} \ln q_t (Y^n_t) \geq H(Y_t) - \frac{\delta}{\ln \frac{1-t(0)+pt(0)}{pt(0)}} \ln \frac{1-q_t(0)}{q_t(0)} \right\} & \mbox{if $q_t(0) > 0.5$}
    \end{array}
  \right.
\end{equation}
where $Y_t$ is a random variable with distribution $q_t$. Consequently, we can apply the left NEP\cite{yang-meng:nep}, chernoff bound, right NEP\cite{yang-meng:nep} with respect to entropy to upper bound $P(B_{t,n,\delta})$ when $q_t(0) <, = , > 0.5$,  respectively.

\begin{figure}[h]
  \centering
  \subfloat[Bounds with $P_m = 10^{-9}$]{\includegraphics[scale=0.4]{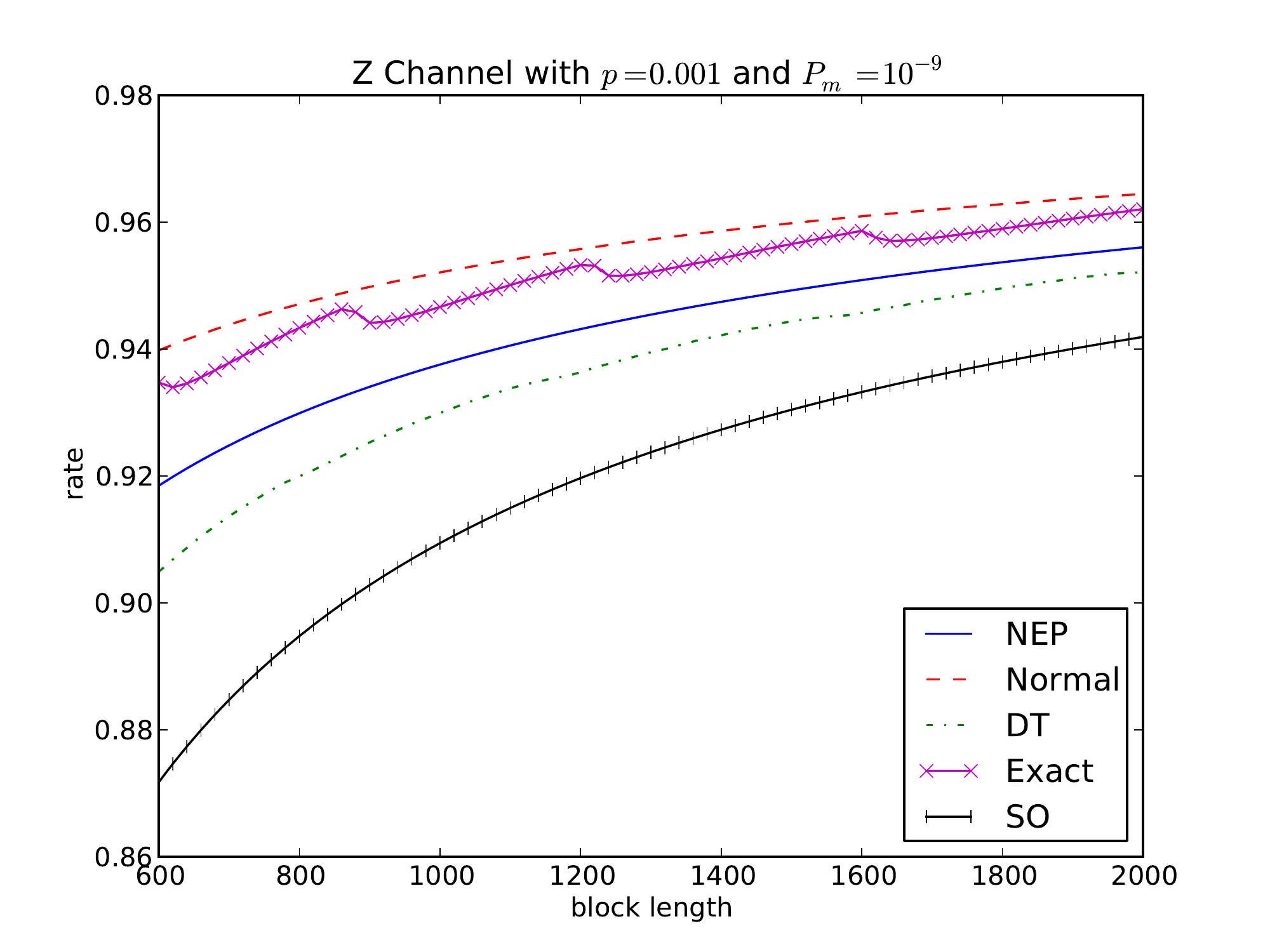}}
  \subfloat[$\delta_{t^*,n} (\epsilon)$ with $P_m = n^{-9}$]{\includegraphics[scale=0.4]{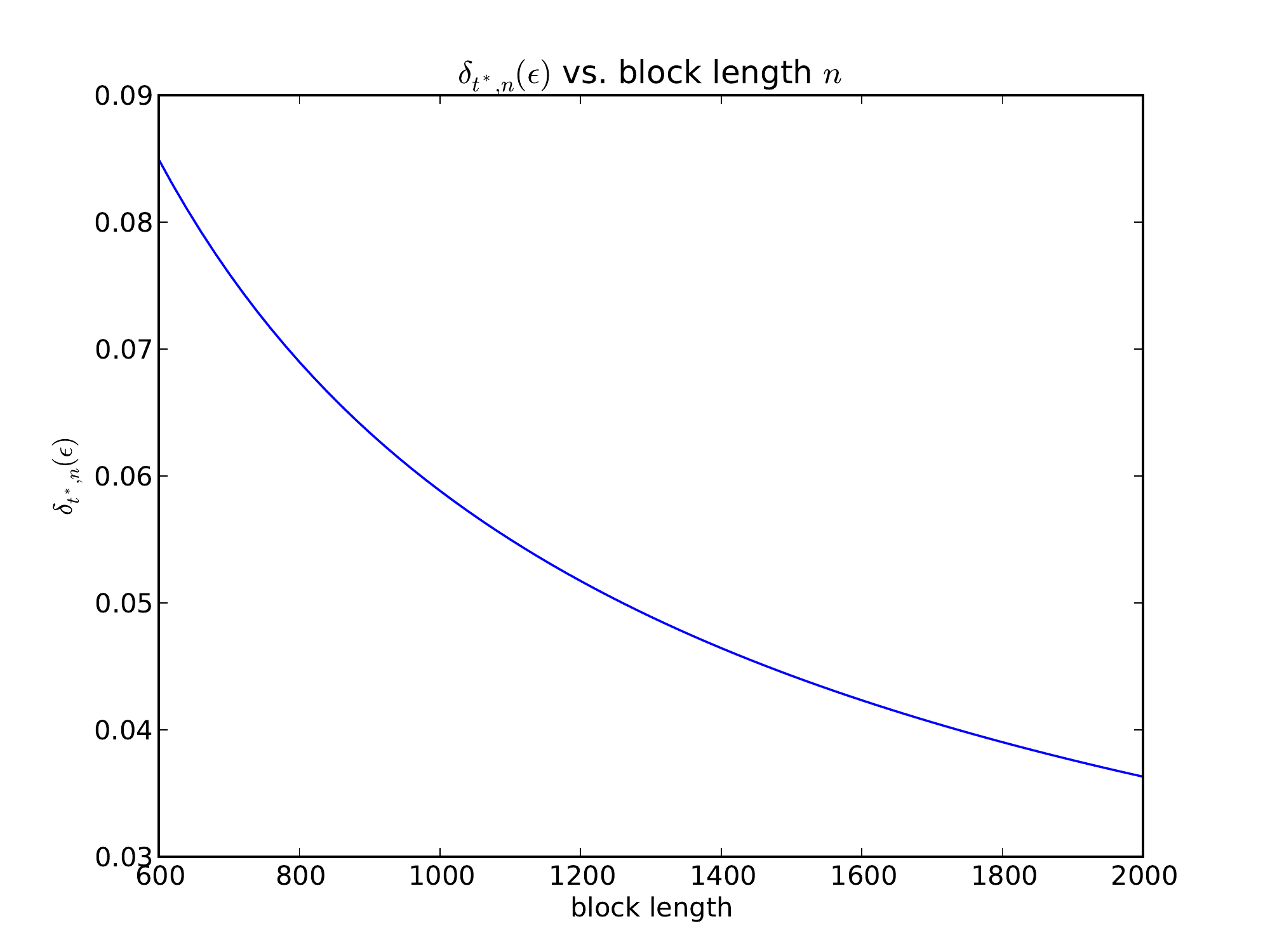}}
  \\
  \subfloat[Bounds with $P_m = g_{t^*;P,n} (\delta)$ and $\delta=0.05$]{\includegraphics[scale=0.4]{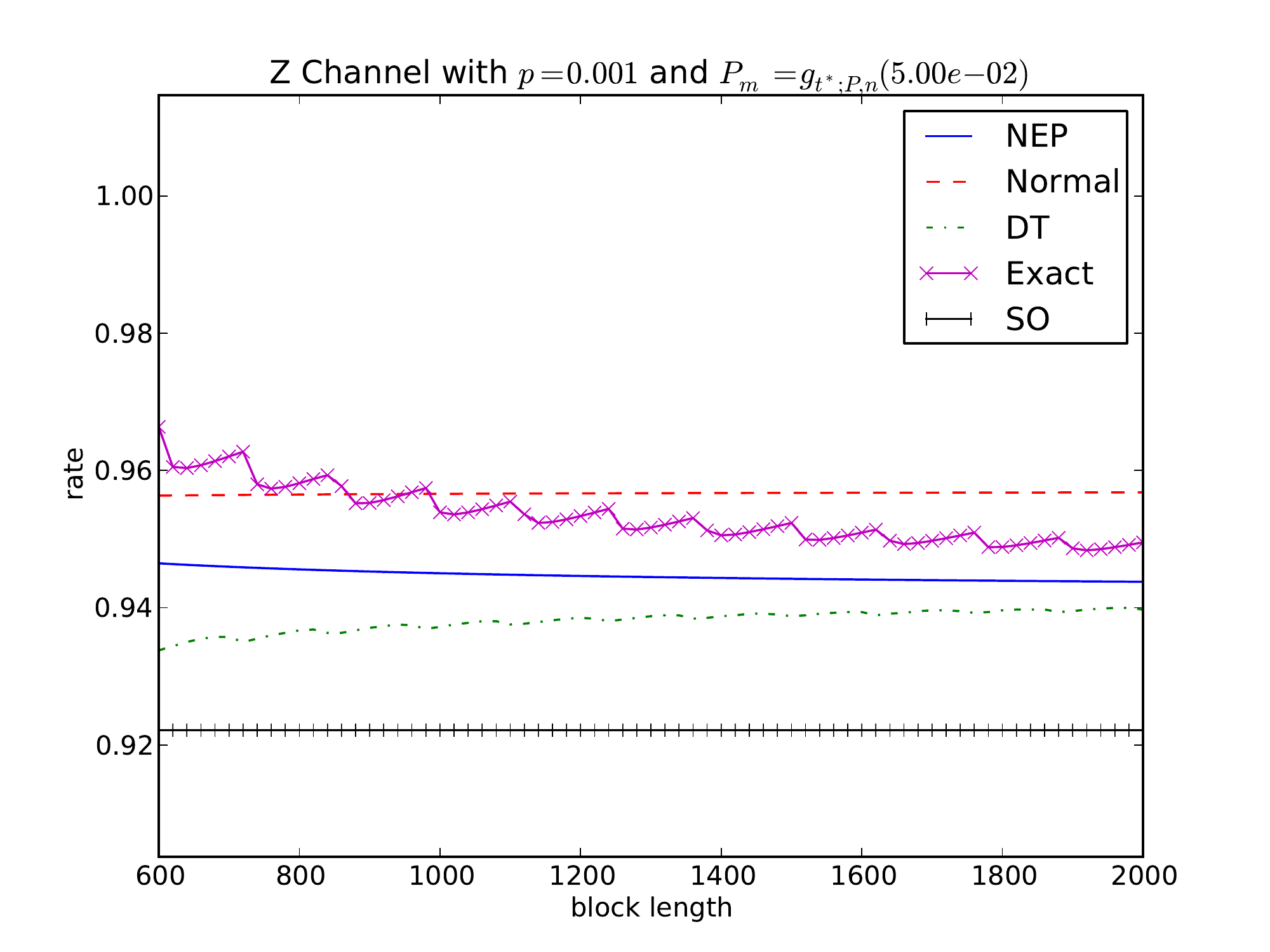}}
  \subfloat[$\log_{10} P_m$ with $P_m = g_{t^*;P,n} (\delta)$ and $\delta=0.05$]{\includegraphics[scale=0.4]{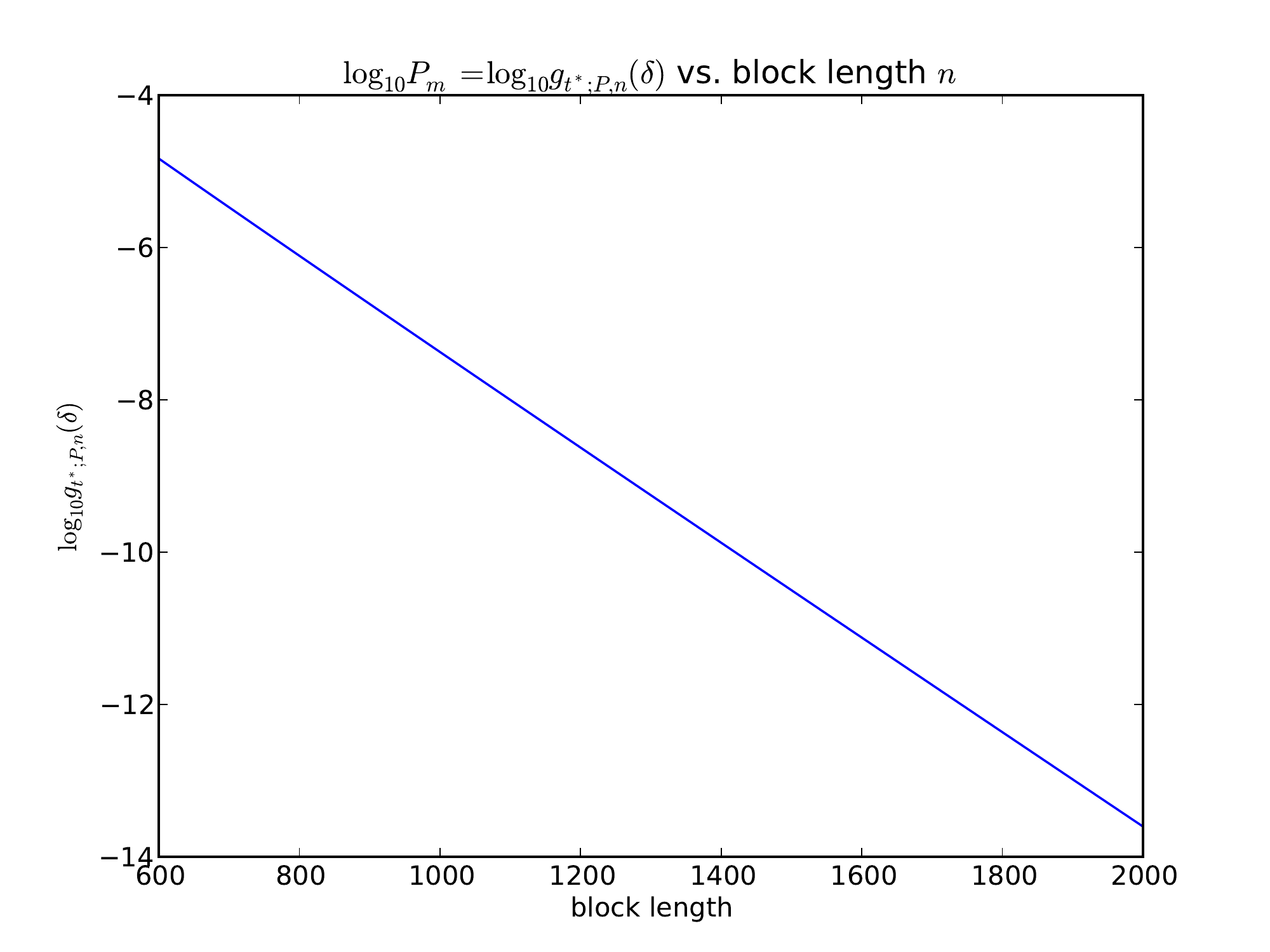}}
  \caption{Comparison of different bounds for Z Channel with $p=0.001$.}
  \label{fig-z}
\end{figure}

\begin{figure}[h]
  \centering
  \subfloat[Bounds with $P_m = 10^{-6}$]{\includegraphics[scale=0.4]{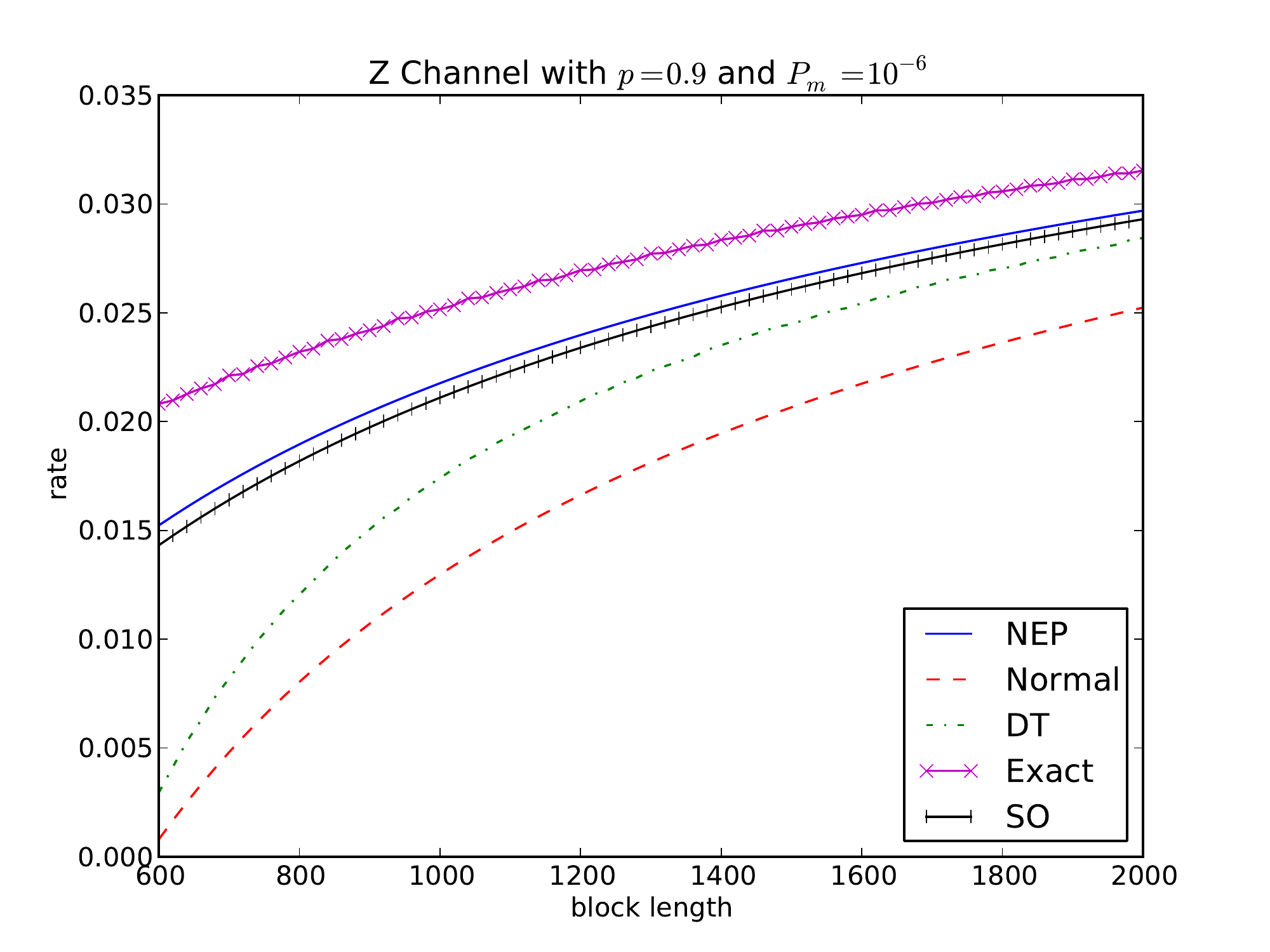}}
  \subfloat[$\delta_{t^*,n} (\epsilon)$ with $P_m = n^{-6}$]{\includegraphics[scale=0.4]{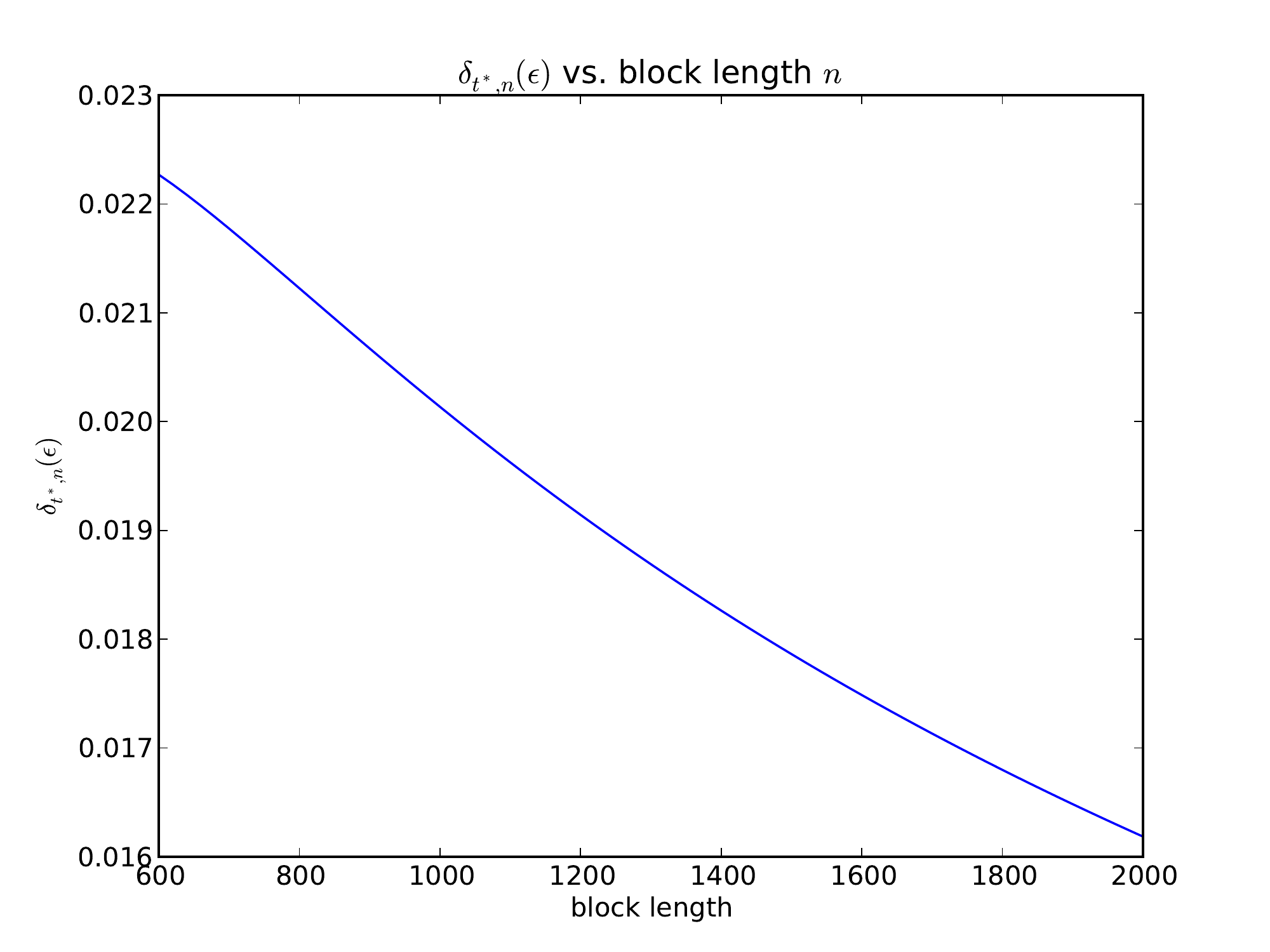}}
  \\
  \subfloat[Bounds with $P_m = g_{t^*;P,n} (\delta)$ and $\delta=0.02$]{\includegraphics[scale=0.4]{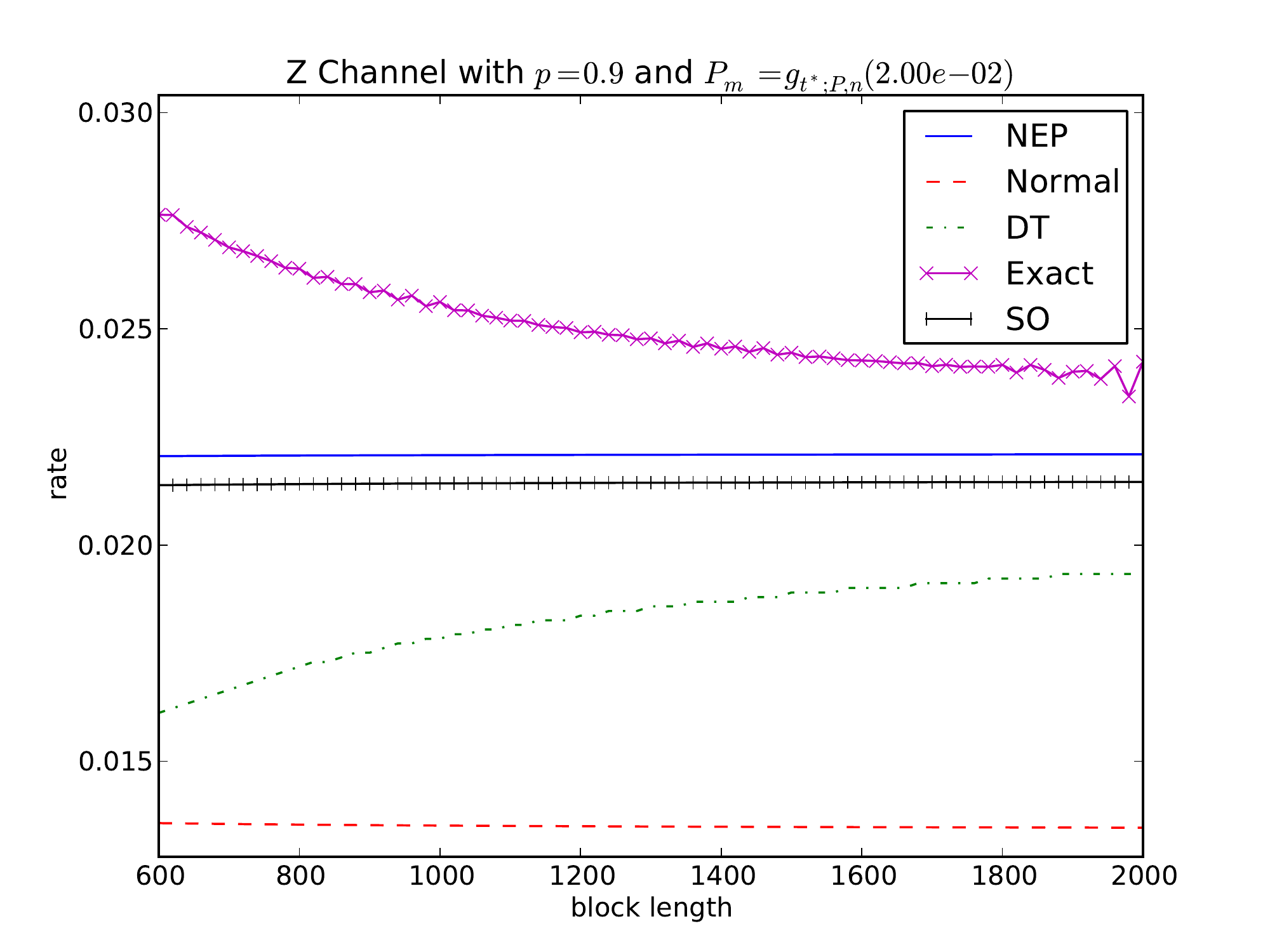}}
  \subfloat[$\log_{10} P_m$ with $P_m = g_{t^*;P,n} (\delta)$ and $\delta=0.02$]{\includegraphics[scale=0.4]{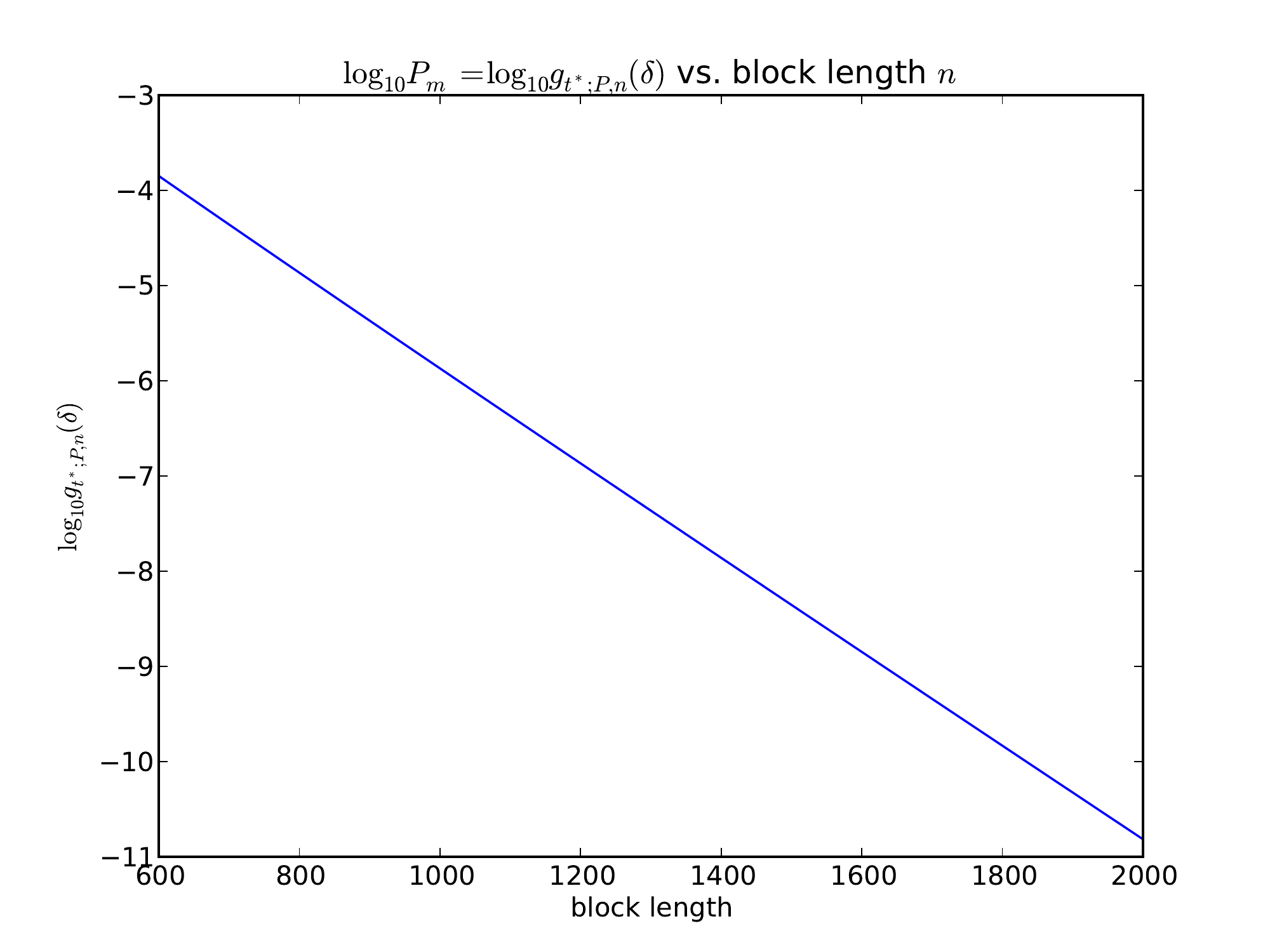}}
  \caption{Comparison of different bounds for Z Channel with $p=0.9$.}
  \label{fig-z-2}
\end{figure}

To provide benchmarks for the comparison of  approximation formulas, exact evaluation of \eqref{eq-thm-maxdsc-0}
(with $|\mathcal{X}|\frac{\ln (n+1)}{n}$ dropped and $t=t^*$) and \eqref{eq-thm-maxdsc-0+}
is provided, which, dubbed ``Exact'', serves as a converse bound, and
Theorem 22 in \cite{Yury-Poor-Verdu-2010} provides an achievable bound, dubbed
``DT'' and given below:
\begin{equation}
  \label{eq-nep-z-5}
  P_m \leq \sum^{m}_{i=0}
    \left(
      \begin{array}{c}
        m \\
        i
      \end{array}
    \right) (1-p)^{m-i} p^i \min \left\{ 1, (M-1)
              \frac{ \left(
                \begin{array}{c}
                  n-m+i \\
                  i
                \end{array}
                     \right)}
                     { \left(
                \begin{array}{c}
                  n \\
                  m
                \end{array}
                     \right)}
    \right\}
\end{equation}
where $M=2^{nR}$ and $m=t^* (0) n$. Figures~\ref{fig-z} and \ref{fig-z-2} again show that the Normal curve is all over the map while the NEP curve always lies in between the DT achievable curve and the Exact converse curve. It is also worth pointing out that if the capacity achieving distribution $t =p_X$ instead of $t^*$ was chosen in the calculation of the Exact and DT bounds, then both of them would be lower, confirming our early discussion that in the practical, non-asymptotic regime, the optimal marginal codeword symbol distribution is not necessarily a capacity achieving distribution.

\section{Conclusion}
\label{sec:conclusion}

In this paper, we have developed a new converse proof technique dubbed the outer mirror image of jar  and used it to establish new non-asymptotic converses for any discrete input memoryless channel with discrete or continuous output. Combining these non-asymptotic converses with  the non-asymptotic achievability proved in \cite{yang-meng:jardecoding} and \cite{yang-meng:isit2012_jardecoding} under jar decoding and with the NEP technique developed recently in \cite{yang-meng:nep},
we have characterized the best coding rate $R_n (\epsilon)$ achievable with finite block length $n$ and error probability $\epsilon$ through introducing a quantity $\delta_{t, n} (\epsilon)$ to measure the relative magnitude of the error probability $\epsilon$ and block length $n$ with respect to a given channel $P$ and an input distribution $t$. We have showed that in the non-asymptotic regime where both $n$ and $\epsilon$ are finite, $R_n (\epsilon)$ has a Taylor-type expansion with respect to $\delta_{t, n} (\epsilon)$, where the first two terms of the expansion are $\max_{t} [ I(t; P) - \delta_{t, n} (\epsilon) ] $, which is equal to $ I(t^*, P) - \delta_{t^*, n} (\epsilon) $ for some optimal distribution $t^*$, and the third order term of the expansion is $O(\delta^2_{t^*, n} (\epsilon)) $ whenever $\delta_{t^*, n} (\epsilon) = \Omega(\sqrt{ \ln n /  n})$.  Based on the new non-asymptotic converses and the Taylor-type expansion of $R_n (\epsilon)$, we have also derived  two approximation formulas (dubbed ``SO'' and ``NEP'') for $R_n (\epsilon)$. These formulas have been further evaluated  and compared against some of the best bounds known so far, as well as the normal approximation revisited recently in the literature. It turns out that while the normal approximation is all over the map, i.e. sometime below achievability and sometime above converse, the SO approximation is much more reliable and stays at the same relative position to achievable and converse bounds; in the meantime, the NEP approximation is the best among the three and always provides an accurate estimation for  $R_n (\epsilon)$.

It is expected that in the non-asymptotic regime where both $n$ and $\epsilon$ are finite, the Taylor-type expansion of $R_n (\epsilon)$ and the NEP approximation formula would play a role similar to that of Shannon capacity \cite{Shannon1948} in the asymptotic regime as $n \to \infty$. For values of $n$ and $\epsilon$ with practical interest for which $\delta_{t^*, n} (\epsilon)$ is not relatively small, the optimal distribution $t^*$ achieving $\max_{t} [ I(t; P) - \delta_{t, n} (\epsilon) ] $ is in general not a capacity achieving distribution except for symmetric channels such as binary input memoryless symmetric channels. As a result, an important implication arising from the Taylor-type expansion of $R_n (\epsilon)$ is that in the practical non-asymptotic regime, the optimal marginal codeword symbol distribution is not necessarily a capacity achieving distribution. Therefore, it will be interesting to examine all practical channel codes proposed so far against the Taylor-type expansion of $R_n (\epsilon)$ and the NEP approximation formula and to see how far their performance is away from that predicted by the Taylor-type expansion of $R_n (\epsilon)$ and the NEP approximation formula. If the performance gap is significant, one way to design a better channel code with practical block length and error probability requirement is to solve the maximization problem  $\max_{t} [ I(t; P) - \delta_{t, n} (\epsilon) ] $, get $t^*$, and then design a code so that its marginal codeword symbol distribution is approximately $t^*$.

Finally, we conclude this paper by saying a few words on non-asymptotic information theory. From the viewpoint of stochastic processes, most classic results in information theory are based, to a large extent, on the strong and weak laws of large numbers and on large deviation theory. For example, most first order asymptotic coding rate results in information theory were established through the applications of asymptotic equipartition properties and typical sequences \cite{cover:informtheory2006}, which in turn depend on the strong and weak laws of large numbers. On other hand, error exponent analysis in both source and channel coding is in the spirit of large deviation theory. The recent second order asymptotic coding rate results \cite{strassen-1962}, \cite{Hayashi-2009}, \cite{Yury-Poor-Verdu-2010} depend heavily on the Berry-Esseen central limit theorem. In the non-asymptotic regime of practical interest, however, none of these probabilistic tools can be applied directly. To fill in this void space, we have developed the NEP in \cite{yang-meng:nep}. Based on the NEP, we have further invented jar decoding in \cite{yang-meng:jardecoding} and presented the outer mirror image of jar converse proof technique in this paper. As demonstrated in this paper along with \cite{yang-meng:jardecoding} and \cite{yang-meng:nep}, the NEP, jar decoding, and the outer mirror image of jar together form a set of essential techniques needed for non-asymptotic information theory. They can also be extended and applied to help develop non-asymptotic multi-user information theory as well.

\appendices
\renewcommand{\theequation}{\Alph{section}.\arabic{equation}}
\setcounter{section}{0} \setcounter{equation}{0} %

\bibliographystyle{IEEEtran}
\bibliography{IEEEabrv,converse}

\end{document}